\documentclass[12pt]{article}

\global\arraycolsep=1pt
\usepackage{epstopdf}
\usepackage{amssymb}
\usepackage{amsmath}
\usepackage {pstricks}
\usepackage{graphicx,color}
\usepackage{mathrsfs}

\usepackage{caption} 
\renewcommand{\thefootnote}{\fnsymbol{footnote}}

\numberwithin{equation}{section}
\begin{document}

\newsavebox{\boxa}
\sbox{\boxa}{\includegraphics[width=83mm,bb=0 0 694 104]{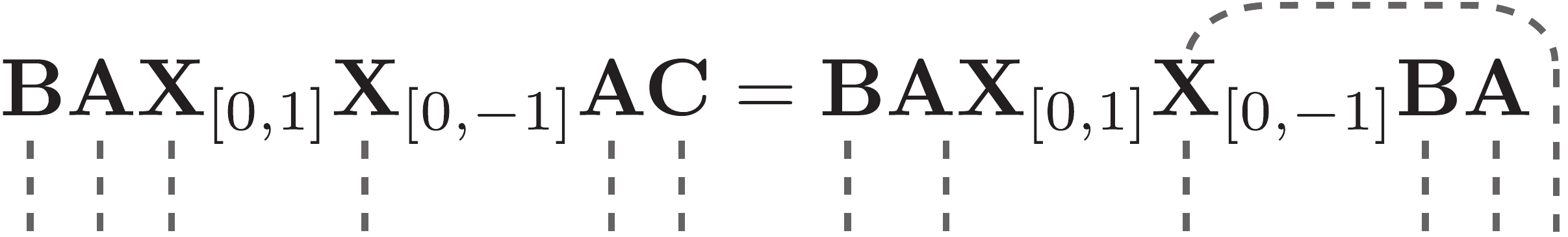}}
\newlength{\bw} 
\settowidth{\bw}{\usebox{\boxa}}
\newsavebox{\boxb}
\sbox{\boxb}{\includegraphics[width=116mm,bb=0 0 998 105]{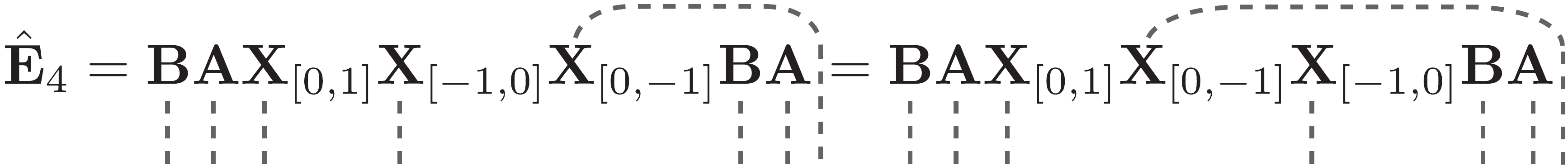}}
\newsavebox{\boxc}
\sbox{\boxc}{\includegraphics[width=64mm,bb=0 0 550 109]{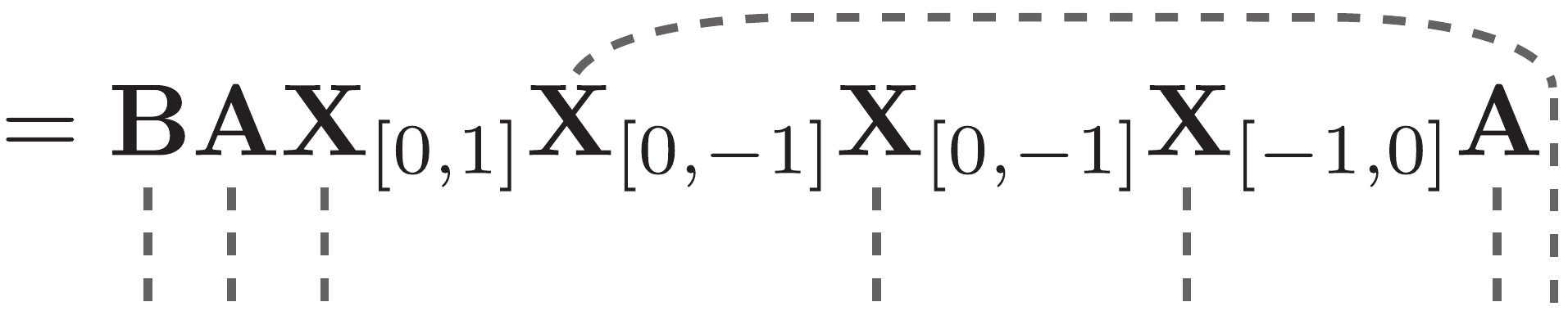}}
\newsavebox{\boxd}
\sbox{\boxd}{\includegraphics[width=84mm,bb=0 0 716 106]{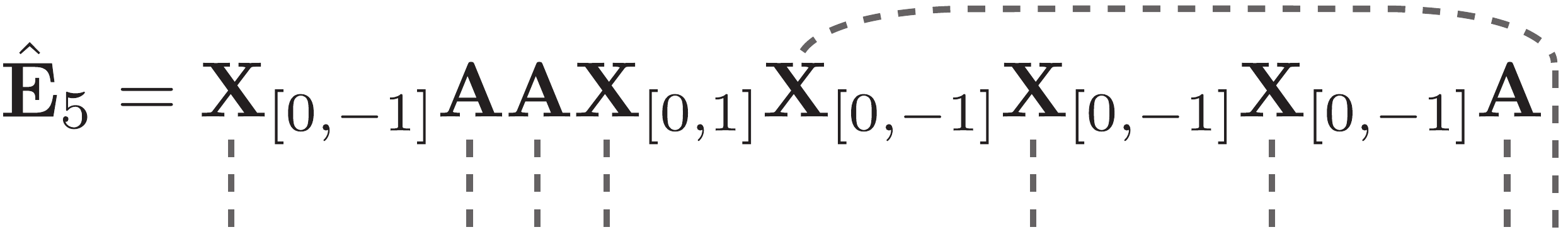}}
\newsavebox{\boxv}
\sbox{\boxv}{\includegraphics[width=24mm,bb=0 0 134 135]{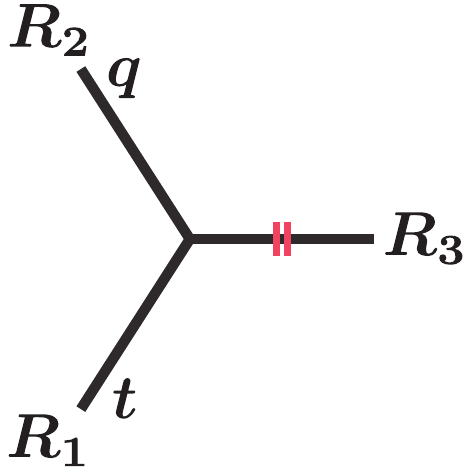}}

%%%%%%%%%%%%%%%%%%%%%%%%%%%%%%%%%%%%%%%%%%%%%%%%%%%%%%%%%%%%%%%%%%%%%%
%
\begin{titlepage}
\begin{flushright}
{RIKEN-MP-83}\\
\end{flushright}
\vspace{0.5cm}
\begin{center}
{\Large \bf 
\textsl{}
\textsc{Seiberg Duality, 
%and
 5d SCFTs
 \\  \vspace{0.3cm}and Nekrasov Partition Functions
 \vspace{0.3cm}
}}
\vskip1.0cm
{\large Masato Taki%\footnote[2]{Email: taki@riken.jp}
}
\vskip 0.8em
{\it
Mathematical Physics Lab., RIKEN Nishina Center,\\ 
Saitama 351-0198, Japan
}%
\\
\vspace{0.3cm}
{\tt taki@riken.jp}
\end{center}
\vskip1.0cm
%%%%%%

%%%%%%%%%
\begin{abstract}
It is known that a 4d $\mathcal{N}=1$ SCFT lives on D3-branes probing a local del Pezzo Calabi-Yau singularity.
The Seiberg (or toric) duality
of this SCFT arises from
the Picard-Lefshetz transformation of the affine $E_N$ 7-brane background
that is associated with the Calabi-Yau threefold. 
In this paper we study the duality of the affine $E_N$ background itself and a 5-brane probing it.
We then find that
many different Type IIB 5-brane webs describe the same SCFT in 5d.
We check this duality by comparing the Nekrasov partition functions
of these 5-brane web configurations.
\end{abstract}
\end{titlepage}

%%%%%%%%%%%%%%%%%%%%%%%%%%%%%%%%%%%%%%%%%%%%%%%%%%%%%%%%%%%%%%%%%%%%%%

\renewcommand{\thefootnote}{\arabic{footnote}} \setcounter{footnote}{0}

%%%%%%%%%%%%%%%%%%%%%%%%%%%%%%%%%%%%%%%%%%%%%%%%%%%%%%%%%%%%%%%%%%%%%%

\section{Introduction}

Developments in string theory have led to the existence of a great number of
non-trivial interacting conformal field theories,
which are not apparent from the framework of perturbative quantum field theory.
Superconformal field theories (SCFTs) in five dimensions (5d)
are typical examples.
In general, 5d gauge theory is non-renormalizable and trivial,
and such a theory cannot be a fundamental microscopic theory.  
However, by employing superstring theory, Seiberg \cite{Seiberg:1996bd} provided evidences
of the existence of many non-Gaussian fixed point in 5d.
The relevant deformation of such CFT
flows to certain 5d $\mathcal{N}=1$ gauge theory,
and actually this gauge theory is non-perturbatively well-defined
in spite of seeming non-renormalizability.

These 5d SCFTs, which are ultraviolet (UV) fixed point theories of 5d $\mathcal{N}=1$ gauge theories,
were studied from various viewpoints:
Type I'/heterotic duality \cite{Seiberg:1996bd,Douglas:1996xp},
M-theory on a Calabi-Yau singularity \cite{Morrison:1996xf,Intriligator:1997pq},
Type IIB 5-brane web configuration \cite{Aharony:1997ju,Aharony:1997bh,Bao:2011rc}
and Type IIB 7-brane background \cite{DeWolfe:1999hj,Yamada:1999xr,Hauer:1999pt,Mohri:2000wu}.
In this paper, we employ the Calabi-Yau compactification, 5-brane web and 7-brane realization
in order to study 5d SCFTs,
and we then find that the branch cut move (the Picard-Lefshetz transformation) \cite{Gaberdiel:1997ud,Gaberdiel:1998mv,DeWolfe:1998zf,DeWolfe:1998eu,DeWolfe:1998pr} 
of 7-brane configurations
leads to non-trivial duality between web configurations
of Calabi-Yau manifolds, and therefore the corresponding 5d SCFTs.
This duality in 5d is deeply related to the Seiberg duality between 4d quiver gauge theories which
are realized as
worldvolume theories of D3-branes on Calabi-Yau singularities.

Let us consider two equivalent Calabi-Yau singularities
that are related through 7-brane move.
It was observed  that two 4d gauge theories on D3-branes probing them are Seiberg-dual
to each other \cite{Feng:2000mi,Feng:2001xr,Hanany:2001py,Feng:2001bn,Feng:2002zw,Feng:2002kk,Franco:2002ae,Feng:2002fv,Franco:2002mu}.
In this sense, our duality between these Calabi-Yau manifolds is a parent of this 4d Seiberg duality.
As we will explain,
a generic dual pair of Calabi-Yau manifolds are not completely equivalent because these compactifications
lead to decoupled extra states \cite{Bergman:2013ala,BMPTY,HKN,Taki:2013vka,Bergman:2013aca}.
In 5d duality, we have to remove this extra contribution to formulate the duality,
but the 4d Seiberg duality is very simple since a D3-brane world-volume theory does not feel these extra degrees of freedom.

In this paper we study the 5d field theories 
arising from M-theory compactified on the local del Pezzo surfaces $dP_{1,2,\cdots,6}$.
In general, a local del Pezzo surface is not toric,
and therefore we do not have efficient way to compute the corresponding partition function.
If a Calabi-Yau is toric, we can utilize the topological vertex formalism
\cite{Nekrasov:2002qd,Iqbal:2002we,Iqbal:2003ix,Aganagic:2003db,Iqbal:2004ne,Awata:2005fa,Iqbal:2007ii,Taki:2007dh,Awata:2008ed,Iqbal:2012mt}
to calculate exactly its partition function.
For a local del Pezzo surface,
we  find local pseudo del Pezzo surfaces $PdP_k^{p=I,II,\cdots}$
that are toric and dual to the local del Pezzo surface $dP_k$.
We then conjecture that the del Pezzo partition function is given by 
the toric partition functions of the pseudo del Pezzo surfaces
through the simple relation 
\begin{align}
Z_{dP_k}=\frac{Z_{PdP_k^{p}}}{Z^{PdP_k^{p}}_{\textrm{extra}}},\quad
p=I,\,II,\,\cdots,
\end{align}
where $p$ labels the corresponding pseudo del Pezzo surfaces.
The point is that the discrepancy between $dP_k$ and $PdP_k^{p}$
is only an overall factor $Z^{PdP_k^{p}}_{\textrm{extra}}$ in all cases.
This extra factor arises from the above-mentioned extra states in the 5d spectrum, which do not transform
correctly under the 5d Lorentz group.
In this paper, we show that this nontrivial relation actually holds for all possible cases $dP_{1,2,\cdots,6}$.

This paper is organized as follows. 
We give a brief review on 5d field theories associated with 5-brane web configurations in section 2. 
In section 3, we study the relation between web configurations by employing 7-brane picture of local Calabi-Yau compactification.
We then conjecture new relation between the Nekrasov partition functions of the Calabi-Yau manifolds associated with  a del Pezzo surface.
In section 4 we check this conjecture based on  instanton expansion.
We conclude in section 5. 
In appendix A and B, we set some conventions, deriving useful formulas.

%%%%%%%%%%%%%%%%%%%%%%%%%%%%%%%%%%%%%%%%%%%%%%%%%%%%%%%%%%%%%%%%%%%%%%
\section{Five-dimensional theories and 5-brane webs}
\begin{figure}[tb]
 \begin{center}
\includegraphics[width=10cm, bb=0 0 546 213]{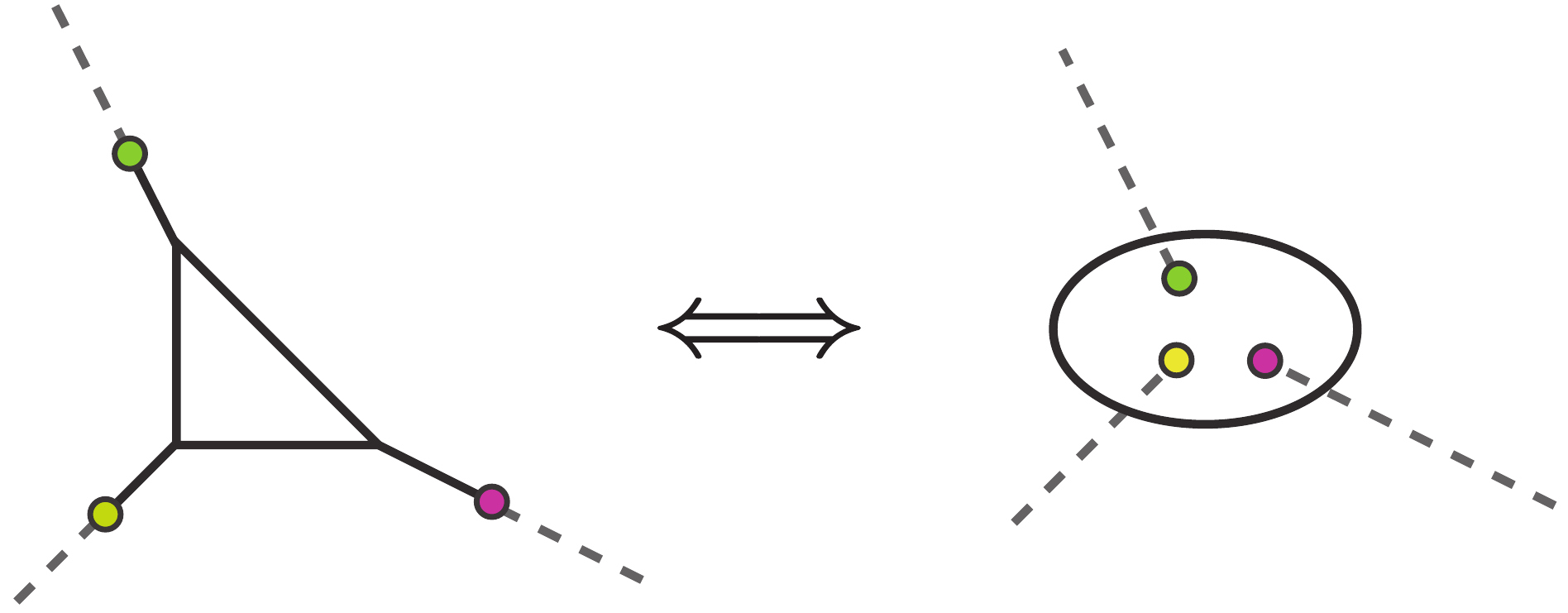}
 \end{center}
\caption{The left hand side is the 5-brane web dual to the local  $\mathbb{CP}^2$ geometry.
We regularize an external leg by terminating it on a 7-brane.
A colored circle is a 7-brane, and a dashed lines is branch cut arising from it.
Moving these three 7-branes inside the 5-brane loop yields
the right hand side through the Hanany-Witten effect.}
 \label{fig;dP0}
\end{figure}
5d $\mathcal{N}=1$ $SU(2)$ gauge theories and their UV fixed point SCFTs
are our main focus in this paper.
We have many stingy realizations of such SCFTs and corresponding gauge theories.
A well-known method to derive these fields theories from a string setup
is using the $(p,q)$ 5-brane web configurations in Type IIB superstring theory  \cite{Aharony:1997ju,Aharony:1997bh,DeWolfe:1999hj}.

By considering $SL(2,\mathbb{Z})$ duality acting on a D5-brane,
we can find there exists a $(p,q)$ 5-brane with generic Ramond-Ramond and NS-NS charges.
Let us consider the planar web configurations of these 5-branes.
We assume that $x^{0,1,\cdots,4}$-dimensions are shared by all 5-branes
and these 5-branes form a planar graph in the $x^{5}$-$x^6$ plane.
A generic web is constructed by gluing trivalent vertex of three $(p_i,q_i)$ 5-branes.
Because of the charge conservation,
we have to impose
\begin{align}
\sum_{i=1}^3 p_i=0=\sum_{i=1}^3 q_i.
\end{align}
To maintain a quarter of the original 32 supercharges,
we have to impose the condition
that the slope of a 5-brane in a planar diagram
is given by its charge vector $(p_i,q_i)$\footnote{In this convention,
we consider the simple choice of the Type IIB coupling $\tau_{\textrm{IIB}}=i$.
 The web diagrams take warped shapes in generic coupling,
 but this is irrelevant to our analysis.
}.
We then find 5d $\mathcal{N}=1$ field theories on their world-volumes.

M-theory compactified on a toric Calabi-Yau manifold also gives 5d $\mathcal{N}=1$  theory.
The toric Calabi-Yau three-folds are specified by the web diagrams
up to $SL(2,\mathbb{Z})$ symmetry of their toric datas.
An important fact is that a 5-brane web
and the toric compactification specified by the same web diagram
lead to the same stringy system because of  string dualities.
We can thus easily recast a 5-brane web system
into the corresponding toric Calabi-Yau compactification.
We therefore consider a web system without distinction
between 5-brane configuration and toric Calabi-Yau geometry.

Let us consider an extension of these 5-brane systems.
In Type IIB superstring theory, there exist 7-branes with generic $(p,q)$ charges
which originate from the $SL(2,\mathbb{Z})$ transformation of a D7-brane.
We can terminate a 5-brane in a web on a 7-brane
stretching to $x^{0,1,\cdots,4,7,8,9}$ directions \cite{DeWolfe:1999hj}.
This modification does not break supersymmetry,
and moreover we can replace all external 5-brane legs 
with finite legs ending on 7-branes without changing the resulting 5d field theory\footnote{As we will see in this paper,
the cases involving adjoining parallel legs are exceptions.}. 
This means that the 5d theory is independent of the lengths of these finite legs.
\textbf{\textit{Figure}$\,$\textit{\ref{fig;dP0}}} illustrates the 5-brane web of the local 
$\mathbb{CP}^2$ Calabi-Yau\footnote{Since a 5-brane web is dual to
the toric Calabi-Yau for the same web diagram
we call the 5-brane web by the name of the Calabi-Yau.}
modified by three 7-branes.
The dashed lines are branch cuts created by 7-branes.
Since 7-branes can move along the corresponding 5-brane legs,
we can move all the 7-branes into the center of the web.
When a 7-brane passes the 5-brane loop,
the anti-Hanani-Witten mechanism occurs and the external leg attached to this 7-brane disappears.
The resulting configuration is illustrated in the right hand side of \textbf{\textit{Figure}$\,$\textit{\ref{fig;dP0}}}.
This system is a 5-brane loop proving a 7-brane background configuration.
This representation as a 7-brane background configuration is not unique
because a 7-brane changes its $(p,q)$-charges when it passes a branch cut of an another 7-brane.
We can therefore find a different 7-brane configuration by changing the ordering of these 7-branes.
This property of branch cut move plays a key role in our analysis.
Some basic facts including the rule of branch cut move
are reviewed in Appendix.A.

\begin{figure}[tbp]
 \begin{center}
  \includegraphics[width=115mm,bb=0 0 727 592]{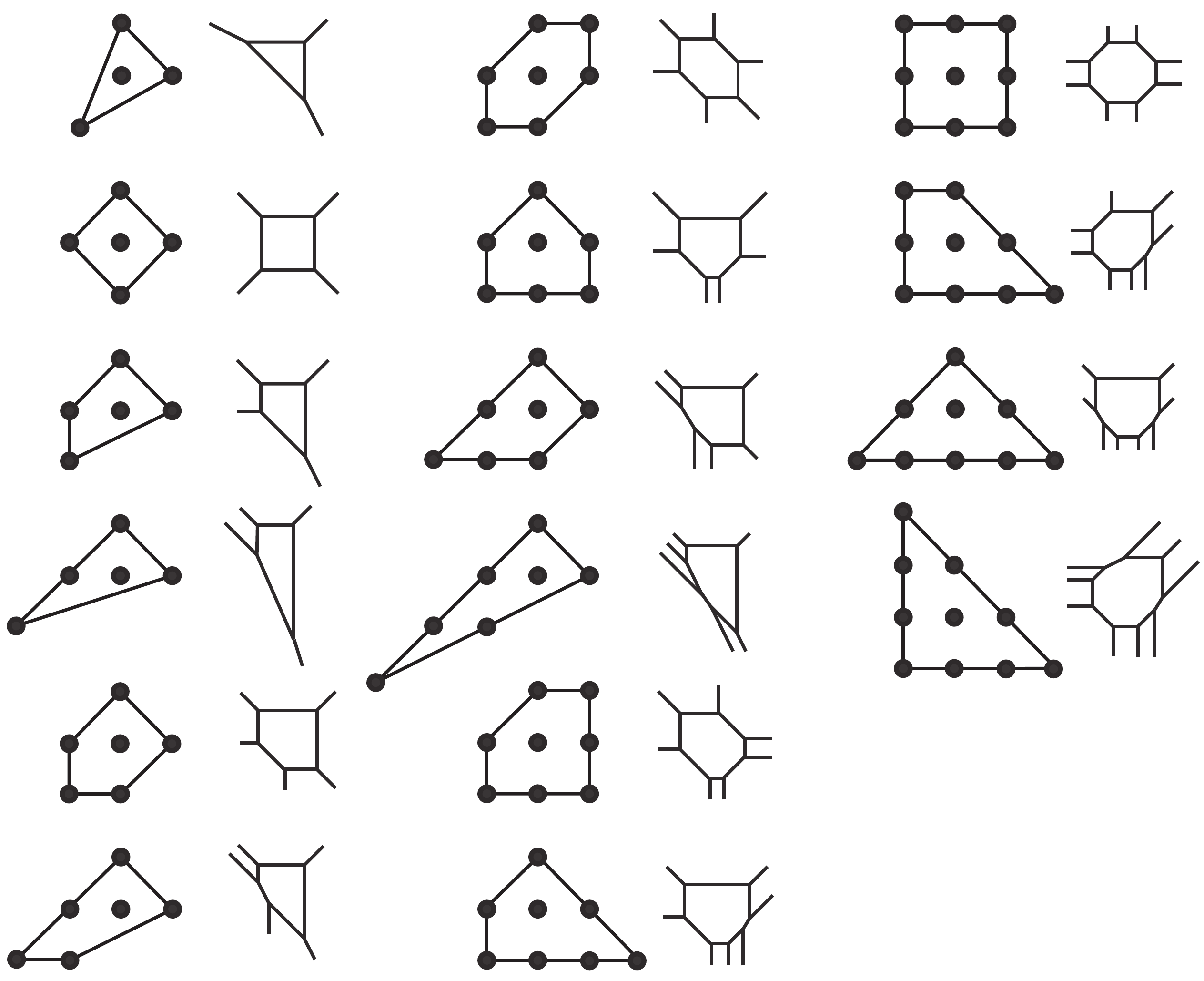}
 \end{center}
 \caption{All the $GL(2,\mathbb{Z})$-inequivalent convex lattice polygons with single internal point
 and their dual web diagrams.}
 \label{fig;grids}
\end{figure}
\section{Toric phases of local (pseudo) del Pezzo surface}
In this paper,
we study the 5-brane web configurations with
single loop.
Since Type IIB superstring theory enjoys $SL(2,\mathbb{Z})$ duality,
we consider only the $SL(2,\mathbb{Z})$-inequivalent configurations.
The reflection is also irrelevant to our analysis,
we consider
the general linear group duality transformation
\begin{align}
GL(2,\mathbb{Z})=\left\{ \left(\begin{array}{ccc}a &\,\,& b \\c &\,\,& d\end{array}\right)\Biggr| 
ad-bc=\pm1,\,\,a,b,c,d\in\mathbb{Z}\right\}.
\end{align}
We can classify the $GL(2,\mathbb{Z})$-inequivalent webs by considering the dual grid diagrams as 
\textbf{\textit{Figure}$\,$\textit{\ref{fig;grids}}}.
There are sixteen inequivalent convex lattice polygons with single internal point,
and this means that there are sixteen inequivalent physical systems in Type IIB.
These web configurations are illustrated in \textbf{\textit{Figure}$\,$\textit{\ref{fig;grids}}}.
The web with three external legs corresponds to the local $\mathbb{CP}^2$ geometry
and does not associated with 5d gauge theory.
We therefore consider the remaining webs.

In the studies on 4d quiver gauge theories
\cite{Feng:2000mi,Feng:2001xr,Hanany:2001py,Feng:2001bn,Feng:2002zw,Feng:2002kk,Franco:2002ae,Feng:2002fv,Franco:2002mu},
it was pointed out that
some of these toric Calabi-Yau manifolds lead to 
Seiberg-dual pair of 4d theories.
Recently, all the 4d $\mathcal{N}=1$ quiver gauge theories associated with these toric manifolds
were determined in \cite{Hanany:2012hi}.
In this paper, we re-examine the relation between these toric manifolds 
from the recent perspectives of 5d SCFTs and related string setups,
and then find new duality between Calabi-Yau compactifications of M-theory.

\subsection{First del Pezzo surfaces $\boldsymbol{dP_1}$ and $\boldsymbol{\widetilde{dP_1}}$}

The first del Pezzo surfaces correspond to 
the 5-brane webs  in \textbf{\textit{Figure}$\,$\textit{\ref{fig;grids}}} with four external legs.
These configurations describe 5d theories whose flavor symmetry is rank-one.
There are three webs,
however, there are only two known 5d SCFT with rank-one flavor symmetry.
In the following we explain the origin of this mismatch 
by showing two of these three webs are actually dual to each other.

\subsubsection*{$\boldsymbol{dP_1}$: local $\boldsymbol{\mathcal{B}_1$ (or $\mathbb{F}_1}$) surface}
\begin{figure}[tbp]
 \begin{center}
  \includegraphics[width=80mm, bb=0 0 451 355]{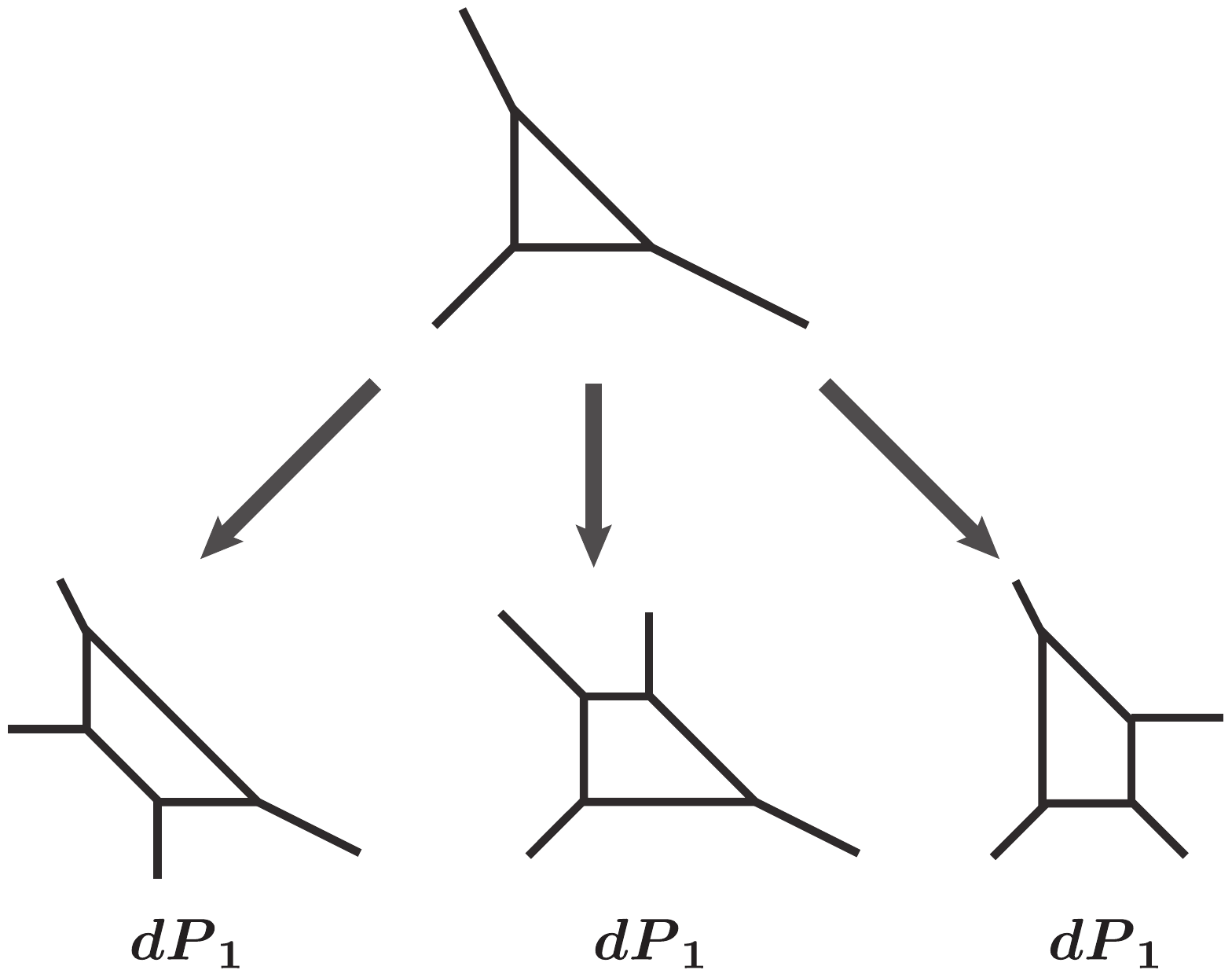}
 \end{center}
 \caption{Blowup in a toric web diagram. The local Calabi-Yau for the
 first del Pezzo  ${\mathcal{B}}_1$ is the one point blowup of the local $\,\mathbb{P}^2$.
 The resulting three toric diagrams are equivalent up to the $SL(2,\mathbb{Z})$ symmetry transformation.}
 \label{fig;dP0dP1}
\end{figure}
The first del Pezzo surface $\mathcal{B}_1$ is the one point blowup of $\mathbb{CP}^1$.
Using the $SL(3,\mathbb{C})$ symmetry on $\mathbb{CP}^1$,
we can move three generic points to the three corners in the toric diagram without loss of generality.
The local  $\mathcal{B}_1$ Calabi-Yau threefold is given by the blowup of the toric web diagram of
the local $\mathbb{CP}^1$ as
 \textbf{\textit{Figure}$\,$\textit{\ref{fig;dP0dP1}}}.
The three choices of  blowup point
lead to three local del Pezzo $dP_1$.
Since a toric diagram specifies the corresponding Calabi-Yau threefold up to the action
of $SL(2,\mathbb{Z})$ symmetry transformation.
The three webs in  \textbf{\textit{Figure}$\,$\textit{\ref{fig;dP0dP1}}} are actually the same geometry,
and so there is only the unique toric phase of the local first del Pezzo surface.
This toric geometry is also known as the local Hirzebruch surface $\mathbb{F}_1$.
The compactification of M-theory on this Calabi-Yau manifold yields the 5d $\hat{E}_1$ SCFT \cite{Seiberg:1996bd,Morrison:1996xf}
whose flavor symmetry is $U(1)$.

\subsubsection*{$\boldsymbol{\widetilde{dP}_1}$: local $\boldsymbol{\tilde{\mathcal{B}}_1$ (or $\mathbb{F}_0}$) surface}
\begin{figure}[tbp]
 \begin{center}
  \includegraphics[width=100mm, bb=0 0 529 207]{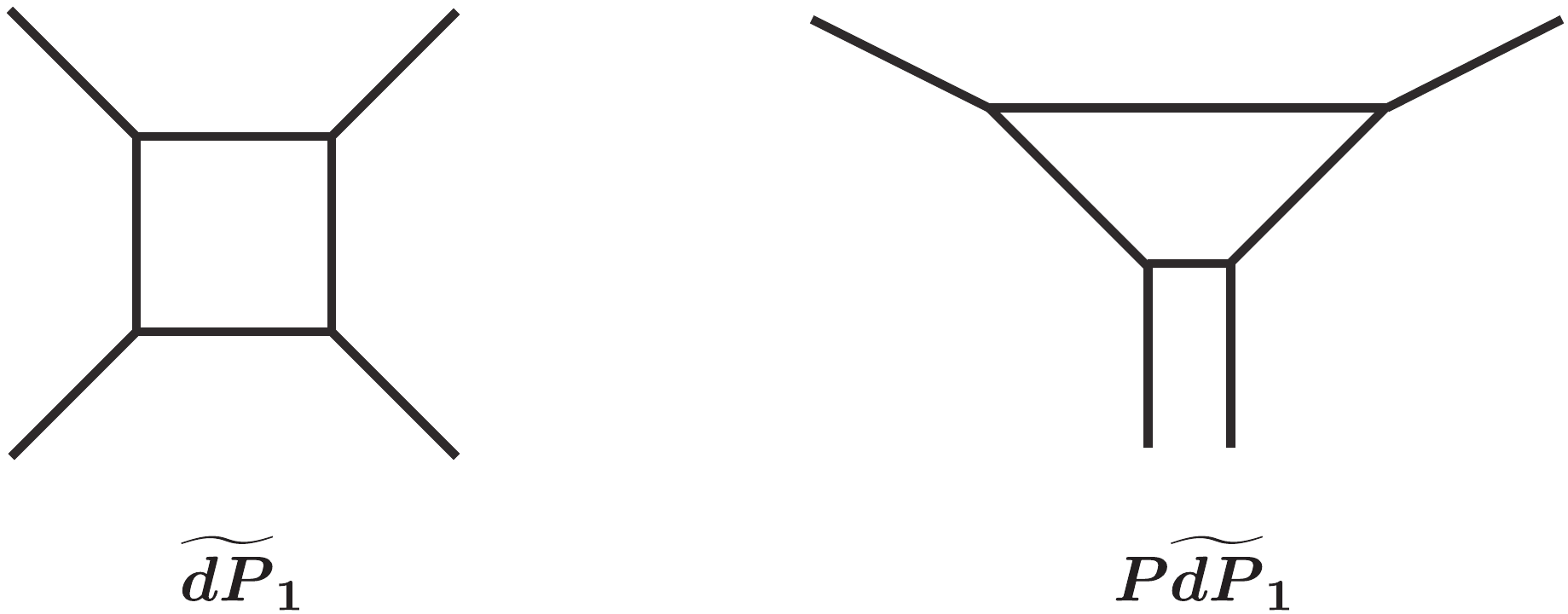}
 \end{center}
 \caption{The two toric phases of the local geometry of the another first del Pezzo $\tilde{\mathcal{B}}_1$.}
 \label{fig;dP1tilphases}
\end{figure}
There is the another class of the first del Pezzo surface $\tilde{\mathcal{B}}_1$
that coincides with the Hirzebruch surface $\mathbb{F}_0=\mathbb{CP}^1\times \mathbb{CP}^1$.
The symbol $\widetilde{dP}_1$ in this paper  denotes the local $\mathbb{F}_0$ geometry
whose web diagram is illustrated in the left side of \textbf{\textit{Figure}$\,$\textit{\ref{fig;dP1tilphases}}}.
The recent work \cite{Bergman:2013ala,Taki:2013vka} showed that
other toric geometry $P\widetilde{dP}_1$
that is the local geometry of $\mathbb{F}_2$ leads to the same compactified string theory as that of $\widetilde{dP}_1$
after removing certain extra contribution.
This conjecture is stated as the following equivalence between their BPS Nekrasov partition functions
\begin{align}
\label{ZdP1til}
Z_{\,\widetilde{dP}_1}(Q_F,u;t,q)=\frac{Z_{\,P\widetilde{dP}_1}(Q_F,u;t,q)}{
Z_{\,\textrm{extra}}^{\,P\widetilde{dP}_1}(u;t,q)}.
\end{align}
$Z_{\,\textrm{extra}}$ is the partition function of the extra contribution given in \cite{Taki:2013vka,Bergman:2013ala}.
$Q_F=e^{2ia}$ is the fugacity associated with the Cartan of the $SU(2)$ gauge group, that is the Coulomb
branch parameter,
and $t$ and $q$ are the exponentiated $\Omega$-background parameters.
The charge associated with the instanton current $J=\ast \textrm{tr}F\wedge F$
is counted by the instanton factor $u$.
The Nekrasov partition function for a toric Calabi-Yau threefold is computed by using
the refined topological vertex formalism \cite{Iqbal:2007ii},
and the algorithm to compute the corresponding extra contribution  $Z_{\,\textrm{extra}}$ for a given toric Calabi-Yau
is given in \cite{BMPTY,HKN,Taki:2013vka}.
We will review the check of the conjecture (\ref{ZdP1til}) in the next section.

This equivalence means that the 5d field theories arise from the two 5-brane web
configurations, namely $\widetilde{dP}_1$ and
$P\widetilde{dP}_1$, are the same quantum field theory
up to essentially decoupled\footnote{Since the factorization (\ref{ZdP1til})
is satisfied by the Nekrasov partition function and superconformal index,
the extra contribution is decoupled from the main 5d theory as far as the BPS sector is concerned.
} extra contributions.
We can actually derive the following equivalence between the superconformal indexes \cite{Kim:2012gu,Iqbal:2012xm}
by using the conjectural relation (\ref{ZdP1til})
\begin{align}
I_{\,\widetilde{dP}_1}=\frac{I_{\,P\widetilde{dP}_1}}{I_{\,\textrm{extra}}}.
\end{align}
The web configurations  therefore lead to 
the same 5d UV fixed point superconformal field theory $E_1$.

\begin{figure}[tb]
 \begin{center}
\includegraphics[width=10cm, bb=0 0 435 236]{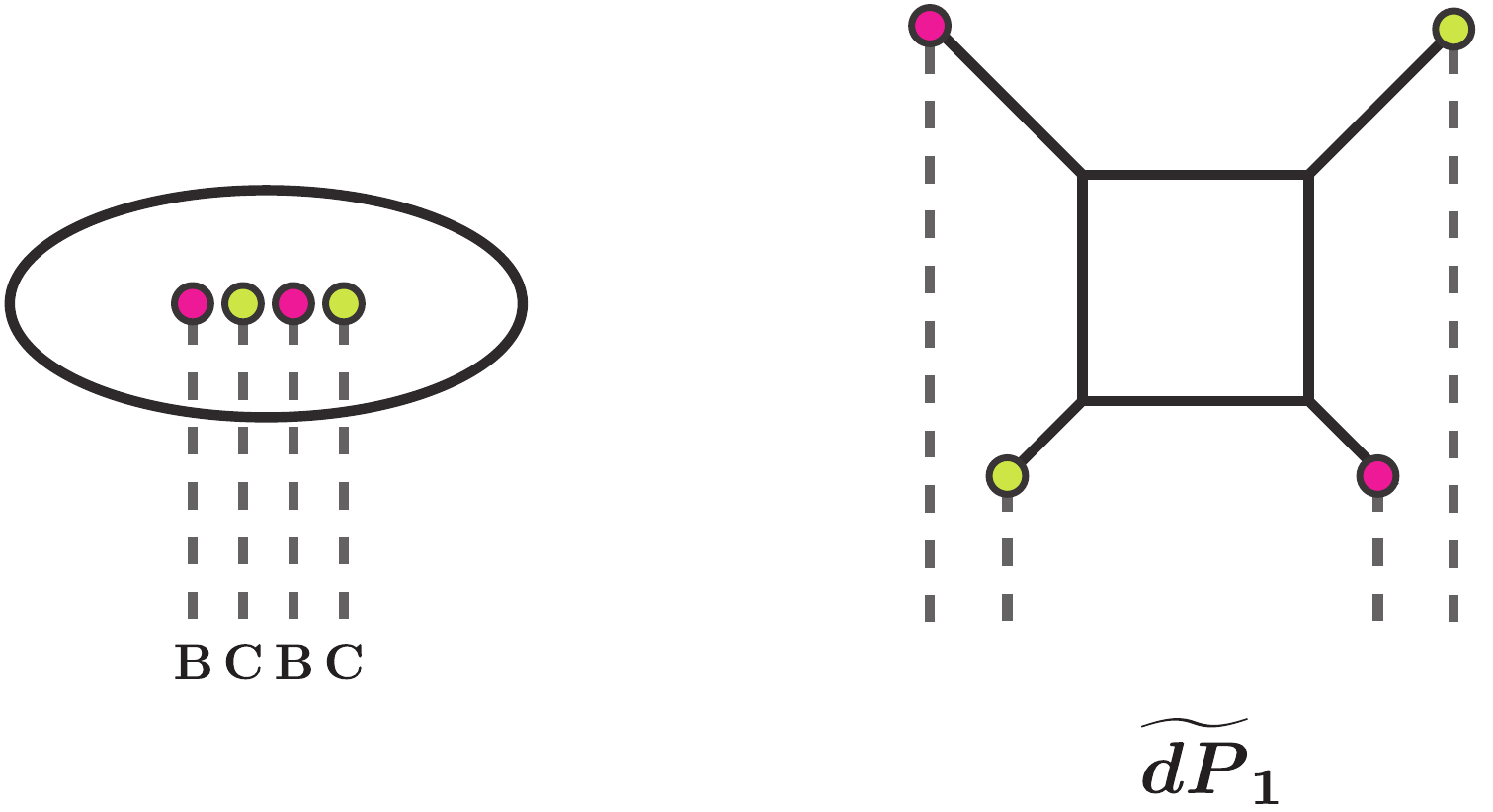}
 \end{center}
\caption{The left hand side is the 5-brane loop probe of $\,\,\bf \hat{E}_1$ 7-brane configuration.
Since all the 7-branes are not collapsible, only a sub-algebra $E_1$ is realized on the 5-brane.
The right hand side is the corresponding web diagram which is obtained by moving a 7-brane the outside of the loop
along the associated geodesic.}
 \label{fig;dPtilI}
\end{figure}
In the following, we will give a physical explanation of this non-trivial equivalence between 
these different Calabi-Yau compactifications and the corresponding 5d effective field theories.
The key is the fact that
a toric Calabi-Yau compactification of M-theory is dual the IIB 5-brane web configuration,
and we can introduce the 7-branes on the edges of the external 5-branes
to make infinitely long external line finite length \cite{DeWolfe:1999hj}.
In the case of the local zeroth Hirzebruch $\widetilde{dP}_1$,
the regularized 5-brane configuration is illustrated in the right  side of  \textbf{\textit{Figure}$\,$\textit{\ref{fig;dPtilI}}}.
We introduce two types of 7-branes $\bf B$ and $\bf C$
to replacing the two types of the infinitely-long 5-branes into  two types of finite 5-branes.
A 7-brane creates a branch cut, and it is illustrated by the short dashed line in \textbf{\textit{Figure}$\,$\textit{\ref{fig;dPtilI}}}.
In this figure we ignore a non-trivial metric created by the 7-brane background because only the asymptotic shape of web
 is important in our analysis.
 
 \begin{figure}[tb]
 \begin{center}
\includegraphics[width=10cm, bb=0 0 452 215]{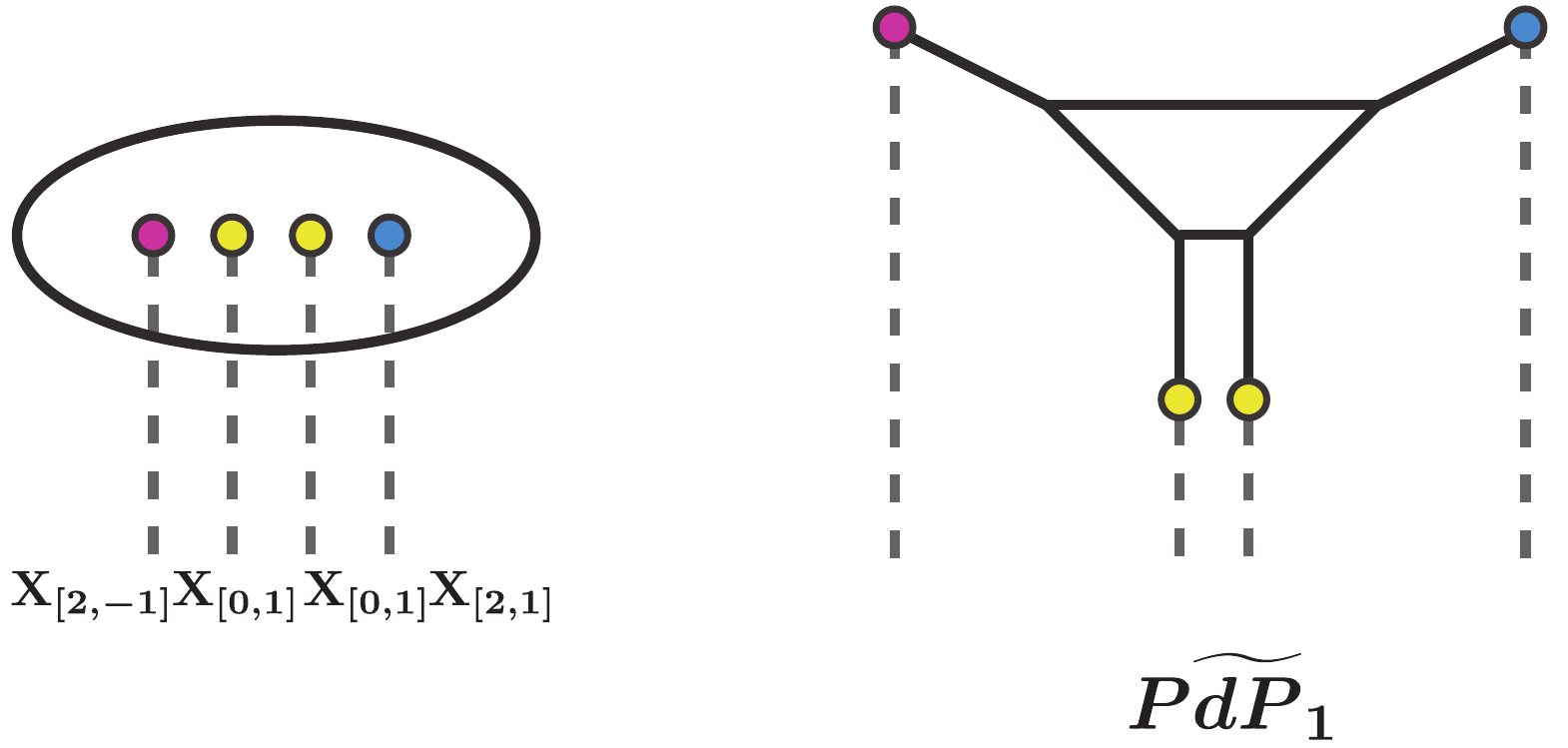}
 \end{center}
\caption{
The left hand side is the 5-brane loop probe of 
a reordered $\,\,\bf \hat{E}_1$ 7-brane configuration.
The right hand side is the corresponding web diagram which is obtained by moving a 7-brane the outside of the loop.
This is the web of $P\widetilde{dP}_1$ modified by four 7-branes.
}
 \label{fig;dPtilII}
\end{figure}
The 5d field theory is independent of the length of the external legs.
We can therefore move these 7-branes  inside of the 5-brane loop.
By using Hanany-Witten effect
in an inverted way,
we can see that the 5-brane prongs disappear when the 7-branes cross the 5-brane loop.
The resulting configuration is illustrated in the left side of  \textbf{\textit{Figure}$\,$\textit{\ref{fig;dPtilII}}}.
This is the 5-brane loop probe of the 7-brane background $\bf BCBC$.
This 7-brane configuration is named as $\hat{\bf E}_1$
\begin{align}
\hat{\bf E}_1
&\equiv
{\bf B}{\bf C}{\bf B}{\bf C}.
\end{align}
In this paper, each branch cut extends downward from the base 7-brane except as otherwise specially provided.
By reordering the 7-branes in this configuration,
we can explain why the two local Hirzebruch surfaces give the same BPS spectrum and 5d field theory.
Since a 7-brane creates the branch cut,
reordering of 7-branes also transforms the $(p,q)$-charges of them.
The basic properties and rules are collected in Appendix.A.
Let us move the 7-branes by using these rules.
Starting with  $\hat{\bf E}_1$ configuration, we find
\begin{align}
\hat{\bf E}_1
&\equiv
{\bf B}{\bf C}{\bf B}{\bf C}
={\bf B}{\bf C}{\bf C}{\bf X}_{[3,1]}
\overset{\scriptscriptstyle T^{-1}}\simeq {\bf X}_{[2,-1]}{\bf X}_{[0,1]}{\bf X}_{[0,1]}{\bf X}_{[2,1]}.
\end{align}
In the last equality we use the $SL(2,\mathbb{Z})$ dual transformtion
\begin{align}
T=\left(\begin{array}{cc}1\,\,\, &0\\ 1 \,\,\,& 1\end{array}\right).
\end{align}
This new configuration is shown in the left hand side of  \textbf{\textit{Figure}$\,$\textit{\ref{fig;dPtilII}}}.
We can move the 7-branes outside of the loop,
and we get the web diagram in the right hand side of  \textbf{\textit{Figure}$\,$\textit{\ref{fig;dPtilII}}}.
This is precisely the (toric) web diagram for the local second Hirzebruch $P\widetilde{dP}_1$ up to the 7-brane regularization.

This geometry $P\widetilde{dP}_1$ is not the genuine del Pezzo surface.
We therefore call the base of this local geometry the pseudo first del Pezzo surface.
This local pseudo del Pezzo surface $P\widetilde{dP}_1$ is a different geometry from the local del Pezzo $\widetilde{dP}_1$,
 however, these systems will be identical once the 7-branes are introduced in the dual 5-brane web picture.

\begin{figure}[tb]
 \begin{center}
\includegraphics[width=16cm, bb=0 0 746 163]{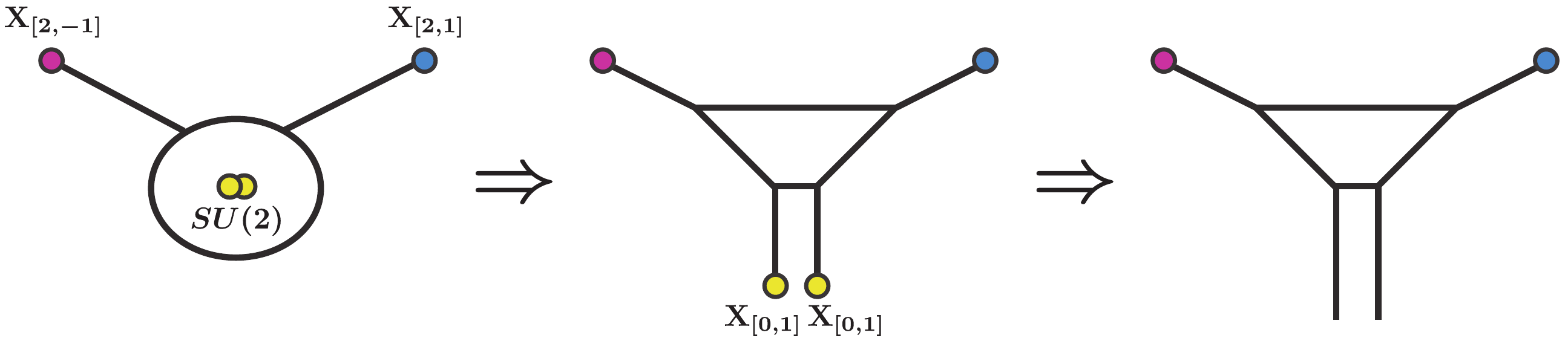}
 \end{center}
\caption{
If two $ {\bf X}_{[2,-1]}$ branes move away to infinity,
two adjoining parallel legs appear as illustrated in the right hand side.
There exist extra states propagating along this infinitely-long additional dimension,
and this resulting system therefore contains surplus spectrum.
}
 \label{fig;dPtilII2}
\end{figure}
We can remove $ {\bf X}_{[2,-1]}$  and $ {\bf X}_{[2,1]}$ branes safely,
however we can not do ${\bf X}_{[0,1]}$ branes.
This is because the attached prongs form the stack of two parallel 5-branes,
and so new six-dimensional states appear once we move the 7-branes to infinity \textbf{\textit{Figure}$\,$\textit{\ref{fig;dPtilII2}}}.
In this sense the two toric phases  $\widetilde{dP}_1$ and $P\widetilde{dP}_1$ are not completely identical,
but the fact pointed out in  \cite{Bergman:2013ala,BMPTY,HKN,Taki:2013vka}
is that the extra states are essentially decoupled from the 5d theory
and we can eliminate the extra contribution as (\ref{ZdP1til}).
By generalizing this idea,
we can find infinitely many dual toric phases which are related by the equivalence relation like (\ref{ZdP1til}).
The phases discussed in the following provide simplest examples of such extension.

The trace of the monodormy matrix around ${\bf X}_{[0,1]}{\bf X}_{[0,1]}$ is
\begin{align}
\textrm{Tr}K_{[0,1]}K_{[0,1]}=2,
\end{align}
and then ${\bf X}_{[0,1]}{\bf X}_{[0,1]}$ pair is collapsible as \textbf{\textit{Figure}$\,$\textit{\ref{fig;dPtilII2}}}.
In this picture this stack of 7-branes gives the enhanced $SU(2)$ symmetry,
and therefore 5d field theory should lose this symmetry once these two branes are removed to obtain  $P\widetilde{dP}_1$
toric geometry.
Actually, the superconformal index $I_{\,P\widetilde{dP}_1}$ does not enjoy $SU(2)$ flavor symmetry,
and we can recover the index $I_{\,\widetilde{dP}_1}$ with $SU(2)$  symmetry by removing the extra contribution.

%%%%%%%%%%%%%%%%%%%%%%%%%%%%%%%%%%%%%%
\subsection{Second del Pezzo surface $\boldsymbol{dP_2}$}

\subsubsection*{Toric del Pezzo phase $\boldsymbol{dP_2}$}

\begin{figure}[tb]
 \begin{center}
\includegraphics[width=14cm, bb=0 0 604 243]{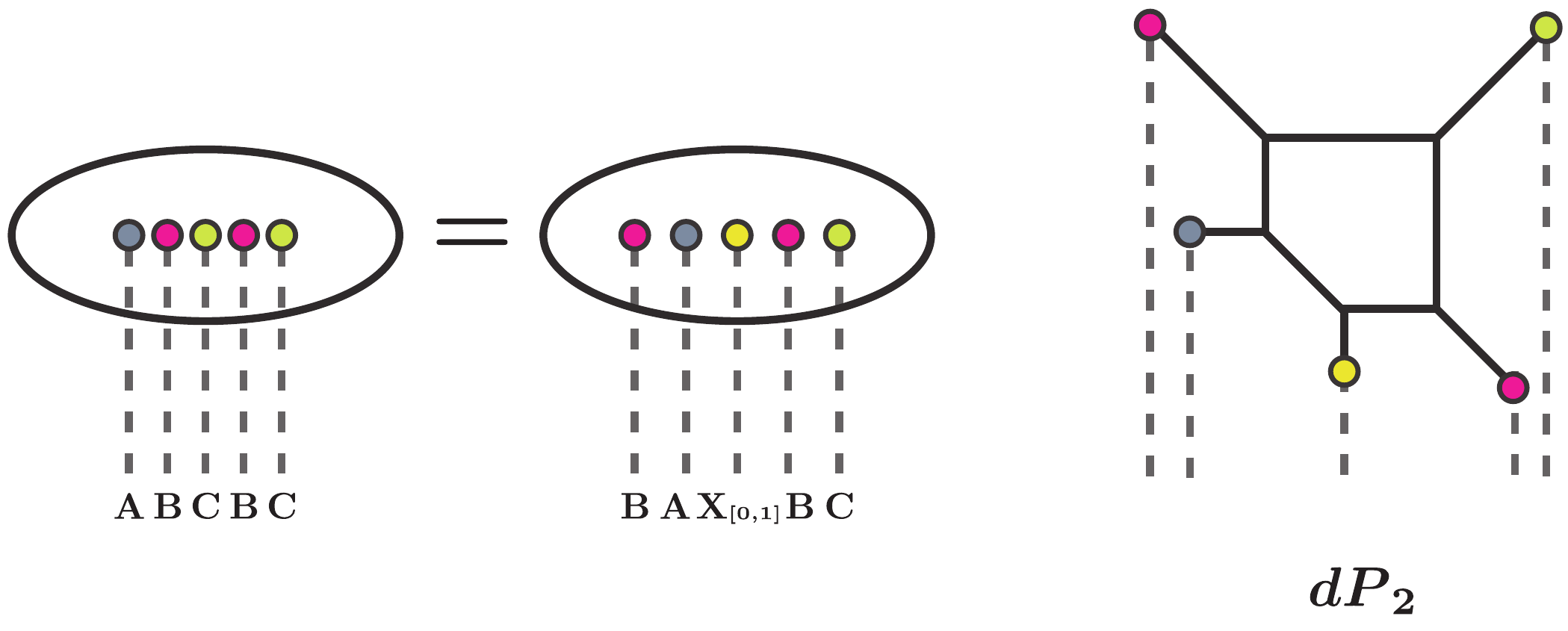}
 \end{center}
\caption{$\hat{\bf E}_2$ configuration
in the left hand side is equivalent to ${\bf B}{\bf A}{\bf X}_{[0,1]}{\bf B}{\bf C}$.
This configuration gives the web of $dP_2$ in the right hand side.}
 \label{fig;dP2I}
\end{figure}

$\hat{\bf E}_2$ configuration is  one-point blowup of $\hat{\bf E}_1$,
and this blowup process is realized by adding an $\bf A$-brane to the original 7-brane configuration.
The resulting 7-brane configuration is converted to  ${\bf B}{\bf A}{\bf X}_{[0,1]}{\bf B}{\bf C}$
by the action of the branch cut move for 7-branes
\begin{align}
\hat{\bf E}_2
&\equiv
{\bf A}{\bf B}{\bf C}{\bf B}{\bf C}
=
{\bf X}_{[0,-1]}{\bf A}{\bf C}{\bf B}{\bf C}
={\bf B}{\bf X}_{[0,-1]}{\bf C}{\bf B}{\bf C}
={\bf B}{\bf C}{\bf A}{\bf B}{\bf C}
\nonumber\\
&={\bf B}{\bf A}{\bf X}_{[0,1]}{\bf B}{\bf C}.
\end{align}
Pulling these five 7-branes out of the 5-brane loop
creates
new 5-brane prongs on the 7-branes
through the Hanany-Witten effect.
The resulting configuration is illustrated in \textbf{\textit{Figure}$\,$\textit{\ref{fig;dP2I}}}.
Moving along geodesics,
all the 7-branes can run away to infinity without changing the 5d theory.
The configuration then consists purely of 5-branes,
and therefore it has dual toric Calabi-Yau geometry.
This toric diagram,
which has completely the same shape as the 5-brane web,
is precisely that of the local second del Pezzo surface.
Therefore, the $\hat{\bf E}_2$ background and the $dP_2$ Calabi-Yau lead to the same 5d theory.

\subsubsection*{Pseudo del Pezzo phase I $\boldsymbol{PdP_2}$}

\begin{figure}[tb]
 \begin{center}
\includegraphics[width=10cm, bb=0 0 415 253]{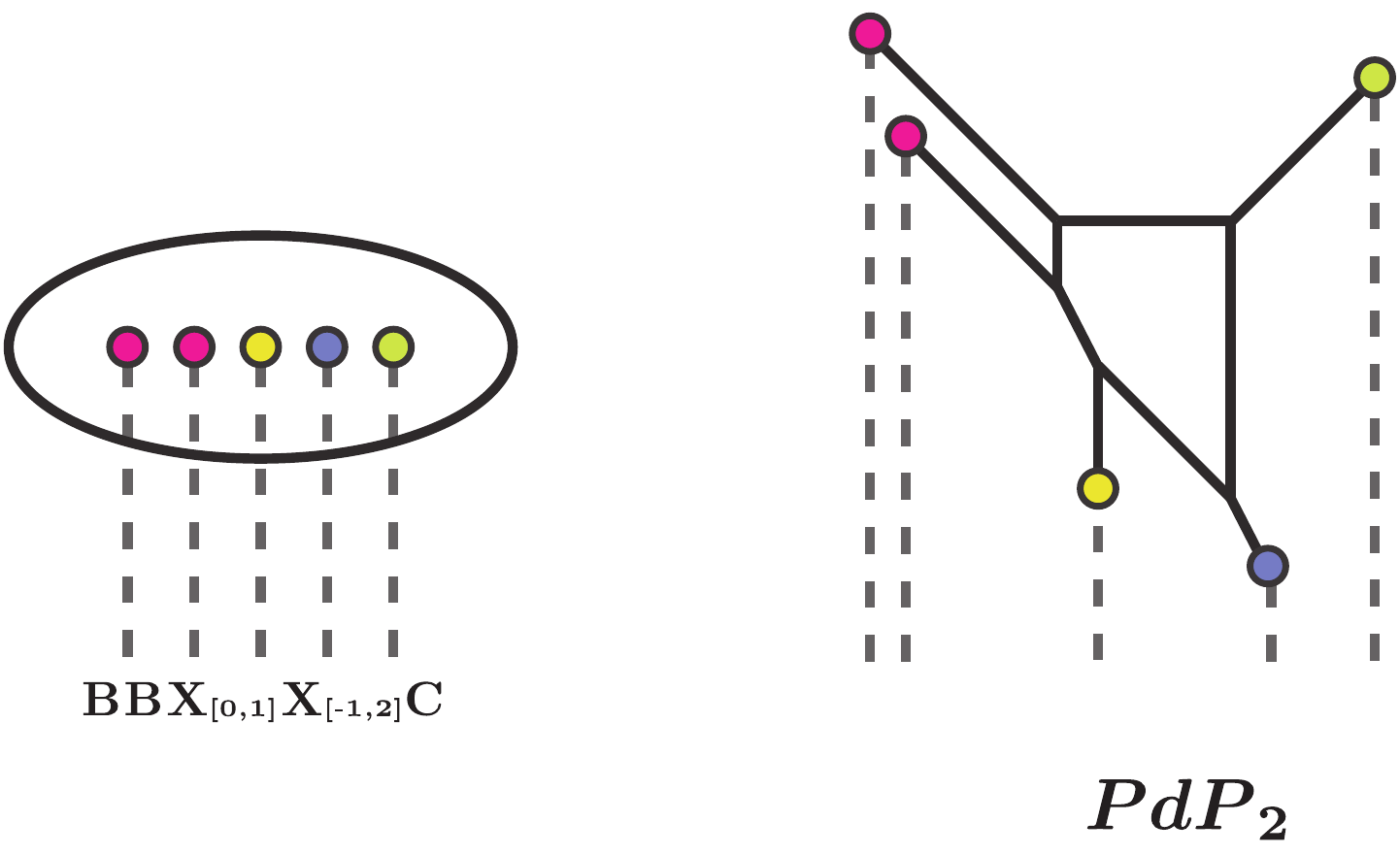}
 \end{center}
\caption{
$\hat{\bf E}_2$ configuration
 is also equivalent to ${\bf B}{\bf B}{\bf X}_{[0,1]}{\bf X}_{[-1,2]}{\bf C}$,
and thus $PdP_2$ describes the same system as $dP_2$ once one add 7-branes to their webs.}
 \label{fig;dP2II}
\end{figure}
We saw that the branch cut move of 7-branes converts 
$\hat{\bf E}_2$ background to $dP_2$ configuration.
Applying successive move,
we can find an another web representation of this configuration.
Let us consider the following branch cut move
\begin{align}
\hat{\bf E}_2=
{\bf B}{\bf A}{\bf X}_{[0,1]}{\bf B}{\bf C}
={\bf B}{\bf A}{\bf B}{\bf X}_{[-1,2]}{\bf C}
={\bf B}{\bf B}{\bf X}_{[0,1]}{\bf X}_{[-1,2]}{\bf C}.
\end{align}
This configuration ${\bf B}{\bf B}{\bf X}_{[0,1]}{\bf X}_{[-1,2]}{\bf C}$
leads to the web diagram for the pseudo del Pezzo phase I of the second del Pezzo surface
as \textbf{\textit{Figure}$\,$\textit{\ref{fig;dP2II}}}.
Since the common 7-brane configuration $\hat{\bf E}_2$ produces the webs
of $dP_2$ and $PdP_2$,
these two webs yield precisely the equivalent  5d theory.
Notice that the web configuration  $PdP_2$ contains the stack of two parallel external legs terminated on ${\bf B}\bf B$.
As we discussed in the previous subsection about $\widetilde{dP}_1$,
we can not move the 7-branes on the parallel stack to infinity without changing the 5d theory.
Therefore, the Nekrasov partition function of $PdP_2$, whose web configuration is obtained by removing all the 7-branes,
can not perfectly coincide with that of $dP_2$ diagram.
The discrepancy between them however takes very simple form as the cases studied in  \cite{Bergman:2013ala,BMPTY,HKN,Taki:2013vka},
and so this mismatch is not essential.
We then expect the following relation between the partition functions of these two phases
\begin{align}
\label{conjecturePdP2}
Z_{\,{dP}_2}(Q_F,u,Q_E;t,q)=\frac{Z_{\,{PdP}_2}(Q_F,u,Q_E;t,q)}{
Z_{\,\textrm{extra}}^{\,{PdP}_2}(u,Q_E;t,q)}.
\end{align}
We will check this equivalence relation between the phases in the next section.

\subsection{Third del Pezzo surface $\boldsymbol{dP_3}$ and $\boldsymbol{PdP_3}$}

\begin{figure}[tb]
\begin{center}
\includegraphics[width=17cm, bb=0 0 818 274]{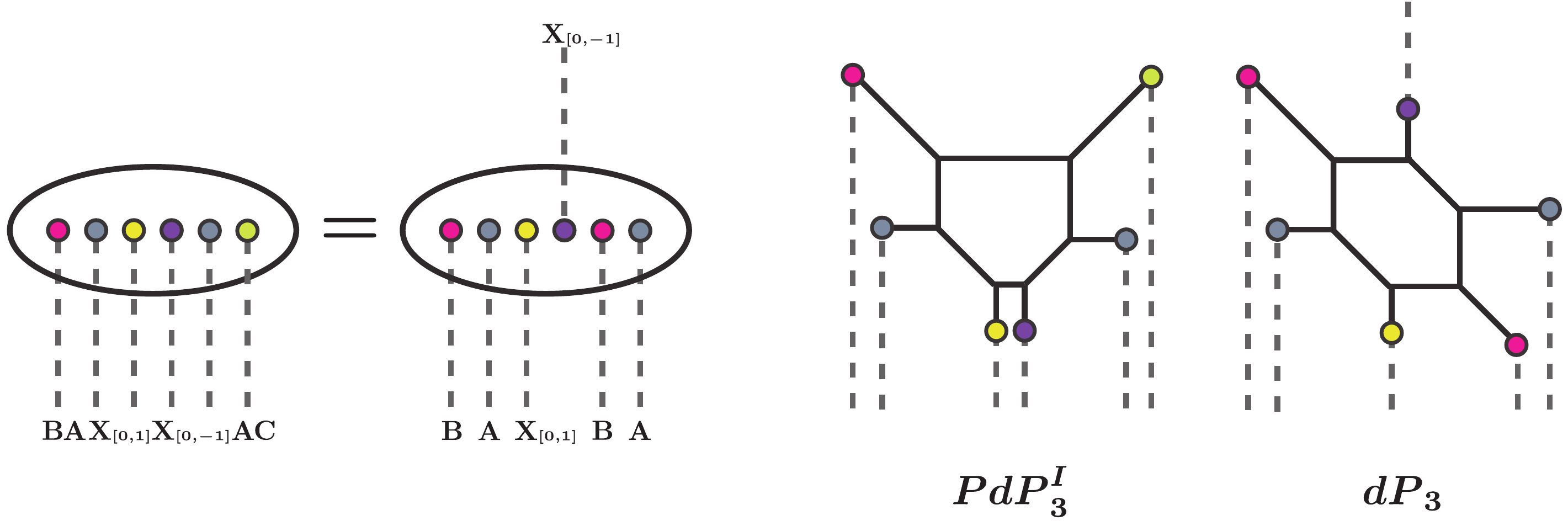}
\end{center}
\caption{
$dP_3$ and  $PdP_3^I$ are related through branch cut move.
Notice that the overall sign of the $[p,q]$ charge is irrelevant and so $\textrm{\bf X}_{[p,q]}=\textrm{\bf X}_{[-p,-q]}$. In
this paper we do not replace $\textrm{\bf X}_{[-p,-q]}$ to $\textrm{\bf X}_{[p,q]}$
just for keeping track of the branch cut moves.
}
\label{fig;dP3I}
\end{figure}
\begin{figure}[tb]
\begin{center}
\includegraphics[width=10cm, bb=0 0 430 239
]{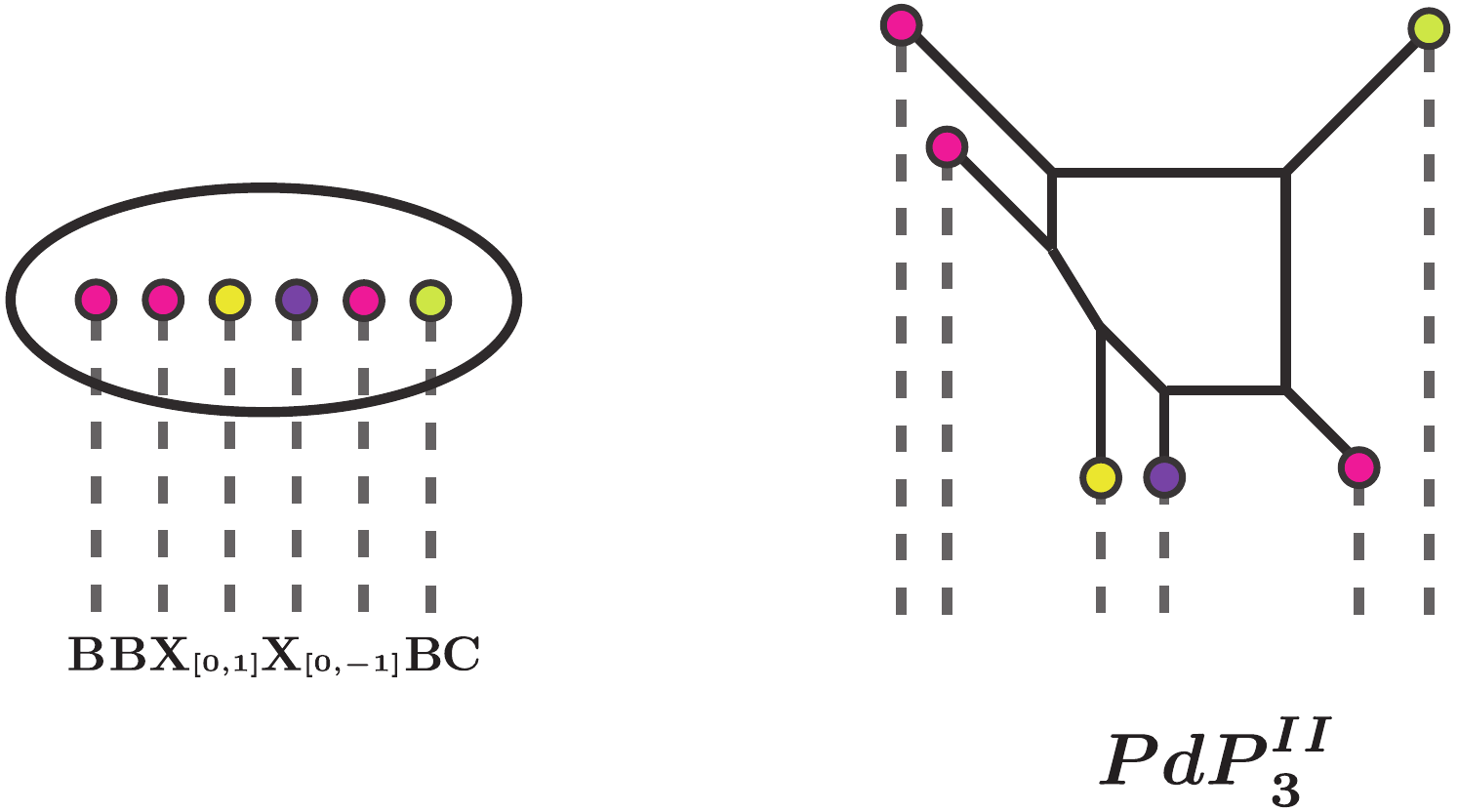}
\end{center}
\caption{$dP_3$ is equivalent to $PdP_3^{II}$
that is associated with 
${\bf B}{\bf A}{\bf X}_{[0,1]}{\bf X}_{[0,-1]}{\bf B}{\bf A}$.}
 \label{fig;dP3II}
\end{figure}

There are four phases of the third del Pezzo surface.
The toric phase for the genuine third del Pezzo surface $dP_3$ is illustrated in the right hand side of
 \textbf{\textit{Figure}$\,$\textit{\ref{fig;dP3II}}}.
The  web diagram next to $dP_3$ in  \textbf{\textit{Figure}$\,$\textit{\ref{fig;dP3I}}}
 is that for the first phase of the pseudo del Pezzo surface $PdP_3^I$.
We will show they are identical once the 7-branes are introduced.

\subsubsection*{Toric $\boldsymbol{dP_3}$ and pseudo del Pezzo phase I $\boldsymbol{PdP_3^I}$}
The third del Pezzo surface is the 5-brane loop probe of the $\hat{\bf E}_3$ 7-brane configuration.
By moving branchs, cuts we obtain
\begin{align}
\label{Ehat3I}
\hat{\bf E}_3
&\equiv
{\bf A}^2{\bf B}{\bf C}{\bf B}{\bf C}\nonumber\\
&=
{\bf A}{\bf B}{\bf A}{\bf X}_{[0,1]}{\bf B}{\bf C}
={\bf B}{\bf X}_{[0,1]}{\bf A}{\bf X}_{[0,1]}{\bf B}{\bf C}
={\bf B}{\bf A}{\bf X}_{[1,-1]}{\bf X}_{[0,1]}{\bf B}{\bf C}
={\bf B}{\bf A}{\bf X}_{[0,1]}{\bf A}{\bf B}{\bf C}\nonumber\\
&
={\bf B}{\bf A}{\bf X}_{[0,1]}{\bf X}_{[0,-1]}{\bf A}{\bf C},
\end{align}
and this 7-brane configuration gives the web of the pseudo del Pezzo $PdP_3^I$.
Moving the branch cut coming from the brane ${\bf X}_{[0,-1]}$ upward, we obtain
\begin{align}
\label{Ehat3I-2}
\parbox{\bw}
{\usebox{\boxa}}\hspace{2mm}.
\end{align}
This representation leads to the web of $dP_3$ as \textbf{\textit{Figure}$\,$\textit{\ref{fig;dP3I}}}.
The two toric phases $dP_3$ and $PdP_3^I$ are therefore the equivalent brane configurations,
and the partition functions become the same function once one factors  the contribution from the stack
of two parallel external branes out
\begin{align}
\label{conjecturePdP3I}
Z_{\,dP_3}=
\frac{
Z_{\,PdP_3^{I}}}
{
Z_{\,\textrm{extra}}^{\,PdP_3^{I}}
},\qquad Z_{\,\textrm{extra}}^{\,PdP_3^I}
=Z_{\,\textrm{parallel two 5-branes}}.
\end{align}
%%%%%%%%%%%%%%%%%
%%%%%%%%%%%%%%%%%
As we will see soon,
this $Z_{\,dP_3}$ partition function has
different expressions based on other pseudo del Pezzo phases.

\subsubsection*{Pseudo del Pezzo phases II,III $\boldsymbol{PdP_3^{II,III}}$}
\begin{figure}[tb]
 \begin{center}
\includegraphics[width=9cm, bb=0 0 398 263]{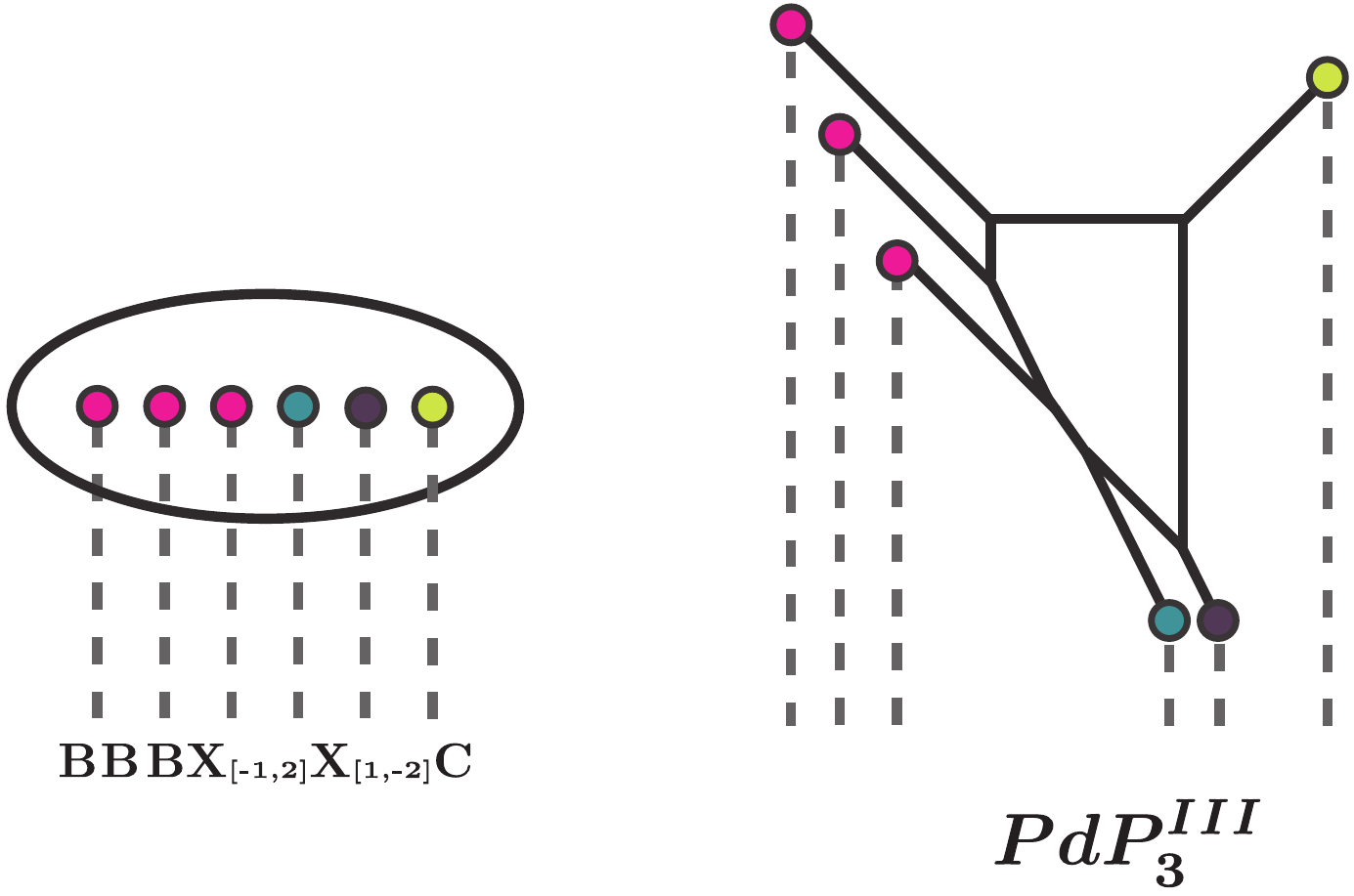}
 \end{center}
\caption{$dP_3$ is equivalent to $PdP_3^{III}$
that is associated with 
${\bf B}
{\bf B}{\bf B}{\bf X}_{[-1,2]}{\bf X}_{[1,-2]}
{\bf C}$.}
 \label{fig;dP3III}
\end{figure}

There are two remaining phases of the third del Pezzo.
\textbf{\textit{Figure}$\,$\textit{\ref{fig;dP3II}}} and \textbf{\textit{Figure}$\,$\textit{\ref{fig;dP3III}}}  are the pseudo third del Pezzo surfaces $PdP_3^{II,III}$.
These phases are also equivalent to the local del Pezzo surface $dP_3$.
We can show this fact by using the 7-brane picture again.

Let us start with the phase II of  the pseudo third del Pezzo $PdP_3^{II}$.
Branch cut move implies the relation
\begin{align}
{\bf A}{\bf X}_{[0,1]}{\bf X}_{[0,-1]}{\bf A}
&=
{\bf A}{\bf X}_{[0,1]}{\bf B}{\bf X}_{[0,-1]}
={\bf A}{\bf X}_{[0,1]}{\bf X}_{[-1,0]}{\bf B}
={\bf A}{\bf X}_{[-1,0]}{\bf X}_{[-1,1]}{\bf B}
\nonumber\\
&=
{\bf A}{\bf X}_{[-1,1]}{\bf X}_{[0,-1]}{\bf B}
={\bf X}_{[0,1]}{\bf A}{\bf X}_{[0,-1]}{\bf B}
={\bf B}{\bf X}_{[0,1]}{\bf X}_{[0,-1]}{\bf B}
.
\end{align}
We can therefore find the following expression for the $\hat{\bf E}_3$ configuration
\begin{align}
\label{7branePdP3II}
\hat{\bf E}_3={\bf B}{\bf A}{\bf X}_{[0,1]}{\bf X}_{[0,-1]}{\bf A}{\bf C}
={\bf B}
{\bf B}{\bf X}_{[0,1]}{\bf X}_{[0,-1]}{\bf B}
{\bf C}.
\end{align}
Since the right hand side of this equation
gives the  $PdP_3^{II}$ configuration as \textbf{\textit{Figure}$\,$\textit{\ref{fig;dP3II}}},
this phase is equivalent to $PdP_3^{I}$ and $dP_3$.
Since there are two stacks of parallel external branes,
the extra contribution to the $PdP_3^{II}$ partition function takes the form
\begin{align}
Z_{\,\textrm{extra}}^{\,PdP_3^{II}}
=Z_{\,\textrm{parallel two 5-branes}}
\times Z_{\,\textrm{parallel two 5-branes}}.
\end{align}
By factoring this extra contribution out, we obtain the $dP_3$ partition function from the  $PdP_3^{II}$ partition function
\begin{align}
\label{conjecturePdP3II}
Z_{\,dP_3}=
\frac{
Z_{\,PdP_3^{II}}}
{
Z_{\,\textrm{extra}}^{\,PdP_3^{II}}
}.
\end{align}

We can also show the equivalence between $dP_3$ and $PdP_3^{III}$.
Branch cut move leads to the relation
\begin{align}
{\bf X}_{[0,1]}{\bf X}_{[0,-1]}{\bf B}
&=
{\bf X}_{[0,1]}{\bf B}{\bf X}_{[1,-2]}
={\bf B}{\bf X}_{[-1,2]}{\bf X}_{[1,-2]}
,
\end{align}
and by applying it to (\ref{7branePdP3II}), we obtain
\begin{align}
\hat{\bf E}_3
={\bf B}
{\bf B}{\bf X}_{[0,1]}{\bf X}_{[0,-1]}{\bf B}
{\bf C}=
{\bf B}
{\bf B}{\bf B}{\bf X}_{[-1,2]}{\bf X}_{[1,-2]}
{\bf C}
.
\end{align}
The right hand side of this equation gives the  $PdP_3^{III}$ configuration as \textbf{\textit{Figure}$\,$\textit{\ref{fig;dP3III}}}.
This phase is therefore equivalent to $PdP_3^{II}$, and therefore to $dP_3$.
 There are a stack of two parallel 5-branes and a stack of three parallel 5-branes,
 and  $PdP_3^{III}$ partition function consequently leads to the $dP_3$ partition function by factoring the following
 extra contribution $Z_{\,\textrm{extra}}^{\,PdP_3^{III}}$ out
\begin{align}
\label{conjecturePdP3III}
Z_{\,dP_3}=
\frac{
Z_{\,PdP_3^{III}}}
{
Z_{\,\textrm{extra}}^{\,PdP_3^{III}}
},\quad Z_{\,\textrm{extra}}^{\,PdP_3^{III}}
=Z_{\,\textrm{parallel two 5-branes}}
\times Z_{\,\textrm{parallel three 5-branes}}.
\end{align}
This equivalence is very non-trivial because the calculation
of the $PdP_3^{III}$ partition function is very hard.
Computing  instanton expansion of this partition function,
our new conjecture is checked in the next section.

\subsection{Fourth del Pezzo surface $\boldsymbol{dP_4}$ and $\boldsymbol{PdP_4}$}
\begin{figure}[tb]
 \begin{center}
\includegraphics[width=10cm, bb=0 0 430 271]{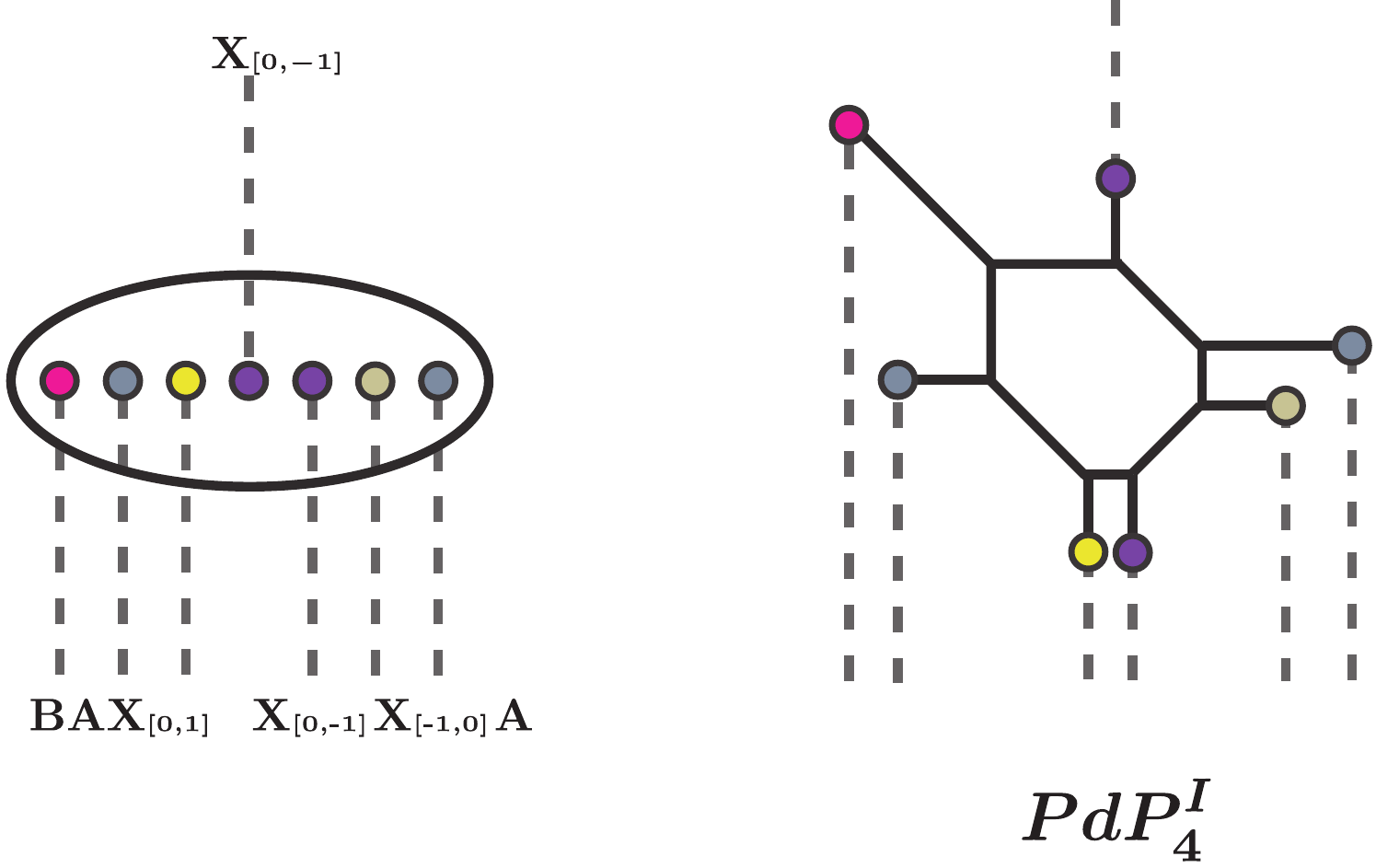}
 \end{center}
\caption{
$\hat{\bf E}_4$ configuration is related to 
${\bf B}{\bf A}{\bf X}_{[0,1]}{\bf X}_{[0,-1]}{\bf X}_{[0,-1]}{\bf X}_{[-1,0]}{\bf C}$,
and therefore $dP_4$ is equivalent to $PdP_4^{I}$.}
 \label{fig;dP4I}
\end{figure}
The local fourth del Pezzo surface $dP_4$ itself is non-toric 
and does not have any 5-brane web description.
We can however construct toric analogue of it by blowing up the toric descriptions of the third pseudo del Pezzo surface $dP_3$,
and there are  two resulting  toric phases $PdP_4^{I,II}$ for the local pseudo del Pezzo surface.
By introducing the 7-brane regularization,
we can show the equivalence between the pseudo del Pezzo surfaces $PdP_4$ and the genuine one $dP_4$.

\subsubsection*{Pseudo del Pezzo phase I $\boldsymbol{PdP_4^I}$}
The local fourth del Pezzo surface is realized by the 5-brane probe of the $\hat{\bf E}_4$ configuration.
By using (\ref{Ehat3I}), we get
\begin{align}
\label{Ehat4I}
\hat{\bf E}_4
&\equiv
{\bf A}^3{\bf B}{\bf C}{\bf B}{\bf C}
=
{\bf A}{\bf B}{\bf A}{\bf X}_{[0,1]}{\bf X}_{[0,-1]}{\bf A}{\bf C}
=
{\bf B}{\bf X}_{[0,1]}{\bf A}{\bf X}_{[0,1]}{\bf X}_{[0,-1]}{\bf A}{\bf C}\nonumber\\
&=
{\bf B}{\bf A}{\bf X}_{[-1,1]}{\bf X}_{[0,1]}{\bf X}_{[0,-1]}{\bf A}{\bf C}
=
{\bf B}{\bf A}{\bf X}_{[0,1]}{\bf X}_{[-1,0]}{\bf X}_{[0,-1]}{\bf A}{\bf C},
\end{align}
and we find the following expression by moving the branch cut coming from
$\textrm{\bf X}_{[0,-1]}$ upward
\begin{align}
&\parbox{\bw}{\usebox{\boxb}}\hspace{27mm}
\nonumber\\
&\quad\,\,\,\parbox{\bw}{\usebox{\boxc}}\hspace{-10mm}.
\end{align}
The last expression immediately gives the pseudo del Pezzo $PdP_4^{I}$
as illustrated in \textbf{\textit{Figure}$\,$\textit{\ref{fig;dP4I}}}.
Therefore this pseudo del Pezzo is equivalent to the genuine del Pezzo $dP_4$
once one introduces the 7-brane regularization to all the external legs.
This pseudo del Pezzo phase contains two stacks of two parallel 5-branes,
and so we can obtain the $dP_4$ partition function by removing the following extra factor
from the $PdP_4^I$ partition function
\begin{align}
\label{conjecturePdP4I}
&Z_{\,{dP}_4}(Q_F,u,Q_{Ef};t,q)=\frac{Z_{\,{PdP_4^{I}}}(Q_F,u,Q_{Ef};t,q)}{Z^{\,PdP_4^{I}}_{\,\textrm{extra}}(u,Q_{Ef};t,q)},\\
&
Z_{\,\textrm{extra}}^{\,PdP_4^{I}}
=Z_{\,\textrm{parallel two 5-branes}}
\times
Z_{\,\textrm{parallel two 5-branes}}.
\end{align}
This conjecture was checked at the level of the superconformal index \cite{BMPTY,HKN},
and the authors showed that the renormalized $PdP_4^I$ partition function leads to
the index with the enhanced $E_4$ symmetry which is expected from property of the del Pezzo surface $dP_4$.
\begin{figure}[tb]
\begin{center}
\includegraphics[width=10cm, bb=0 0 448 239]{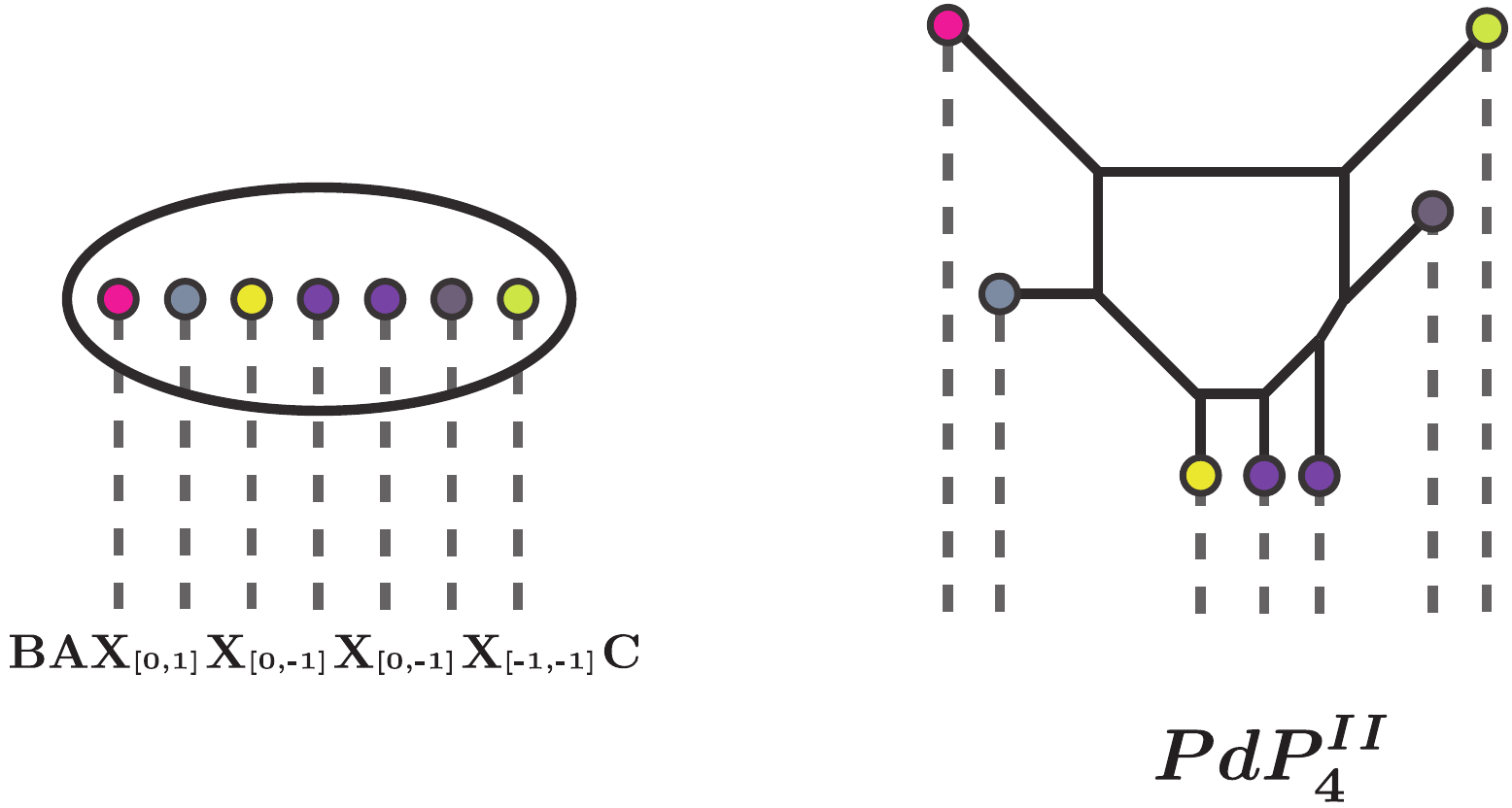}
\end{center}
\caption{$dP_4$ is equivalent to $PdP_4^{II}$.
This 5-brane web is associated with the 7-brane configuration 
${\bf B}{\bf A}{\bf X}_{[0,1]}{\bf X}_{[0,-1]}{\bf X}_{[0,-1]}{\bf X}_{[-1,-1]}{\bf C}$.}
\label{fig;dP4II}
\end{figure}

\subsubsection*{Pseudo del Pezzo phase II $\boldsymbol{PdP_4^{II}}$}
For the fourth del Pezzo surface,
we can employ the additional toric phase $PdP_4^{II}$ illustrated in \textbf{\textit{Figure}$\,$\textit{\ref{fig;dP4II}}}.
This pseudo del Pezzo surface is also equivalent to the del Pezzo $dP_4$ as follows.
By using the relation (\ref{Ehat4I}),
we find
\begin{align}
\label{Ehat4II}
\hat{\bf E}_4
&
=
{\bf B}{\bf A}{\bf X}_{[0,1]}{\bf X}_{[-1,0]}{\bf X}_{[0,-1]}{\bf A}{\bf C}
=
{\bf B}{\bf A}{\bf X}_{[0,1]}{\bf X}_{[0,-1]}{\bf X}_{[-1,-1]}{\bf A}{\bf C}\nonumber\\
&=
{\bf B}{\bf A}{\bf X}_{[0,1]}{\bf X}_{[0,-1]}{\bf X}_{[0,-1]}{\bf X}_{[-1,-1]}{\bf C}.
\end{align}
The last line is precisely the 7-brane configuration that leads to $PdP_4^{II}$ as \textbf{\textit{Figure}$\,$\textit{\ref{fig;dP4II}}}.

This pseudo del Pezzo surface contains a stack of two parallel 5-branes and a stack of three parallel 5-branes,
and therefore the $dP_4$ partition function is the $PdP_4^{II}$ partition function divided by
the following extra factor
\begin{align}
\label{conjecturePdP4II}
&Z_{\,{dP}_4}(Q_F,u,Q_{Ef};t,q)=\frac{Z_{\,{PdP_4^{II}}}(Q_F,u,Q_{Ef};t,q)}{Z^{\,PdP_4^{II}}_{\,\textrm{extra}}(u,Q_{Ef};t,q)},\\
&
Z_{\,\textrm{extra}}^{\,PdP_4^{II}}
=Z_{\,\textrm{parallel two 5-branes}}
\times
Z_{\,\textrm{parallel three 5-branes}}.
\end{align}
This new conjecture is checked in the next section.

\subsection{$\boldsymbol{PdP_5}$}
\begin{figure}[tb]
\begin{center}
\includegraphics[width=10cm, bb=0 0 441 260]{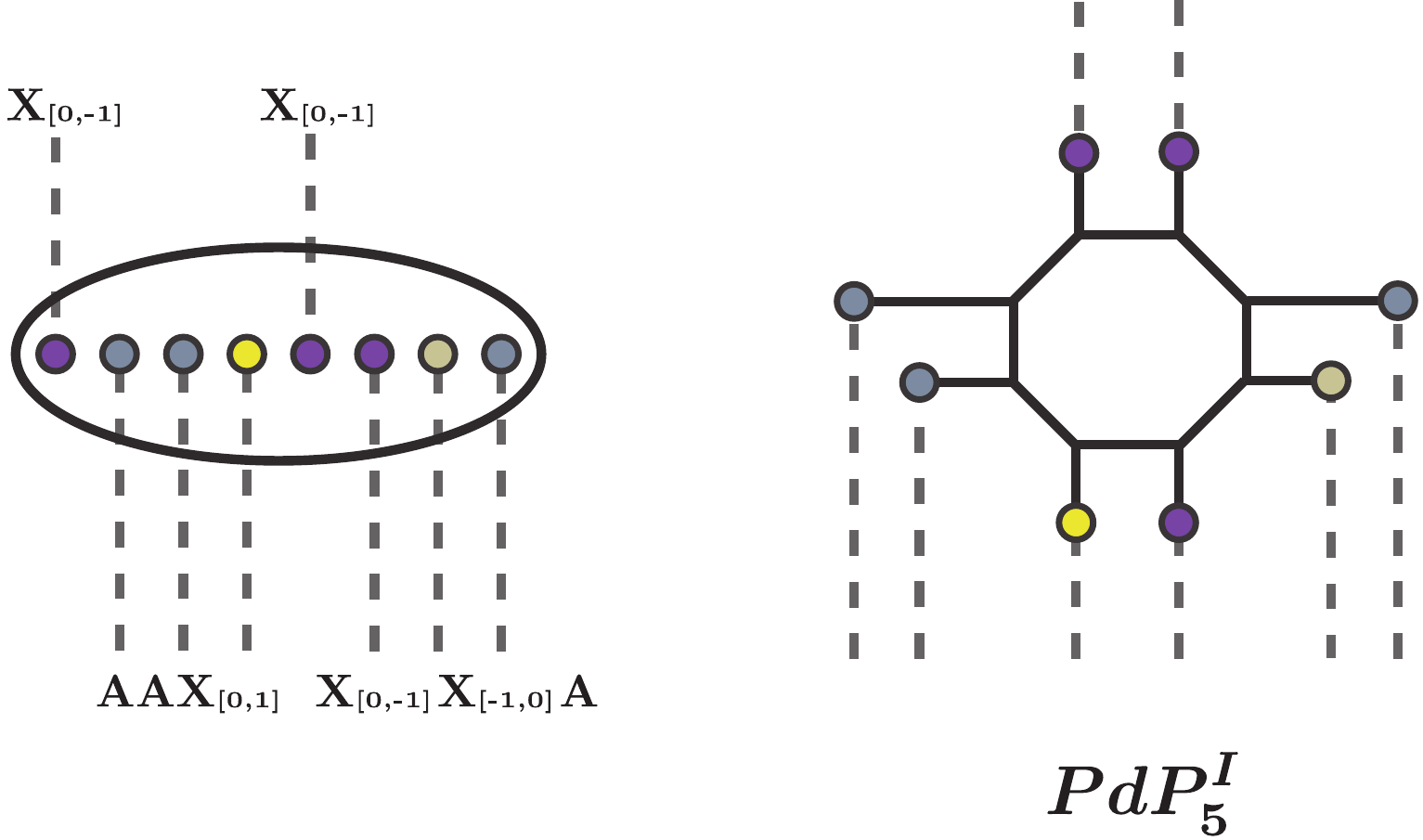}
\end{center}
\caption{Moving branch cuts of $\hat{\textrm{\bf E}}_5$ configuration,
we obtain the configuration on the left side.
The pseudo del Pezzo surface illustrated on the right side
is thus equivalent to $dP_5$.}
\label{fig;dP5I}
\end{figure}
The local fifth del Pezzo surface $dP_4$ is not toric,
and therefore it does not have  5-brane web description.
We can, however, construct pseudo del Pezzo surfaces as the previous cases,
and there are  three toric pseudo del Pezzo phases $PdP_5^{I,II,II}$.
By introducing the 7-brane regularization,
we can show that these pseudo del Pezzo surfaces lead to the same compactification of superstring theory as that on the non-toric del Pezzo $dP_5$.

\subsubsection*{Pseudo del Pezzo phase I $\boldsymbol{PdP_5^I}$}
Let us start with the pseudo del Pezzo surface $PdP_5^I$ illustrated in \textbf{\textit{Figure}$\,$\textit{\ref{fig;dP5I}}}.
We can regularize the corresponding 5-brane web by introducing two types of 7-branes
$\textrm{\bf A}=\textrm{\bf X}_{[-1,0]}$ and $\textrm{\bf X}_{[0,1]}=\textrm{\bf X}_{[0,-1]}$.
We can show that this configuration is equivalent to the local  fifth del Pezzo surface associated with $\hat{\textrm{\bf E}}_5$ configuration
\begin{align}
\label{Ehat5I}
\hat{\bf E}_5
&\equiv
{\bf A}^4{\bf B}{\bf C}{\bf B}{\bf C}
=
{\bf A}{\bf B}{\bf A}{\bf X}_{[0,1]}{\bf X}_{[-1,0]}{\bf X}_{[0,-1]}{\bf A}{\bf C}
=
{\bf X}_{[0,-1]}{\bf A}{\bf A}{\bf X}_{[0,1]}{\bf X}_{[-1,0]}{\bf X}_{[0,-1]}{\bf A}{\bf C}.
\end{align}
We use (\ref{Ehat4I}) to show the second equality.
By moving the branch cut as (\ref{Ehat3I-2}), we obtain
\begin{align}
\parbox{\bw}{\usebox{\boxd}}\,\,\,,\hspace{-5mm}
\end{align}
and this expression leads to the pseudo del Pezzo surface as \textbf{\textit{Figure}$\,$\textit{\ref{fig;dP5I}}}.
This toric phase contains four stacks of four parallel 5-branes and the extra contribution from the non-full spin content is
\begin{align}
\label{conjecturePdP5I}
&Z_{\,{dP}_5}(Q_F,u,Q_{Ef};t,q)=\frac{Z_{\,{PdP_5^{I}}}(Q_F,u,Q_{Ef};t,q)}{Z^{\,PdP_5^{I}}_{\,\textrm{extra}}(u,Q_{Ef};t,q)},\\
&
Z_{\,\textrm{extra}}^{\,PdP_5^{I}}
=Z_{\,\textrm{parallel two 5-branes}}
\times
Z_{\,\textrm{parallel two 5-branes}}\nonumber\\
&\qquad\qquad\times
Z_{\,\textrm{parallel two 5-branes}}
\times
Z_{\,\textrm{parallel two 5-branes}}.
\end{align}
By removing this extra contribution,
we can obtain the Nekrasov partition function of the local del Pezzo surface $dP_5$
from the pseudo del Pezzo partition function.
This conjecture was checked in \cite{BMPTY,HKN} based on the explicit computation.

\begin{figure}[tb]
\begin{center}
\includegraphics[width=10cm, bb=0 0 432 239]{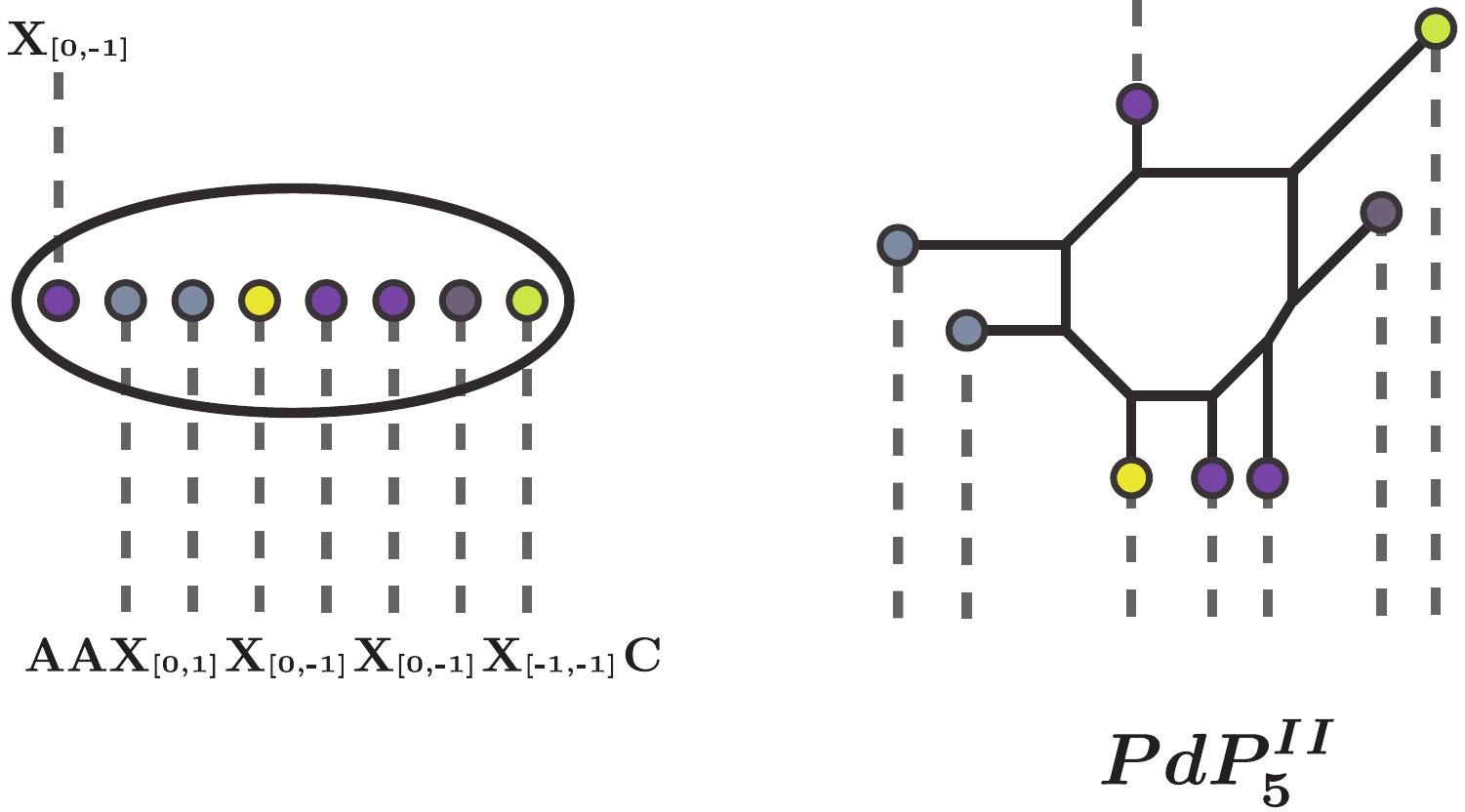}
\end{center}
\caption{$\hat{\textrm{\bf E}}_5$ configuration leads to
 the 7-brane configuration on the left side. $PdP_5^{II}$ is therefore equivalent to $dP_5$.
 }
\label{fig;dP5II}
\end{figure}
\subsubsection*{Pseudo del Pezzo phase II $\boldsymbol{PdP_5^{II}}$}
There are another toric phase of the del Pezzo $dP_5$.
\textbf{\textit{Figure}$\,$\textit{\ref{fig;dP5II}}} is the second phase $PdP_5^{II}$.
This phase is also equivalent to $\hat{\bf E}_5$ because branch cut move implies
\begin{align}
\hat{\bf E}_5
=
{\bf A}{\bf B}{\bf A}{\bf X}_{[0,1]}{\bf X}_{[0,-1]}{\bf X}_{[0,-1]}{\bf X}_{[-1,-1]}{\bf C}
=
{\bf X}_{[0,-1]}{\bf A}{\bf A}{\bf X}_{[0,1]}{\bf X}_{[0,-1]}{\bf X}_{[0,-1]}{\bf X}_{[-1,-1]}{\bf C}.
\end{align}
To show the first equality we use the relation
(\ref{Ehat4II}).
The last 7-brane configuration leads to the second toric phase as illustrated in \textbf{\textit{Figure}$\,$\textit{\ref{fig;dP5II}}}.
This web configuration contains three stacks of infinitely-long external 5-branes once one remove the 7-brane regularization.
 The $PdP_5^{II}$ partition function consequently coincides with the $dP_5$ partition function after removing the extra contribution from the stacks as
\begin{align}
\label{conjecturePdP5II}
&Z_{\,{dP}_5}(Q_F,u,Q_{Ef};t,q)=\frac{Z_{\,{PdP_5^{II}}}(Q_F,u,Q_{Ef};t,q)}{Z^{\,PdP_5^{II}}_{\,\textrm{non-ful}}(u,Q_{Ef};t,q)},\\
&Z_{\,\textrm{extra}}^{\,PdP_5^{II}}
=
Z_{\,\textrm{parallel three 5-branes}}
\times Z_{\,\textrm{parallel two 5-branes}}
\times Z_{\,\textrm{parallel two 5-branes}}.
\end{align}
This new conjecture is checked in the next section.

\begin{figure}[tb]
\begin{center}
\includegraphics[width=10cm, bb=0 0 453 239]{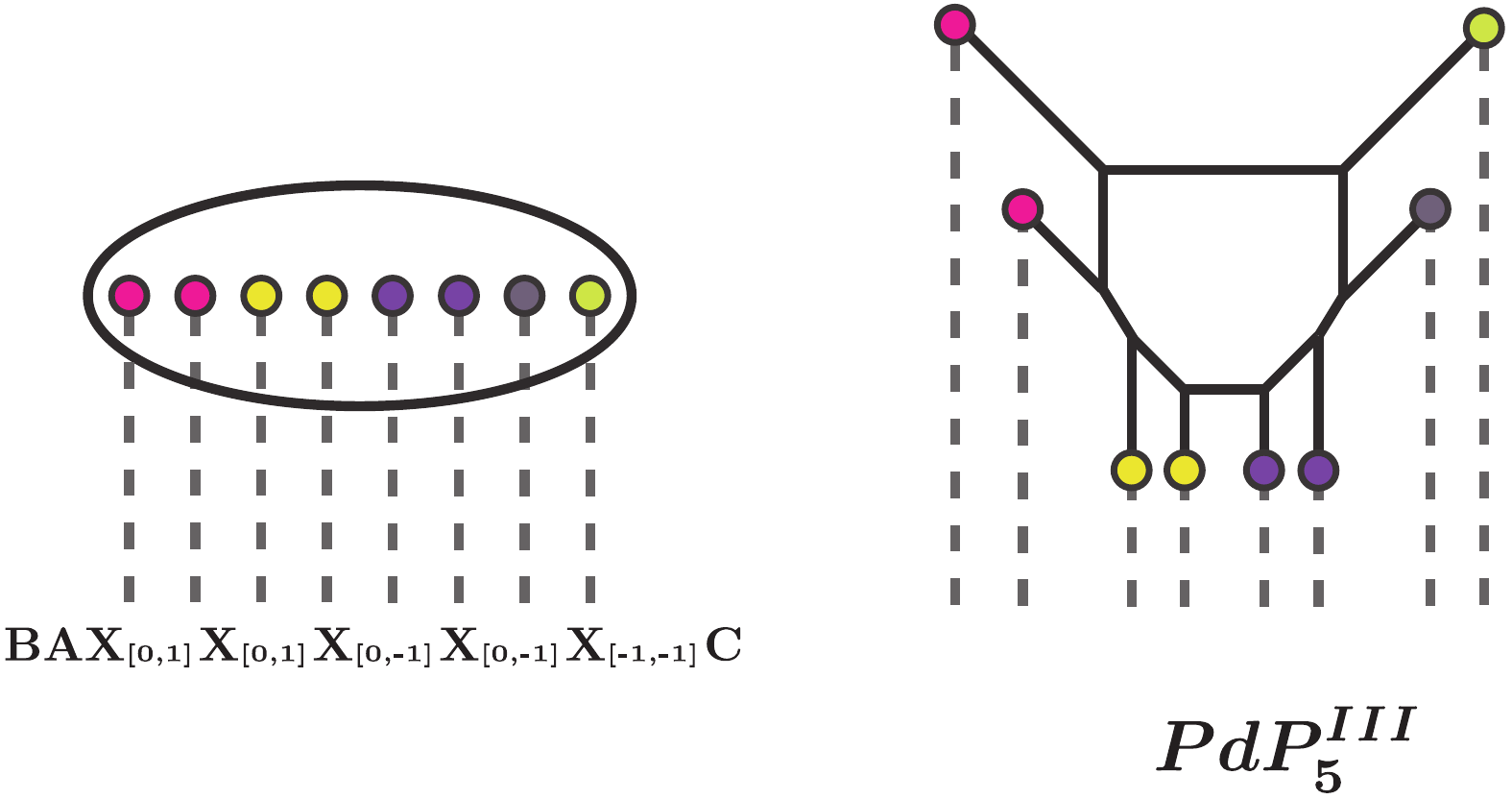}
\end{center}
\caption{$\hat{\textrm{\bf E}}_5$ configuration leads to
 the 7-brane configuration on the left side. $PdP_5^{III}$ is therefore equivalent to $dP_5$.}
\label{fig;dP5III}
\end{figure}
\subsubsection*{Pseudo del Pezzo phase III $\boldsymbol{PdP_5^{III}}$}
The third toric phase $PdP_5^{III}$ is illustrated in \textbf{\textit{Figure}$\,$\textit{\ref{fig;dP5III}}}.
The following relation follows from  branch cut move
\begin{align}
\hat{\bf E}_5
&=
{\bf A}{\bf B}{\bf A}{\bf X}_{[0,1]}{\bf X}_{[0,-1]}{\bf X}_{[0,-1]}{\bf X}_{[-1,-1]}{\bf C}
=
{\bf B}{\bf X}_{[0,1]}{\bf A}{\bf X}_{[0,1]}{\bf X}_{[0,-1]}{\bf X}_{[0,-1]}{\bf X}_{[-1,-1]}{\bf C}
\nonumber\\
&={\bf B}{\bf B}{\bf X}_{[0,1]}{\bf X}_{[0,1]}{\bf X}_{[0,-1]}{\bf X}_{[0,-1]}{\bf X}_{[-1,-1]}{\bf C},
\end{align}
and the last line leads to the $PdP_5^{III}$ brane web configuration as \textbf{\textit{Figure}$\,$\textit{\ref{fig;dP5III}}}.
Therefore this phase $PdP_5^{III}$ is also equivalent to the genuine local del Pezzo surface in the presence of the 7-brane regularization.

By removing the 7-branes, we obtain the 5-brane web and can compute the corresponding partition function
by using the refined topological vertex formalism.
Once one takes away the 7-brane regularization,
the stacks of the parallel external 5-branes develop  extra contribution.
The $PdP_5^{III}$ partition function therefore coincides with the $dP_5$ partition function
after eliminating the extra contribution
\begin{align}
\label{conjecturePdP5III}
&Z_{\,{dP}_5}(Q_F,u,Q_{Ef};t,q)=\frac{Z_{\,{PdP_5^{III}}}(Q_F,u,Q_{Ef};t,q)}{Z^{\,PdP_5^{II}}_{\,\textrm{extra}}(u,Q_{Ef};t,q)},\\
&Z_{\,\textrm{extra}}^{\,PdP_5^{III}}
=
Z_{\,\textrm{parallel four 5-branes}}
\times Z_{\,\textrm{parallel two 5-branes}}
\times Z_{\,\textrm{parallel two 5-branes}}.
\end{align}
This new conjecture is also checked in the next section.

\subsection{$\boldsymbol{PdP_6}$}

\begin{figure}[tb]
\begin{center}
\includegraphics[width=10cm, bb=0 0 451 254]{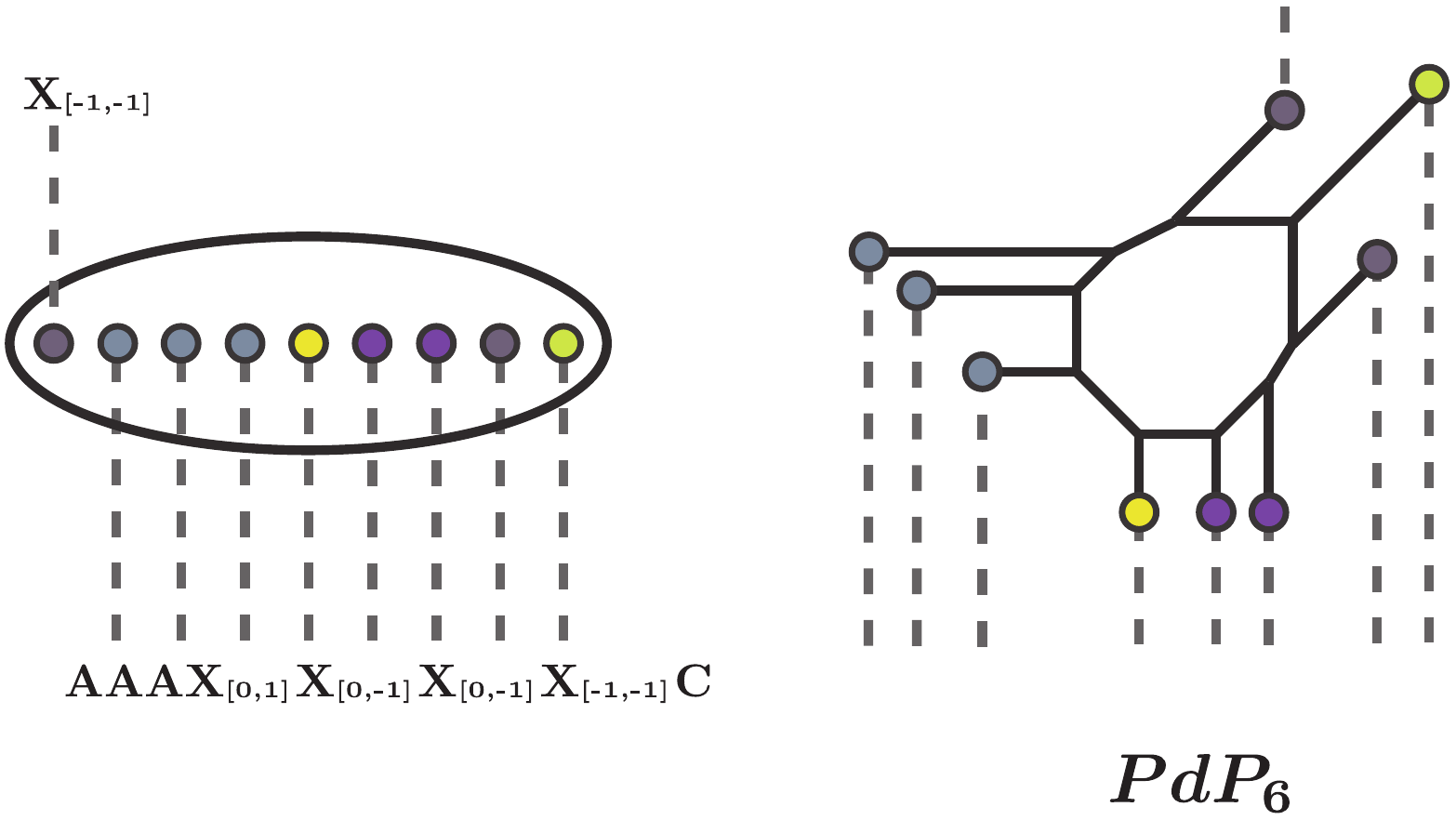}
\end{center}
\caption{Moving branch cuts of $\hat{\textrm{\bf E}}_6$ configuration,
we obtain the configuration on the left side.
The pseudo del Pezzo surface $PdP_6$ illustrated on the right side
is thus equivalent to $dP_6$.}
\label{fig;dP6}
\end{figure}
The non-toric local fifth del Pezzo surface $dP_5$ does not have any 5-brane web description.
There are however the toric pseudo del Pezzo phase $PdP_6$ illustrated in \textbf{\textit{Figure}$\,$\textit{\ref{fig;dP6}}}.
We can relate the pseudo del Pezzo phase $PdP_6$ to the genuine del Pezzo $dP_6$
by moving branch cuts of the corresponding 7-brane background
\begin{align}
\hat{\bf E}_6
&\equiv{\bf A}^5{\bf B}{\bf C}{\bf B}{\bf C}
=
{\bf A}{\bf X}_{[0,-1]}{\bf A}{\bf A}{\bf X}_{[0,1]}{\bf X}_{[0,-1]}{\bf X}_{[0,-1]}{\bf X}_{[-1,-1]}{\bf C}\nonumber\\
&=
{\bf X}_{[-1,-1]}{\bf A}{\bf A}{\bf A}{\bf X}_{[0,1]}{\bf X}_{[0,-1]}{\bf X}_{[0,-1]}{\bf X}_{[-1,-1]}{\bf C}.
\end{align}
The last line leads to the web of $PdP_5$ as \textbf{\textit{Figure}$\,$\textit{\ref{fig;dP6}}}.

Since the pseudo del Pezzo configuration contains  three stacks of three parallel 5-branes,
the $PdP_6$ partition function  coincides with the $dP_6$ partition function
after removing the extra contribution arising from these stacks
\begin{align}
&Z_{\,{dP}_6}(Q_F,u,Q_{Ef};t,q)=\frac{Z_{\,{PdP_6}}(Q_F,u,Q_{Ef};t,q)}{Z^{\,PdP_6}_{\,\textrm{extra}}(u,Q_{Ef};t,q)},\\
&Z_{\,\textrm{extra}}^{\,PdP_6}
=
Z_{\,\textrm{parallel three 5-branes}}
\times Z_{\,\textrm{parallel three 5-branes}}
\times Z_{\,\textrm{parallel three 5-branes}}.
\end{align}
This relation was conjectures in \cite{BMPTY,HKN}. The authors checked it by computing the corresponding superconformal index
and confirming the enhancement of the $E_6$ flavor symmetry expected from the geometry of $dP_6$.
This SCFT with $E_6$ flavor symmetry is precisely 
the 5d uplift of  Gaiotto's $T^3$ theory \cite{Benini:2009gi,Gaiotto:2009we}.

\section{Nekrasov partition functions of del Pezzo suraces}

\begin{figure}[tb]
 \begin{center}
  \includegraphics[width=140mm, bb=0 0 581 197]{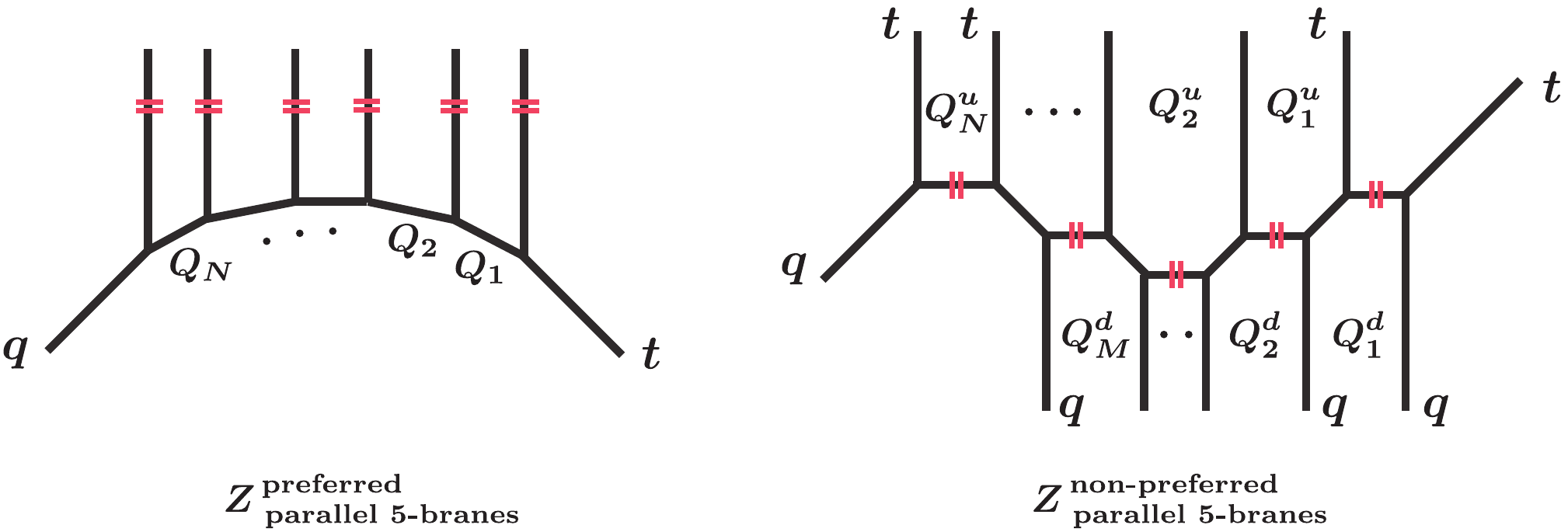}
 \end{center}
 \caption{The extra contributions of the non-full spin content.}
 \label{fig;nonfull}
\end{figure}
In this section,
we check explicitly our conjecture
that
the Nekrasov partition functions of the toric phases of a local del Pezzo surface
lead to the same partition function that describes the local del Pezzo surface
once the non-full spin content part is removed
\begin{align}
Z_{dP_k}=\frac{Z_{PdP_k^{p}}}{Z^{PdP_k^{p}}_{\textrm{extra}}},\quad
p=I,\,II,\,III,\,\cdots,
\end{align}
where $Z^{PdP_k^{p}}_{\textrm{extra}}$ is the product of all the 
contributions from the stacks of parallel external legs.
$p$ labels the toric phases associated with the local del Pezzo surface $dP_k$.
We can compute this extra part as follows.
There are two types of stacks of parallel external 5-branes as illustrated in \textbf{\textit{Figure}$\,$\textit{\ref{fig;nonfull}}}:
one is the non-preferred external legs, and the other is the preferred external legs.
These sub-diagrams in a full web give the extra contribution in question.
The topological string partition functions of these sub-diagrams are then the extra contributions from the non-full spin content.
Assuming the slicing invariance \cite{Iqbal:2007ii},
the refined topological vertex formalism gives
the following explicit expressions \cite{BMPTY,HKN}
\begin{align}
&Z_{\,\textrm{parallel $N-1$ 5-branes}}^{\,\textrm{preferred}}(Q_1,\cdots,Q_N;t,q)
=M(Q_1,\cdots,Q_N;t,q),\\
&\rule{0pt}{3ex}Z_{\,\textrm{parallel $N-1$ and $M-1$ 5-branes}}^{\,\textrm{non-preferred}}(Q^u_1,\cdots,Q^u_N,Q^d_1,\cdots,Q^d_M;t,q)\nonumber\\
&\rule{0pt}{3ex}\qquad\qquad\qquad\qquad\qquad\qquad\quad
=\frac{M(Q^u_1,\cdots,Q^u_N;t,q)\,
M(Q^d_M,\cdots,Q^d_1;q,t)}{M^{\,\textrm{pert.}}(Q^u_1,\cdots,Q^u_N,Q^d_1,\cdots,Q^d_M;t,q)},\\
&\rule{0pt}{4ex}M(Q_1,\cdots,Q_N;t,q)=\prod_{i,j=1}^\infty
\prod_{1\leq \ell\leq m\leq N}
\frac{1}{1-Q_\ell Q_{\ell+1}\cdots Q_m\,t^{i}q^{j-1}},
\end{align}
where $M^{\,\textrm{pert.}}(Q^u_1,\cdots,Q^u_N,Q^d_1,\cdots,Q^d_M;t,q)$ is the factors appearing in the numerator 
that are shared with the perturbative partition function of a full web diagram.
This rule means that we do not take the finite legs that form the loop in this full web into consideration.
We remove this factor $M^{\,\textrm{pert.}}$ because it can be collected into the perturbative part
of the vector multiplet contribution that associated with the web loop.

Since these partition functions of the extra contributions are not invariant under the replacement of $t$ with $q$,
the corresponding spectrum does not form any representation of the little group of 5d Lorentz group
$SU(2)_L\times SU(2)_R$.
This means that the full spectrum contains extra states which do not form any full spin content of the Lorentz group.
To obtain proper spectrum in 5d, we have to remove such pathological contribution associated with
a non-compact direction in the corresponding web configuration.
This is the practical reason why we need to factor the non-full spin part out of a Nekrasov partition function.

\subsection{The two toric phases for $\boldsymbol{E_1}$ theory}

Let us start with the simplest and non-trivial example.
In the previous section,
we claimed that the two Calabi-Yau manifolds $\widetilde{dP}_1$ and  $P\widetilde{dP}_1$
lead to the same $E_1$ SCFT.
We can check this claim by comparing their Nekrasov partition functions.

\begin{figure}[tbp]
 \begin{center}
  \includegraphics[width=120mm, bb=0 0 548 253]{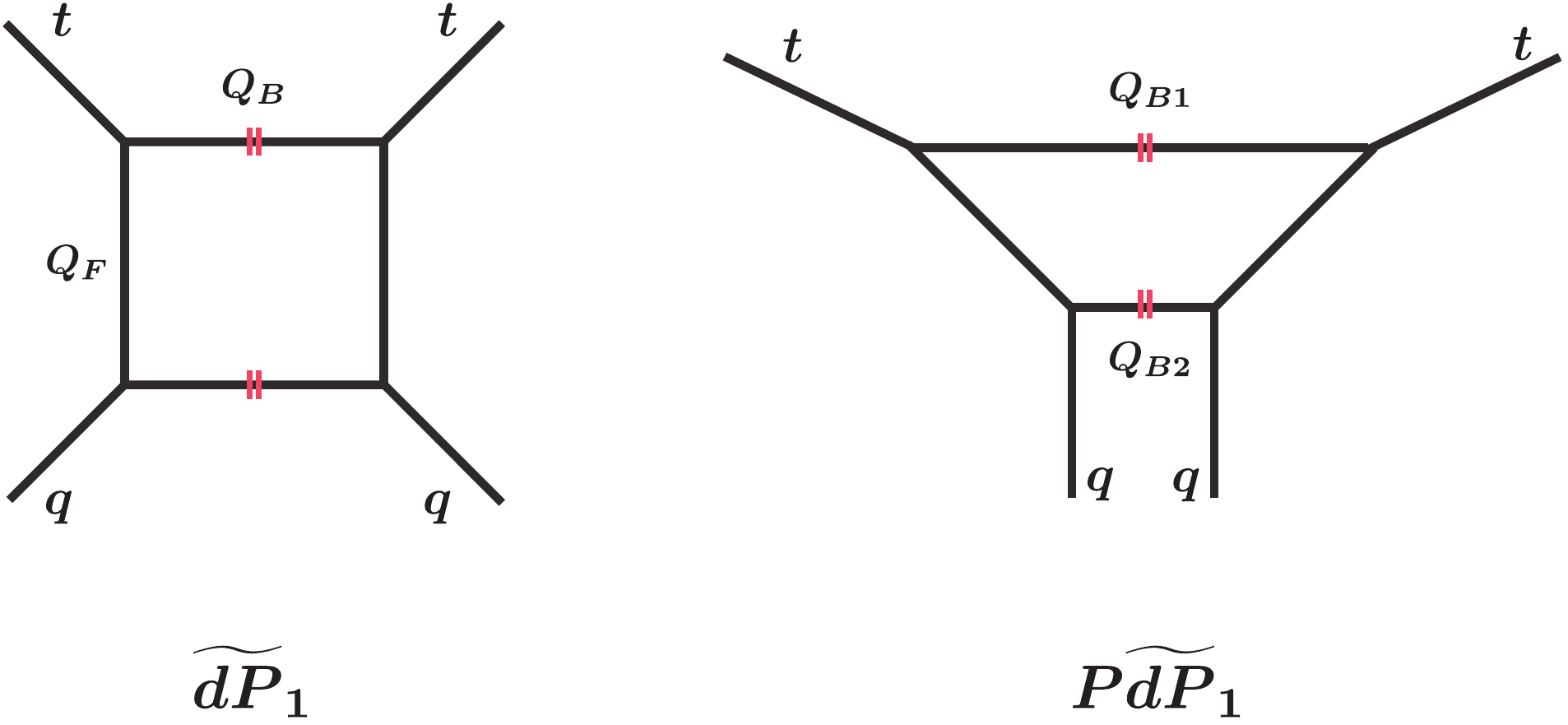}
 \end{center}
 \caption{The two toric diagrams associated with the $\hat{\bf E}_1$ configuration.}
 \label{fig;PdP1}
\end{figure}
Since these two geometries  share the same perturbative partition function,
we compare their instanton partition functions.
Using formulas in Appendix.B, we can compute the partition function of $\widetilde{dP}_1$
\begin{align}
Z_{\,\widetilde{dP}_1}(u,Q_F;t,q)=
\sum_{R_{1,2}}
(-Q_B)^{\vert\vec{R}\vert}\,
f_{R_1}^{-1}(t,q)\,f_{R_2}(t,q)
K_{R_1R_2}^{\,[{\boldsymbol{1}}]}(Q_F;t,q)
K_{R_2^TR_1^T}^{\,[{\boldsymbol{1}}]}(Q_F;q,t),
\end{align}
where the instanton factor is $u=Q_BQ_F^{-1}$.
The instanton part is then
\begin{align}
Z_{\,\widetilde{dP}_1}^{\,\textrm{inst.}}(u,Q_F;t,q)=
\sum_{R_{1,2}}
\left(u\frac{q}{t}\right)^{\vert\vec{R}\vert  }
Z^{\,\textrm{vect.}}_{\,\vec{R}}
(Q_{F};t,q).
\end{align}
This is the Nekrasov instanton partition function of
5d $SU(2)$ pure Yang-Mills theory.

We can also compute the partition function of $P\widetilde{dP}_1$ 
\begin{align}
&Z_{\,P\widetilde{dP}_1}(u,Q_F;t,q)\nonumber\\
&=
\sum_{R_{1,2}}
(-Q_{B1})^{\vert{R_1}\vert}\,(-Q_{B2})^{\vert{R_2}\vert}\,
f^{-3}_{R_1}(t,q)\,f_{R_2}^{-1}(t,q)
K_{R_1R_2}^{\,[{\boldsymbol{1}}]}(Q_F;t,q)
K_{R_2^TR_1^T}^{\,[{\boldsymbol{1}}]}(Q_F;q,t),
\end{align}
where the K\"ahler parameters associated with the base 2-cycle are
\begin{align}
Q_{B1}=uQ_F^2,\quad Q_{B1}=u.
\end{align}
The instanton partition function is then
\begin{align}
Z_{\,P\widetilde{dP}_1}^{\,\textrm{inst.}}(u,Q_F;t,q)=
\sum_{R_{1,2}}
\left(u\frac{q}{t}\right)^{\vert\vec{R}\vert  }
Z^{\,\textrm{CS,}m=2}_{\,\vec{R}}(Q_F;t,q)\,
Z^{\,\textrm{vect.}}_{\,\vec{R}}
(Q_{F};t,q).
\end{align}
This is the Nekrasov instanton partition function of
5d $SU(2)$ Yang-Mills theory with the non-zero Cern-Simons level $m=2$.

We can easily compute the one-instanton parts of these partition functions
that are the terms proportional to $u^1$
\begin{align}
&Z_{\,\widetilde{dP}_1}^{\,1\textrm{-inst.}}(Q_F;t,q)=\frac{q}{t}
\frac{1+\frac{q}{t}}
{(1-q)(1-t^{-1})(1-Q_Ft^{-1}q)(1-Q_F^{-1}t^{-1}q)},
\nonumber\\
&Z_{\,P\widetilde{dP}_1}^{\,1\textrm{-inst.}}(Q_F;t,q)=
\left(\frac{q}{t}\right)^2
\frac{Q_F+1+\frac{1}{Q_F}-\frac{q}{t}}
{(1-q)(1-t^{-1})(1-Q_Ft^{-1}q)(1-Q_F^{-1}t^{-1}q)}.
\end{align}
They are obviously different rational functions.
We, however, expect that the nontrivial relation (\ref{ZdP1til}) holds for them
and there is certain simple connection between them.  
Remarkably, the $Q_F$-dependence disappears once one computes
the difference 
\begin{align}
Z_{\,\widetilde{dP}_1}^{\,1\textrm{-inst.}}(Q_F;t,q)
-
Z_{\,P\widetilde{dP}_1}^{\,1\textrm{-inst.}}(Q_F;t,q)
=
-\frac{q}
{(1-q)(1-t)}.
\end{align}
This term precisely cancels the contribution from the non-full spin content because
the instanton expansion of the inversed extra contribution is
\begin{align}
\label{ZextraPdP1}
\frac{1}{Z_{\,\textrm{extra}}(u;q,t)}
&=\prod_{i,j=1}^\infty
({1-u\,t^{i-1}q^{j}})\nonumber\\
&=1-u\frac{q}
{(1-q)(1-t)}
+u^2\frac{q^2(t+q)}
{(1-q)^2(1-t)^2(1+q)(1+t)}
+\cdots.
\end{align}
This result confirms the one-instanton part of the conjectural relation
\begin{align}
Z_{\,\widetilde{dP}_1}^{\,\textrm{inst.}}(u,Q_F;t,q)
=\frac{Z_{\,P\widetilde{dP}_1}^{\,\textrm{inst.}}(u,Q_F;t,q)}
{Z_{\,\textrm{extra}}(u;q,t)}.
\end{align}
We can also prove it at the two-instanton level by showing
the two-instanton part of the relation (\ref{ZdP1til})
\begin{align}
&Z_{\,\widetilde{dP}_1}^{\,2\textrm{-inst.}}(Q_F;t,q)
-
Z_{\,P\widetilde{dP}_1}^{\,2\textrm{-inst.}}(Q_F;t,q)\nonumber\\
&=
-\frac{q}
{(1-q)(1-t)}
Z_{\,P\widetilde{dP}_1}^{\,1\textrm{-inst.}}(Q_F;t,q)+
\frac{q^2(t+q)}
{(1-q)^2(1-t)^2(1+q)(1+t)}.
\end{align}
The $Q_F$-dependence in the denominator is cancelled out again,
and this difference in the two-instanton part can be collected into the expected extra factor (\ref{ZextraPdP1}).
This test based on instanton expansion tells us that
some unknown mathematical structure simplifies and factorizes the discrepancy between 
these two different partition functions.

\subsection{The two toric phases for $\boldsymbol{E_2}$ theory}

\begin{figure}[tbp]
 \begin{center}
  \includegraphics[width=130mm, bb=0 0 582 250]{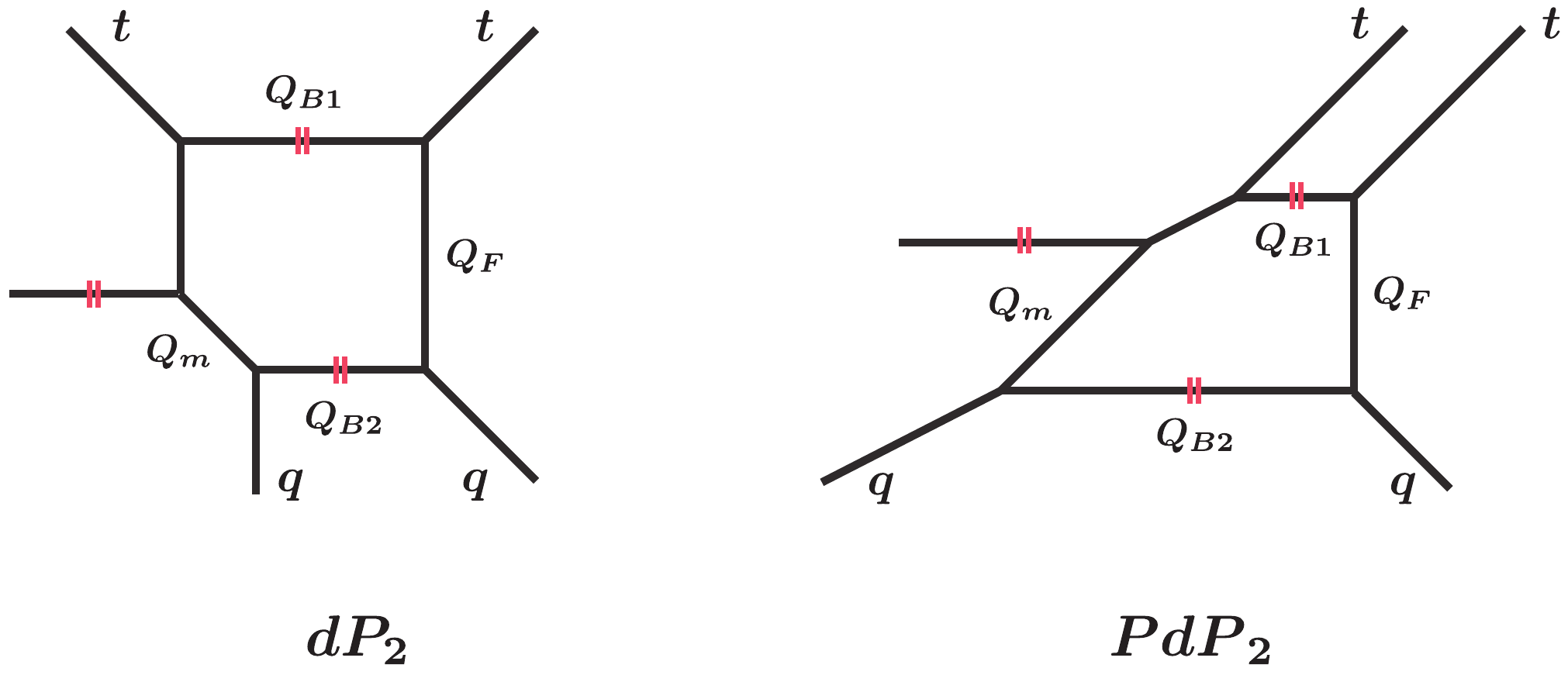}
 \end{center}
 \caption{The two toric diagrams associated with the $\hat{\bf E}_2$ configuration.}
 \label{fig;dP2}
\end{figure}
As we showed in the previous section,
there are two toric descriptions of the local del Pezzo $dP_2$.
Since $dP_2$ is toric,
we can compute its partition function directly by using the refined topological vertex formalism.
The web-diagram of $dP_2$  is illustrated in the left hand side
of \textbf{\textit{Figure}$\,$\textit{\ref{fig;dP2}}},
and this phase $dP_2$ was studied in \cite{BMPTY,HKN}.
We review their result for readers convenience.
Its partition function is
\begin{align}
Z_{\,{dP}_2}(u,Q_F,Q_m;t,q)=
&\sum_{R_{1,2}}
(-Q_{B1})^{\vert{R_1}\vert}\,(-Q_{B2})^{\vert{R_2}\vert}\nonumber\\&\times
f_{R_1}^{-1}(t,q)\,
K_{R_1R_2}^{\,[{\boldsymbol{0}},\boldsymbol{0}]}(Q_F,Q_m;t,q)
K_{R_2^TR_1^T}^{\,[{\boldsymbol{1}}]}(Q_F;q,t),
\end{align}
where the K\"ahler parameters for the base direction are
\begin{align}
Q_{B1}=uQ_F,\quad Q_{B2}=\frac{uQ_F}{Q_m}.
\end{align}
The perturbative and instanton parts are therefore
\begin{align}
&Z_{\,{dP}_2}^{\,\textrm{pert.}}(u,Q_F,Q_m;t,q)=
Z^{\,\textrm{vect.}}_{\,\textrm{pert.}}(Q_{F};t,q)
Z^{\,\textrm{matt.}}_{\,\textrm{pert.}}(Q_{F},Q_m;t,q),
\\
&\label{Z-dP2}Z_{\,{dP}_2}^{\,\textrm{inst.}}(u,Q_F,Q_m;t,q)=
\sum_{R_{1,2}}
\left(u\frac{q}{t}\right)^{\vert\vec{R}\vert  }
Z^{\,\textrm{vect.}}_{\,\vec{R}}(Q_{F};t,q)
Z^{\,\textrm{matt.}}_{\,\vec{R}}
(Q_{F},Q_m;t,q).
\end{align}
The full partition function is the product of these two functions.
This is the Nekrasov partition function of
5d $SU(2)$ gauge theory with single fundamental matter multiplet.

\subsubsection*{Pseudo del Pezzo Phase: $\boldsymbol{PdP_2}$}
We have the another description of the same system that is
based on the pseudo del Pezzo $PdP_2$.
The local pseudo del Pezzo surface 
$PdP_2$  is illustrated in the right hand side
of  \textbf{\textit{Figure}$\,$\textit{\ref{fig;dP2}}},
and its partition function is
\begin{align}
Z_{\,{PdP}_2}(u,Q_F,Q_m;&t,q)=
\sum_{R_{1,2}}
(-Q_{B1})^{\vert{R_1}\vert}\,(-Q_{B2})^{\vert{R_2}\vert}\nonumber\\&\times
f_{R_1}(t,q)\,f^2_{R_2^T}(t,q)
K_{R_1R_2}^{\,[{\boldsymbol{0}},\boldsymbol{0}]}(Q_F,Q_m;t,q)
K_{R_2^TR_1^T}^{\,[{\boldsymbol{1}}]}(Q_F;q,t).
\end{align}
The K\"ahler parameters $Q_{B1,2}$ are
\begin{align}
Q_{B1}=u,\quad Q_{B2}=\frac{uQ_F^2}{Q_m}.
\end{align}
The instanton part of this partition function takes the form 
\begin{align}
Z_{\,{PdP}_2}^{\,\textrm{inst.}}(u,Q_F,Q_m;t,q)=
\sum_{R_{1,2}}&
\left(u\frac{q}{t}\right)^{\vert\vec{R}\vert  }
Z^{\,\textrm{CS,}m=-2}_{\,\vec{R}}(Q_F;t,q)\nonumber\\
&\times 
\label{Z-PdP2}Z^{\,\textrm{vect.}}_{\,\vec{R}}(Q_{F};t,q)
Z^{\,\textrm{matt.}}_{\,\vec{R}}
(Q_{F},Q_m;t,q).
\end{align}
This web configuration contains the extra contribution that is raised from single stack of two parallel
legs in \textbf{\textit{Figure}$\,$\textit{\ref{fig;dP2}}}
\begin{align}
Z_{\,\textrm{extra}}^{\,{PdP}_2}(u;t,q)=M(u;t,q).
\end{align}
The partition function (\ref{Z-PdP2}) is different from that of $dP_2$ (\ref{Z-dP2})
because this $PdP_2$ theory 
has the non-vanishing Chern-Simons level $m=2$ as the case of $P\widetilde{dP}_1$.

Recall that branch cut move of the 7-brane configuration for $\hat{{\bf E}}_2$
leads to the following conjectural relation between these two descriptions
\begin{align}
\label{conjdP2}
Z_{\,{dP}_2}(u,Q_F,Q_m;t,q)
=\frac{Z_{\,{PdP}_2}^{\,\textrm{inst.}}(u,Q_F,Q_m;t,q)}{Z^{\,{PdP}_2}_{\,\textrm{extra}}(u;t,q)}.
\end{align}
Let us check this conjecture.
Since the perturbative partition function is the same for these two phases,
we need to check the instanton part of this relation.
The one-instanton partition functions, which are the first order of $u$-expansion, are
\begin{align}
&Z_{\,{dP}_2}^{\,1\textrm{-inst.}}=
\frac{q}{t}
\frac{1+\frac{q}{t}
-\frac{1+Q_F}{Q_m}\sqrt{\frac{q}{t}}}
{(1-q)(1-t^{-1})(1-Q_Ft^{-1}q)(1-Q_F^{-1}t^{-1}q)},
\nonumber\\
&Z_{\,{PdP}_2}^{\,1\textrm{-inst.}}=
\frac{q}{t}
\frac{Q_F+1+
\frac{1}{Q_F}-\frac{t}{q}
-\frac{1+Q_F}{Q_m}\sqrt{\frac{q}{t}}
}
{(1-q)(1-t^{-1})(1-Q_Ft^{-1}q)(1-Q_F^{-1}t^{-1}q)},
\end{align}
and after computing the difference between them,
dependence on $Q_F$ and $Q_m$ disappears
\begin{align}
Z_{\,{dP}_2}^{\,1\textrm{-inst.}}
-
Z_{\,{PdP}_2}^{\,1\textrm{-inst.}}=-
\frac{t}{(1-q)(1-t)}.
\end{align}
This is precisely the one-instanton part of the inversed extra factor $\frac{1}{Z_{\,\textrm{extra}}^{\,{PdP}_2}}$,
and thus we can confirm our conjectural relation (\ref{conjdP2}).
Higher instanton check is also straightforward.
We can  check the two instanton part of the relation (\ref{conjdP2}) for instance.

\subsection{The four toric phases for $\boldsymbol{E_3}$ theory}

\begin{figure}[thbp]
 \begin{center}
  \includegraphics[width=50mm, bb=0 0 211 203]{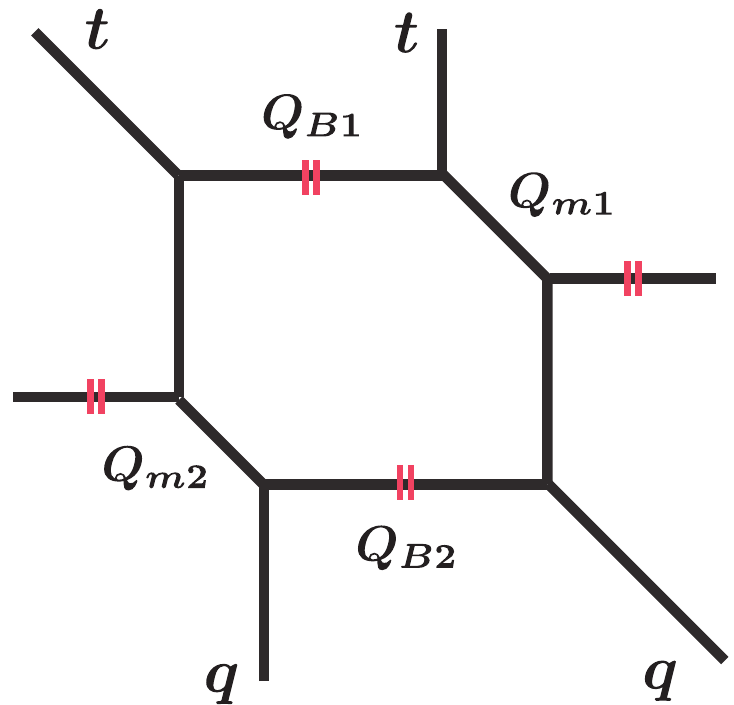}
 \end{center}
 \caption{The $dP_3$ toric diagram.}
 \label{fig;dP3web}
\end{figure}
Let us move on to the case of the third del Pezzo.
In this case we have four toric phases,
where one of them is the toric local del Pezzo $dP_3$,
and we expect the following relation
\begin{align}
\label{conjdP3}
Z_{dP_3}=\frac{Z_{PdP_3^{p}}}{Z^{PdP_3^{p}}_{\textrm{extra}}},\quad
p=I,\,II,\,III.
\end{align}
The web diagram of $dP_3$ is given in \textbf{\textit{Figure}$\,$\textit{\ref{fig;dP3web}}}.
The refined topological vertex formalism gives to the following partition function of $dP_3$
\begin{align}
&Z_{\,{dP}_3}(u,Q_F,Q_{m1,2};t,q)\nonumber\\&=
\sum_{R_{1,2}}
(-Q_{B1})^{\vert{R_1}\vert}\,(-Q_{B2})^{\vert{R_2}\vert}\,
K_{R_1R_2}^{\,[{\boldsymbol{0}},\boldsymbol{0}]}(Q_F,Q_{m2};t,q)
K_{R_2^TR_1^T}^{\,[{\boldsymbol{0}},\boldsymbol{0}]}(Q_F,Q_{m1};q,t),
\end{align}
where the K\"ahler parameters for the base direction are
\begin{align}
Q_{B1}=\frac{uQ_F}{Q_{m1}},\quad Q_{B2}=\frac{uQ_F}{Q_{m2}}.
\end{align}
The perturbative and the instanton partition functions are therefore
\begin{align}
&Z_{\,{dP}_3}(Q_F,Q_{m1,2};t,q)\rule{0pt}{3ex}=Z_{\,{dP}_3}^{\,\textrm{pert.}}(Q_F,Q_{m1,2};t,q)\,Z_{\,{dP}_3}^{\,\textrm{inst.}}(Q_F,Q_{m1,2};t,q),\\
&Z_{\,{dP}_3}^{\,\textrm{pert.}}(Q_F,Q_{m1,2};t,q)=Z_{\,\textrm{pert.}}^{\,\textrm{vect.}}(Q_F;t,q)
Z_{\,\textrm{pert.}}^{\,\textrm{matt.}}(Q_F,Q_{m1};t,q)
Z_{\,\textrm{pert.}}^{\,\textrm{matt.}}(Q_F,Q_{m2};t,q),\\
&Z_{\,{dP}_3}^{\,\textrm{inst.}}(u,Q_F,Q_{m1,2};t,q)=
\sum_{R_{1,2}}
\left(u\frac{q}{t}\right)^{\vert\vec{R}\vert  }
Z^{\,\textrm{vect.}}_{\,\vec{R}}(Q_{F};t,q)\nonumber\\
&\qquad\qquad\qquad\qquad\qquad
\times {Z'}^{\textrm{matt.}}_{\vec{R}}
(Q_{F},Q_{m1};t,q)
Z^{\,\textrm{matt.}}_{\,\vec{R}}
(Q_{F},Q_{m2};t,q).
\end{align}
Let us compare this partition function with those of pseudo del Pezzo surfaces.
In addition to the three phases of the pseudo del Pezzo surfaces $PdP_3^p$,
we have double assignments of the preferred direction in the cases $p=I,\,II$.
There are thus five patterns of  topological string partition function,
and we show that these partition functions lead to that of $dP_3$ through
the relation (\ref{conjdP3}).

\subsubsection*{Pseudo del Pezzo Phase I: $\boldsymbol{PdP_3^I}$}

\begin{figure}[tbp]
 \begin{center}
  \includegraphics[width=120mm, bb=0 0 530 245]{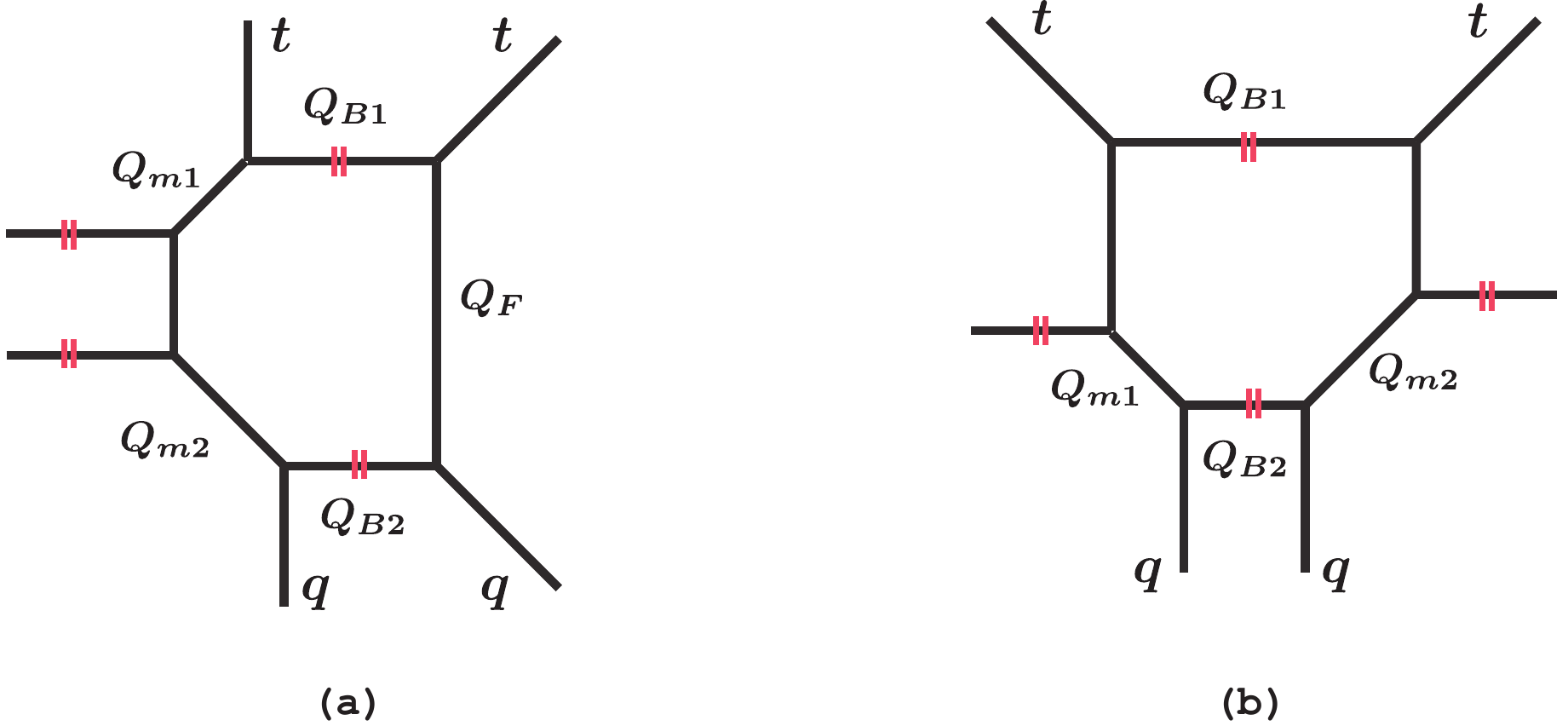}
 \end{center}
 \caption{The two choices of the preferred direction of the $PdP_3^I$ toric diagram.}
 \label{fig;PdP3I}
\end{figure}

The first phase $PdP_3^{I}$ is illustrated in \textbf{\textit{Figure}$\,$\textit{\ref{fig;PdP3I}}},
and we can find two choices {\bf (a,b)} of the preferred direction
that is denoted by red double lines.
This simple case $PdP_3^{I}$ was already studied in \cite{BMPTY,HKN},
but we review computation for readers convenience.
These two choices {\bf (a,b)} lead to different partition functions,
however they reduce to the same $dP_3$ partition function after removing the extra contributions
arising from their non-full spin contents. 
Let us star with the case {\bf (a)}.
The refined topological vertex formalism gives to the following partition function of $PdP_3^{I(a)}$
\begin{align}
&Z_{\,{PdP}_3^{I(a)}}(u,Q_F,Q_{m1,2};t,q)\nonumber\\&=
\sum_{R_{1,2}}
(-Q_{B1})^{\vert{R_1}\vert}\,(-Q_{B2})^{\vert{R_2}\vert}\,
K_{R_1R_2}^{\,[{\boldsymbol{0}},\boldsymbol{-1},\boldsymbol{0}]}(Q_F,Q_{m2},Q_{m1};t,q)
K_{R_2^TR_1^T}^{\,[{\boldsymbol{1}}]}(Q_F;q,t),
\end{align}
where the K\"ahler parameters $Q_{B1,2}$ are
\begin{align}
Q_{B1}=\frac{uQ_F}{Q_{m1}},\quad Q_{B2}=\frac{uQ_F}{Q_{m2}}.
\end{align}
Since the parallel external legs are horizontal,
their contribution is independent of the instanton factor $u$.
This extra factor therefore makes an effect only on the perturbative part,
and the partition function is the product of the following perturbative and instanton contributions 
\begin{align}
&Z_{\,{PdP}_3^{I(a)}}^{\,\textrm{pert.}}(Q_F,Q_{m1,2};t,q)=M\left(\frac{Q_F}{Q_{m1}Q_{m2}};t,q\right)\,Z_{\,{dP}_3}^{\,\textrm{pert.}}(Q_F,Q_{m1,2};t,q),
\\
&Z_{\,{PdP}_3^{I(a)}}^{\,\textrm{inst.}}(u,Q_F,Q_{m1,2};t,q)=Z_{\,{dP}_3}^{\,\textrm{inst.}}(u,Q_F,Q_{m1,2};t,q).
\end{align}
Because there is only single stack of parallel legs,
the factor $M({Q_F}({Q_{m1}Q_{m2}})^{-1};t,q)$
precisely gives the full extra contribution $Z^{\,{PdP}_3^{I(a)}}_{\,\textrm{extra}}$ coming from the non-full spin content.
We  can consequently prove the relation (\ref{conjdP3}) for this case at all order in the instanton expansion.

The next case {\bf (b)} comes with difficulty of proof of the relation
because the extra factor involves $u$-dependence and it affects instanton expansion drastically.
We will provide one-instanton check of the relation in the following.
The partition function of {\bf (b)} is
\begin{align}
Z_{\,{PdP}_3^{I(b)}}(u,Q_F,Q_{m1,2};t,q)=&
\sum_{R_{1,2}}
(-Q_{B1})^{\vert{R_1}\vert}\,(-Q_{B2})^{\vert{R_2}\vert}\,
f_{R_1^T}(q,t)\,f_{R_2^T}(q,t)\nonumber\\&\times
K_{R_1R_2}^{\,[{\boldsymbol{0}},\boldsymbol{0}]}\left(Q_F,{Q_{m1}};t,q\right)
K_{R_2^TR_1^T}^{\,[{\boldsymbol{0}},\boldsymbol{0}]}(Q_F,\frac{Q_F}{Q_{m2}};q,t),
\end{align}
where the K\"ahler parameters $Q_{B1,2}$ are
\begin{align}
Q_{B1}={uQ_F},\quad Q_{B2}=\frac{uQ_F}{Q_{m1}Q_{m2}}.
\end{align}
The perturbative partition function coincides with that of $dP_3$,
but
the instanton part takes the different form 
\begin{align}
&Z_{\,{PdP}_3^{I(b)}}^{\,\textrm{pert.}}(Q_F,Q_{m1,2};t,q)=Z_{\,{dP}_3}^{\,\textrm{pert.}}(Q_F,Q_{m1,2};t,q)
\\
&Z_{\,{PdP}_3^{I(b)}}^{\,\textrm{inst.}}(u,Q_F,Q_{m1,2};t,q)\nonumber\\
&
\label{Zinst-PdP3Ib}=\sum_{R_{1,2}}
\left(u\frac{q}{t}\right)^{\vert\vec{R}\vert  }
Z^{\,\textrm{vect.}}_{\,\vec{R}}(Q_{F};t,q)
Z^{\,\textrm{matt.}}_{\,\vec{R}}
(Q_{F},Q_{m1};t,q)
Z^{\,\textrm{matt.}}_{\,\vec{R}}
(Q_{F},Q_{m2};t,q).
\end{align}
Since the extra contribution arises from the downward parallel two legs in \textbf{\textit{Figure}$\,$\textit{\ref{fig;PdP3I}}},
its contribution is
\begin{align}
Z_{\,\textrm{extra}}^{\,{PdP}_3^{I(b)}}
=M(Q_{B2};q,t),
\end{align}
and this function depends on $u$ through $Q_{B2}$.

It is straightforward to compute the one-instanton partition functions of $PdP_3^{I(a,b)}$
by using above results
\begin{align}
&Z_{\,{dP}_3}^{\,1\textrm{-inst.}}(Q_F,Q_{m1,2};t,q)
=Z_{\,{PdP}_3^{I(a)}}^{\,1\textrm{-inst.}}(Q_F,Q_{m1,2};t,q)\nonumber\\
&\rule{0pt}{3ex}=
\frac{q}{t}
\frac{\left(1+\frac{q}{t}\right)\left( 1+\frac{Q_F}{Q_{m1}Q_{m2}} \right)-
\left( \frac{1}{Q_{m1}} + \frac{1}{Q_{m2}}  \right)\left( 1+{Q_F} \right)\sqrt{\frac{q}{t}}
}
{(1-q)(1-t^{-1})(1-Q_Ft^{-1}q)(1-Q_F^{-1}t^{-1}q)}
,\\
&Z_{\,{PdP}_3^{I(b)}}^{\,1\textrm{-inst.}}(Q_F,Q_{m1,2};t,q)\nonumber\\
&
\rule{0pt}{3ex}=
\frac{q}{t}
\frac{1+\frac{q}{t}-
\left( \frac{1}{Q_{m1}} + \frac{1}{Q_{m2}}  \right)\left( 1+{Q_F} \right)\sqrt{\frac{q}{t}}
+\frac{1}{Q_{m1}Q_{m2}}(Q_F^2+Q_F+1)\frac{q}{t}-\frac{Q_F}{Q_{m1}Q_{m2}}\left( \frac{q}{t} \right)^2
}
{(1-q)(1-t^{-1})(1-Q_Ft^{-1}q)(1-Q_F^{-1}t^{-1}q)}
.
\end{align}
The difference between these partition functions is the following simple function
\begin{align}
Z_{\,{dP}_3}^{\,1\textrm{-inst.}}-Z_{\,{PdP}_3^{I(b)}}^{\,1\textrm{-inst.}}
=-\frac{Q_F}{Q_{m1}Q_{m2}}
\frac{q}{(1-q)(1-t)}.
\end{align}
This is precisely the one-instanton part of our conjecture
\begin{align}
\label{conjPdP3Ib}
Z_{\,{dP}_3}=\frac{Z_{\,{PdP}_3^{I(b)}}}{M\left(\frac{uQ_F}{Q_{m1}Q_{m2}};q,t \right)}.
\end{align}

\begin{figure}[tbp]
 \begin{center}
  \includegraphics[width=120mm, bb=0 0 549 284]{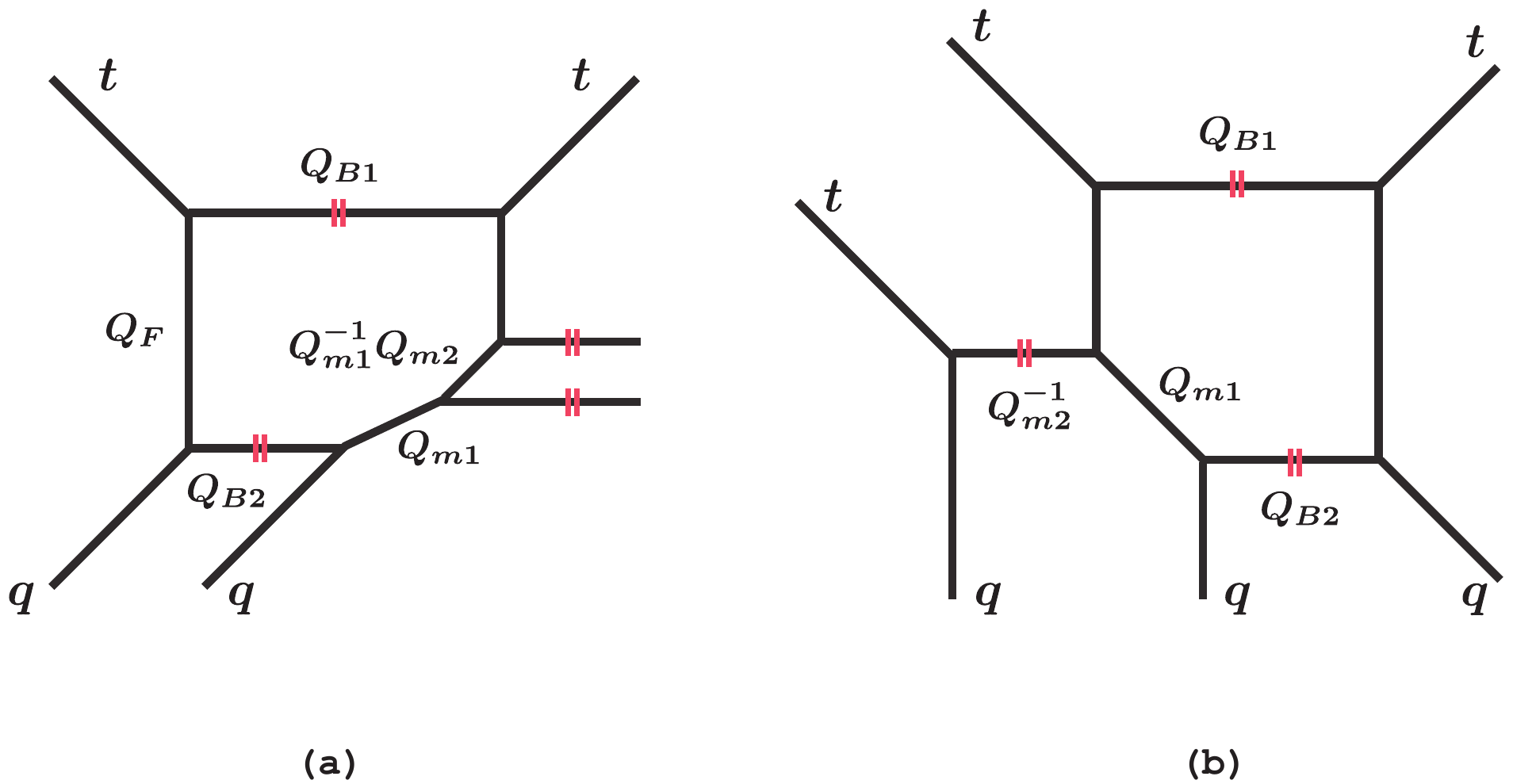}
 \end{center}
 \caption{The two choices of the preferred direction of the $PdP_3^{II}$ toric diagram.}
 \label{fig;PdP3II}
\end{figure}
\subsubsection*{Pseudo del Pezzo Phase II: $\boldsymbol{PdP_3^{II}}$}
The second phase of $PdP_3$ has two choices of preferred direction,
and these two cases are illustrated in \textbf{\textit{Figure}$\,$\textit{\ref{fig;PdP3II}}}.
Since the diagrams {\bf (a)} and {\bf(b)} are related though $SL(2,\mathbb{Z})$ transformation and flop transition\footnote{We employ
the flop transition \cite{Taki:2008hb,Iqbal:2012mt}
to avoid using the new vertex function \cite{Iqbal:2012mt}.},
these two setups describe the same toric manifold.
We start with the case {\bf (a)}.
The refined topological vertex formalism yields the following expression 
\begin{align}
Z_{\,{PdP}_3^{II(a)}}(u,Q_F,Q_{m1,2};t,q)&=
\sum_{R_{1,2}}
(-Q_{B1})^{\vert{R_1}\vert}\,(-Q_{B2})^{\vert{R_2}\vert}\,
f^{-1}_{R_1}(q,t)\,f^{-1}_{R_2}(q,t)\nonumber\\&
\quad\times K_{R_1R_2}^{\,[{\boldsymbol{1}}]}(Q_F;t,q)
K_{R_2^TR_1^T}^{\,[{\boldsymbol{0}},\boldsymbol{-1},\boldsymbol{0}]}\left(Q_F,\frac{Q_F}{Q_{m2}},Q_{m1};q,t\right).
\end{align}
The parameters $Q_{B1,2}$ are given by $u$ as
\begin{align}
Q_{B1}=uQ_F,\quad
Q_{B2}=\frac{uQ_F}{Q_{m1}Q_{m2}},
\end{align}
and then the partition function coming from
two stacks
takes the following form
\begin{align}
Z_{\,{PdP}_3^{II(a)}}=&
M
\left(\frac{Q_{m2}}{Q_{m1}};q,t
\right)\,
Z_{\,\textrm{pert}}^{\,\textrm{vect}}(Q_F)\,
Z_{\,\textrm{pert}}^{\,\textrm{matter}}(Q_{m1})\,
Z_{\,\textrm{pert}}^{\,\textrm{matter}}(Q_{m2})\nonumber\\
&\times\sum_{R_{1,2}}
\left(u\frac{q}{t} \right)^{\vert \vec{R}\vert}
\,Z_{\,\vec{R}}^{\,\textrm{vect.}}(Q_F)\,
Z_{\,\vec{R}}^{\,\textrm{matt.}}(Q_{m1})\,
Z_{\,\vec{R}}^{\,\textrm{matt.}}(Q_{m2}).
\end{align}
Notice that the instanton part of this result is equal to (\ref{Zinst-PdP3Ib}).
The extra contribution is given by
\begin{align}
Z^{\,{PdP}_3^{II(a)}}_{\,\textrm{extra}}=M\left(\frac{Q_{m2}}{Q_{m1}};q,t
\right)\,
M\left(\frac{uQ_F}{Q_{m1}Q_{m2}};q,t
\right),
\end{align}
and then we can show that this partition function is equivalent to that of $PdP_3^{I(b)}$ after removing the extra contributions as follows
\begin{align}
\frac{Z_{\,{PdP}_3^{I(b)}}}
{Z^{\,{PdP}_3^{I(b)}}_{\,\textrm{extra}}}
=\frac{Z_{\,{PdP}_3^{II(a)}}}
{Z^{\,{PdP}_3^{II(a)}}_{\,\textrm{extra}}}.
\end{align}
We therefore reduce the conjecture (\ref{conjdP3}) in this case
to that for the $PdP_3^{I}$ phase (\ref{conjPdP3Ib}).

\begin{figure}[tbp]
 \begin{center}
  \includegraphics[width=50mm, bb=0 0 242 207]{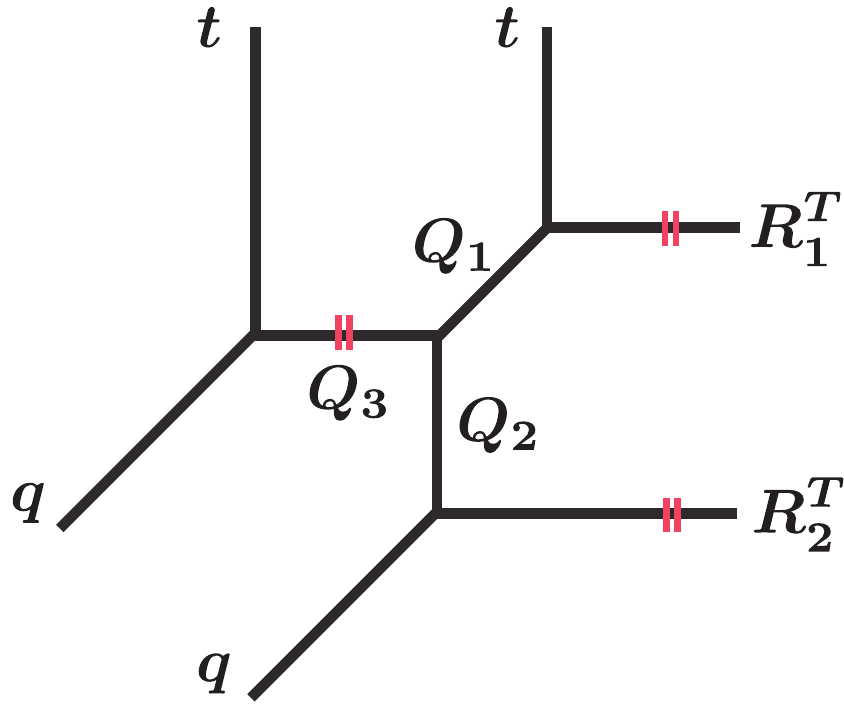}
 \end{center}
 \caption{The partition function of $T^2$ geometry with two non-empty Young diagrams on two adjoining external legs.}
 \label{fig;T2}
\end{figure}

Let us move on to the second choice of the preferred direction of  $PdP_3^{II}$ diagram
that is illustrated in {\bf (b)} of \textbf{\textit{Figure}$\,$\textit{\ref{fig;PdP3II}}}.
This case is very non-trivial because
the corresponding topological string partition function
is given by gluing a strip geometry \cite{Iqbal:2004ne}
and the $T^2$ geometry which is a typical off-strip geometry \textbf{\textit{Figure}$\,$\textit{\ref{fig;T2}}}.
We therefore need to compute the topological string partition function of $T^2$ with two parallel external legs
with non-empty Young diagrams.
Unfortunately, it is very hard to compute exactly a partition function of such an off-strip geometry.
This is because in this computation we confront certain summation over the Young diagrams
that we can not evaluate by any combinatorial formula in existence.
We hence compute this partition function up to certain order as a power series in a exponentiated K\"ahler parameter $Q_3$.

A noteworthy exception is the case with empty Young diagrams $R_1=R_2=\emptyset$ in \textbf{\textit{Figure}$\,$\textit{\ref{fig;T2}}},
and we can find the following closed expression.
The partition function in this case was recently computed in \cite{BMPTY}\footnote{The unrefined version of $T^2$ partition function
was computed in \cite{Sulkowski:2006jp}.}
\begin{align}
\label{T2}
Z_{\,T^2}(Q_1,Q_2,Q_3,t,q)
=
\prod_{i,j=1}^\infty
\frac{
\left(1-Q_1Q_2Q_3t^{i-\frac{1}{2}}q^{j-\frac{1}{2}}\right)
\prod_{\ell=1}^3
\left(1-Q_{\ell}t^{i-\frac{1}{2}}q^{j-\frac{1}{2}}\right)
}{
\left(1-Q_1Q_2t^{i-1}q^{j}\right)
\left(1-Q_2Q_3t^{i-1}q^{j}\right)
\left(1-Q_1Q_3t^{i}q^{j-1}\right)
}.
\end{align}
This closed expression was first observed in \cite{Kozcaz:2010af}.

The topological string partition function with generic assignment of Young diagrams \textbf{\textit{Figure}$\,$\textit{\ref{fig;T2}}}
is far more complicated.
The topological vertex formalism yields
\begin{align}
&K^{\,T^2}_{\,\vec{R}}
(Q_1,Q_2,Q_3,t,q)\nonumber\\
&=
\sum_{Y_{1,2,3}}
\prod_{\ell=1}^3(-Q_\ell)^{\vert Y_\ell\vert}\,
C_{Y_1^T\emptyset R_1^T}(q,t)\,
C_{\emptyset Y_2^T R_2^T}(q,t)\,
C_{\emptyset\emptyset Y_3^T}(q,t)\,
C_{Y_1Y_2Y_3}(t,q).
\end{align}
In contrast to the cases of strip geometries,
there is no formulas to calculate all three summations over the Young diagrams $Y_{1,2,3}$. 
Using the Cauchy formulas reduces it to the following expression with a remaining summation over $Y_3$
\begin{align}
K^{\,T^2}_{\,\vec{R}}
%(Q_1,Q_2,Q_3,t,q)
&=K_{\,\vec{R}}^{\,[{\boldsymbol 1}]}(Q_1Q_2,t,q)
\sum_{Y_3}(-Q_3)^{\vert Y_3\vert}\,
t^{\frac{\parallel Y_3^T\parallel^2}{2}}
q^{\frac{\parallel Y_3\parallel^2}{2}}
\widetilde{Z}_{Y_3}(t,q)\,
\widetilde{Z}_{Y^T_3}(q,t)\nonumber\\
&\times\prod_{i,j=1}^\infty
\left(1-Q_1t^{-R_{1j}^T+i-\frac{1}{2}}q^{-Y_{3i}+j-\frac{1}{2}}\right)
\left(1-Q_1t^{-Y_{3j}^T+i-\frac{1}{2}}q^{-R_{2i}+j-\frac{1}{2}}\right).
\end{align}
The extra contribution coming from non-full spin content on $T^2$ local geometry 
is included in the expression (\ref{T2}) as $M(Q_2Q_3;q,t)M(Q_1Q_3;t,q)$,
and thus we normalize the partition function by this function $Z_{T^2}$.
The topological string partition function is then
\begin{align}
&K^{\,T^2}_{\,\vec{R}}
(Q_1,Q_2,Q_3,t,q)\nonumber\\&
\label{KT2}
=\frac{Z_{\,T^2}(Q_1,Q_2,Q_3,t,q)}{M(Q_1Q_2;q,t)}
{K_{\,\vec{R}}^{\,[{\boldsymbol 1}]}(Q_1Q_2,t,q)}
P_{\,\vec{R}}(Q_1,Q_2,Q_3,t,q).
\end{align}
The remarkable characteristics of this function is that
$P_{\,\vec{R}}(Q_1,Q_2,Q_3,t,q)$ is a polynomial in $Q_3$ even though this function is defined as a ratio of infinite power series in $Q_3$
as was first observed in \cite{HKN,BMPTY}.
Let us consider simplest case $\vec{R}=([1],\emptyset)$
\begin{align}
&P_{([1],\emptyset)}(Q_1,Q_2,Q_3,t,q)
%\nonumber\\&=\rule{0pt}{3ex}
=\frac{
\prod_{i,j}
\left(1-Q_2Q_3t^{i-1}q^{j}\right)
\left(1-Q_1Q_3t^{i}q^{j-1}\right)}{\prod_{i,j}
\left(1-Q_1Q_2Q_3t^{i-\frac{1}{2}}q^{j-\frac{1}{2}}\right)
\left(1-Q_{3}t^{i-\frac{1}{2}}q^{j-\frac{1}{2}}\right)}
\nonumber\\&\times\rule{0pt}{3ex}
\sum_{Y}(-Q_3)^{\vert Y\vert}\,t^{\frac{\parallel Y^T\parallel^2}{2}}
q^{\frac{\parallel Y\parallel^2}{2}}
\widetilde{Z}_{Y}(t,q)\,
\widetilde{Z}_{Y^T}(q,t)
\nonumber\\
&\rule{0pt}{3ex}\times\prod_{s\in Y}
\left(1-Q_1t^{-\ell_{[1]}-\frac{1}{2}}q^{-a_Y-\frac{1}{2}}\right)
\left(1-Q_2t^{\ell_{\emptyset}+\frac{1}{2}}q^{a_Y+\frac{1}{2}}\right)
\prod_{s=(1,1)}
\left(1-Q_1t^{\ell_Y+\frac{1}{2}}q^{a_{[1]}+\frac{1}{2}}\right).
\end{align}
Computing this summation up to few order in $Q_3$,
we can confirm that this function is actually the following simple linear function of $Q_3$
because of certain cancellation mechanism 
\begin{align}
\label{Poly}
P_{([1],\emptyset)}(Q_1,Q_2,Q_3,t,q)
=1-\left(Q_1+Q_1Q_2Q_3\right)\sqrt{\frac{q}{t}}+Q_1Q_3.
\end{align}

Using this result on $T^2$ partition function,
we can compute the partition function of $PdP_3^{II(b)}$.
The refined topological vertex gives
\begin{align}
Z_{\,PdP_3^{II(b)}}&=\sum_{R_{1,2}}
(-Q_{B1})^{\vert R_1\vert}(-Q_{B2})^{\vert R_2\vert}
f_{R_1}^{-1}(t,q)\nonumber\\
&\times
K^{\,T^2}_{\,(R_1,R_2)}
(Q_1,Q_2,Q_3,t,q)\,
K_{\,(R_2^T,R_1^T)}^{\,[{\boldsymbol 1}]}(Q_F,q,t).
\end{align}
We introduce the following parametrization
\begin{align}
Q_F=Q_1Q_2,\quad
Q_2=Q_{m1},\quad
Q_3=\frac{1}{Q_{m2}}.
\end{align}
The partition function then takes the following form
\begin{align}
Z_{\,PdP_3^{II(b)}}&=
M(Q_2Q_3;q,t)\,M(Q_1Q_3;t,q)\,
Z_{\,\textrm{pert}}^{\,\textrm{vect}}(Q_F)\,
Z_{\,\textrm{pert}}^{\,\textrm{matter}}(Q_{m1})\,
Z_{\,\textrm{pert}}^{\,\textrm{matter}}(Q_{m2})
\nonumber\\
&\times
\sum_{R_{1,2}}
\left(u\frac{q}{t} \right)^{\vert \vec{R}\vert}
\,Z_{\,\vec{R}}^{\,\textrm{vect}}(Q_F)
(Q_{m1})^{-\vert R_2\vert}\,
f^{-1}_{R_2}(t,q)\,
P_{\,\vec{R}}(Q_{1,2,3};t,q).
\end{align}
Using (\ref{Poly}) gives the following one-instanton partition function
\begin{align}
Z_{\,{PdP}_3^{II(b)}}^{\,1\textrm{-inst.}}
=
\frac{q}{t}
\frac{\left(1+\frac{q}{t}\right)\left( 1+\frac{Q_F}{Q_{m1}Q_{m2}} \right)-
\left( \frac{1}{Q_{m1}} + \frac{1}{Q_{m2}}  \right)\left( 1+{Q_F} \right)\sqrt{\frac{q}{t}}
}
{(1-q)(1-t^{-1})(1-Q_Ft^{-1}q)(1-Q_F^{-1}t^{-1}q)}
=Z_{\,{dP}_3}^{\,1\textrm{-inst.}}.
\end{align}
We can expect this equality is valid to all order in the instanton expansion of $Z_{\,{PdP}_3^{II(b)}}^{\,\textrm{inst.}}=Z_{\,{dP}_3}^{\,\textrm{inst.}}$.
Moreover the extra contribution $M(Q_2Q_3;q,t)\,M(Q_1Q_3;t,q)$ does not make an effect on the instanton part.
Therefore, this result provides one-instanton check of the relation
\begin{align}
Z_{\,{dP}_3}=\frac{Z_{\,{PdP}_3^{II(b)}}}{Z^{\,{PdP}_3^{II(b)}}_{\,\textrm{extra}}}.
\end{align}

\subsubsection*{Pseudo del Pezzo Phase III: $\boldsymbol{PdP_3^{III}}$}
\begin{figure}[tbp]
 \begin{center}
  \includegraphics[width=82mm, bb=0 0 417 206]{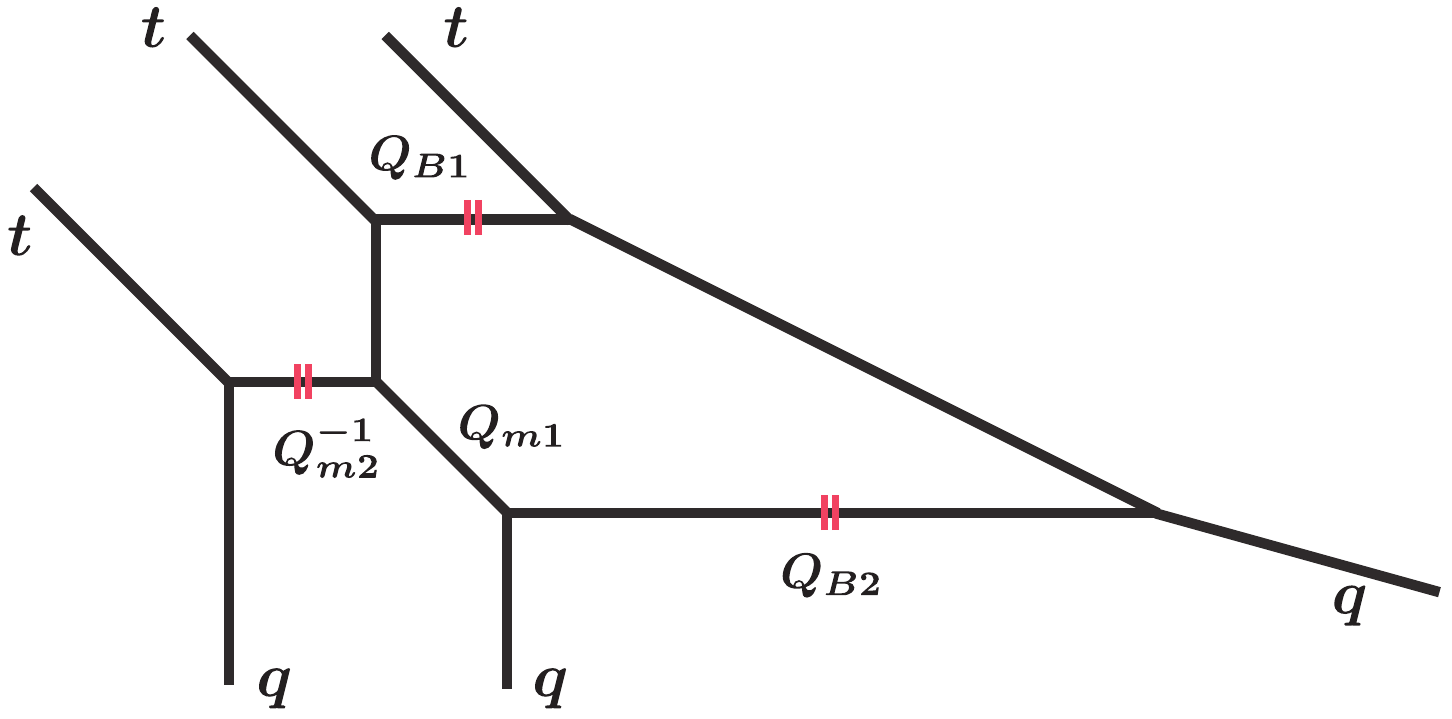}
 \end{center}
 \caption{The $PdP_3^{III}$ toric diagram.}
 \label{fig;PdP3III}
\end{figure}
The third phase is also nontrivial since it involves $T^2$ geometry as a toric sub-diagram.
The web diagram and the assignment of the preferred direction
are illustrated in \textbf{\textit{Figure}$\,$\textit{\ref{fig;PdP3III}}}.
We can decompose the web into $T^2$ and a strip geometry in the right hand side.
Gluing the topological string partition functions of the $T^2$ and  strip sub-geometries
yields the the partition function of $PdP_3^{III}$
\begin{align}
Z_{\,PdP_3^{III}}&=\sum_{R_{1,2}}
(-Q_{B1})^{\vert R_1\vert}(-Q_{B2})^{\vert R_2\vert}
f_{R_1}(t,q)\,\left(f_{R_2}(t,q)\right)^2\nonumber\\
&\times
K^{\,T^2}_{\,(R_1,R_2)}
(Q_F,Q_1,Q_2,Q_3,t,q)\,
K_{\,(R_2^T,R_1^T)}^{\,[{\boldsymbol 1}]}(Q_1Q_2,q,t).
\end{align}
The instanton factor $u$ is given by
\begin{align}
Q_{B1}=u,\quad Q_{B2}=uQ_1(Q_2)^2.
\end{align}
The partition function then takes the following form
\begin{align}
&Z_{\,PdP_3^{III}}
=
M(Q_2Q_3;q,t)\,M(Q_1Q_3;t,q)\,
Z_{\,\textrm{pert}}^{\,\textrm{vect}}(Q_1Q_2)\,
Z_{\,\textrm{pert}}^{\,\textrm{matter}}(Q_2)\,
Z_{\,\textrm{pert}}^{\,\textrm{matter}}(Q_3^{-1})
\nonumber\\
&\times
\sum_{R_{1,2}}
\left(u\frac{q}{t} \right)^{\vert \vec{R}\vert}
\,Z_{\,\vec{R}}^{\,\textrm{vect}}(Q_1Q_2)
(Q_1Q_2)^{-\vert R_1\vert}\,
(Q_2)^{\vert R_2\vert}
\,f^2_{R_1}(t,q)\,
f_{R_2}(t,q)\,
P_{\,\vec{R}}(Q_{1,2,3};t,q).
\end{align}
Let us compare this result with the del Pezzo partition function.
To ensure the coincidence between perturbative partition functions,
the Coulomb branch and mass parameters in the Nekrasov partition function are introduced as
\begin{align}
Q_F=Q_1Q_2,\quad Q_{2}=Q_{m1},\quad Q_{3}=\frac{1}{Q_{m2}}.
\end{align}
Then the one instanton part is given by
\begin{align}
Z_{\,PdP_3^{III}}^{\,1{\textrm{-inst.}}}=\frac{q}{t}\frac{
\left( 1+\frac{Q_F}{Q_{m1}Q_{m2}}\right)
\left( -\frac{t}{q}+Q_F+1+\frac{1}{Q_F}\right)
-\sqrt{\frac{q}{t}}\left(\frac{1}{Q_{m1}} +\frac{1}{Q_{m2}} \right)(1+Q_F)
}{(1-q)(1-t^{-1})(1-Q_F^{-1}t^{-1}q)(1-Q_Ft^{-1}q)}.
\end{align}
It is easy to see that the difference between the following partition functions takes simple form
\begin{align}
Z_{\,dP_3}^{\,1{\textrm{-inst.}}}-Z_{\,PdP_3^{III}}^{\,1{\textrm{-inst.}}}
=
-\left(1+\frac{Q_F}{Q_{m1}Q_{m2}}\right)
\frac{t}{(1-q)(1-t)}.
\end{align}
This provides the one instanton confirmation of the expected relation
\begin{align}
Z_{\,dP_3}^{\,{\textrm{inst.}}}
=\frac{Z_{\,PdP_3^{III}}^{\,{\textrm{inst.}}}}{
M(u;t,q)\,M\left(\frac{uQ_F}{Q_{m1}Q_{m2}};t,q\right)}.
\end{align}
Meanwhile it is easy to show that the perturbative part satisfies
\begin{align}
Z_{\,dP_3}^{\,{\textrm{pert.}}}
=\frac{Z_{\,PdP_3^{III}}^{\,{\textrm{pert.}}}}{
M\left(\frac{Q_F}{Q_{m1}Q_{m2}};t,q\right)
\,M\left(\frac{Q_{m1}}{Q_{m2}};q,t\right)
}.
\end{align}
These results are consistent with our expression of the extra part of the partition function 
\begin{align}
Z^{\,PdP_3^{III}}_{\,{\textrm{extra}}}=
M\left(\frac{Q_{m1}}{Q_{m2}};q,t\right)\,M\left(\frac{Q_F}{Q_{m1}Q_{m2}};t,q\right)
\,M(u;t,q)\,M\left(\frac{uQ_F}{Q_{m1}Q_{m2}};t,q\right).
\end{align}
The relation (\ref{conjecturePdP3III}) is consequently satisfied 
in one-instanton level.

\subsection{The toric phases for $\boldsymbol{E_4}$ theory}

Since the local del Pezzo surface $dP_4$
is non-toric,
we can not compute its partition function directly by using the refined topological vertex formalism.
However we have two pseudo del Pezzo descriptions of $dP_4$,
and we can expect that the $dP_4$ partition function coincides with those of toric $PdP_4$s
after removing extra contribution.

\subsubsection*{Pseudo del Pezzo Phase I: $\boldsymbol{PdP_4^{I}}$}
\begin{figure}[tbp]
 \begin{center}
  \includegraphics[width=47mm, bb=0 0 245 246]{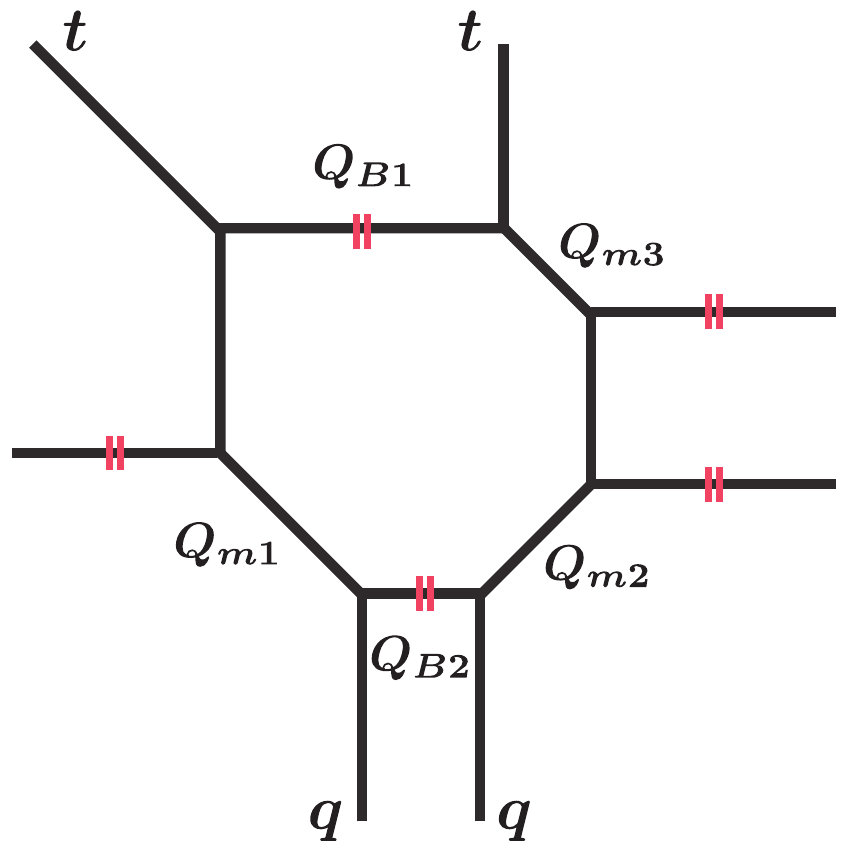}
 \end{center}
 \caption{The $PdP_4^{I}$ toric diagram.}
 \label{fig;PdP4I}
\end{figure}
The first phase $PdP_4^I$ of $dP_4$ is illustrated in \textbf{\textit{Figure}$\,$\textit{\ref{fig;PdP4I}}}.
This simple case was already studied in \cite{BMPTY,HKN},
but we review computation for readers convenience.
The gauge theory parameters are given by 
\begin{align}
Q_{B1}=\frac{uQ_F}{Q_{m3}},\quad Q_{B2}=\frac{uQ_F}{Q_{m1}Q_{m2}}.
\end{align}
Using the refined topological vertex formalism gives the following expression of the  $PdP_4^I$  partition function
\begin{align}
Z_{\,PdP_4^{I}}
=&\sum_{R_{1,2}}
(-Q_{B1})^{\vert R_1\vert}(-Q_{B2})^{\vert R_2\vert}
\left(f_{R_2}(t,q)\right)^{-1}
\nonumber\\&
\rule{0pt}{3ex}\times
K^{\,[{\boldsymbol 0},{\boldsymbol{-1}}]}_{\,(R_1,R_2)}
(Q_F,Q_{m1};t,q)\,
K_{\,(R_2^T,R_1^T)}^{\,[{\boldsymbol 0},{\boldsymbol{-1}},{\boldsymbol 0}]}(Q_F,Q_{m3},Q_{m2};q,t)
=Z_{\,PdP_4^{I}}^{\,{\textrm{pert.}}}\,
Z_{\,PdP_4^{I}}^{\,{\textrm{inst.}}},
\end{align}
where the perturbative and instanton parts are
\begin{align}
&Z_{\,PdP_4^{I}}^{\,{\textrm{pert.}}}
=M\left(\frac{Q_{F}}{Q_{m2}Q_{m3}};q,t\right)\,
Z_{\,\textrm{pert}}^{\,\textrm{vect}}(Q_F)\,
\prod_{f=1}^3Z_{\,\textrm{pert}}^{\,\textrm{matter}}(Q_{mf}),
\\
&Z_{\,PdP_4^{I}}^{\,{\textrm{inst.}}}
=
\sum_{R_{1,2}}
\left(u\frac{q}{t} \right)^{\vert \vec{R}\vert}
\,Z_{\,\vec{R}}^{\,\textrm{vect}}(Q_F)\,
{Z}^{\textrm{matt.}}_{\vec{R}}
(Q_{m1})\,
Z^{\,\textrm{matt.}}_{\,\vec{R}}
(Q_{m2})\,
{Z'}^{\textrm{matt.}}_{\vec{R}}
(Q_{m3})
.
\end{align}

The web of the toric geometry $PdP_4^{I}$ has two stacks of two parallel external legs.
The extra contribution  thus takes the following form
\begin{align}
Z^{\,PdP_4^{I}}_{\,{\textrm{extra}}}=
M\left(\frac{uQ_F}{Q_{m1}Q_{m2}};q,t\right)\,M\left(\frac{Q_F}{Q_{m2}Q_{m3}};q,t\right).
\end{align}
Since the factor $M\left({Q_F}/{Q_{m2}Q_{m3}};t,q\right)$ appears in the overall coefficient and does not depend on the
instanton factor $u$,
we can recast our conjecture in terms of the instanton partition functions
\begin{align}
\label{conjPdP4I}
Z_{\,dP_4}^{\,{\textrm{inst.}}}=
\frac{Z_{\,PdP_4^{I}}^{\,{\textrm{inst.}}}}{M\left(\frac{uQ_F}{Q_{m1}Q_{m2}};q,t\right)}.
\end{align}
This relation was checked in \cite{BMPTY,HKN} by comparing it with 
the Nekrasov partition function 
and the superconformal index of the corresponding $Sp(1)$ gauge theory \cite{Kim:2012gu} .
In this article we employ an another point of view:
this is not the unique topological string expression of the  $dP_4$  partition function
because there are other toric phases.
We will  compare the above partition function with those of other phases in the following.
This computation provides an another check of our conjecture.

\subsubsection*{Pseudo del Pezzo Phase II: $\boldsymbol{PdP_4^{II}}$}

\begin{figure}[tbp]
 \begin{center}
  \includegraphics[width=120mm, bb=0 0 584 273]{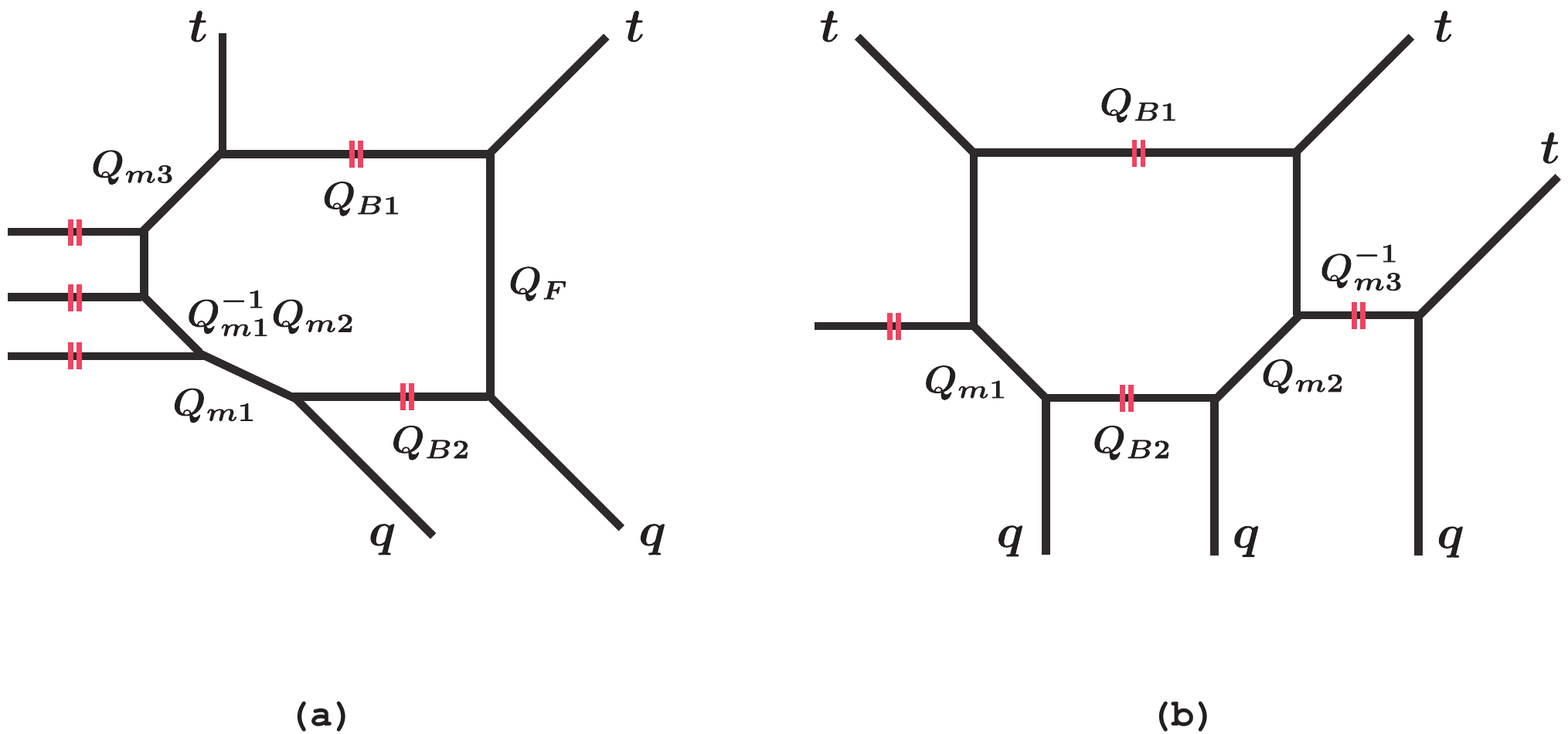}
 \end{center}
 \caption{The two choices of the preferred direction of the $PdP_4^{II}$ toric diagram.}
 \label{fig;PdP4II}
\end{figure}
There are two choices of the preferred direction of the 
web-diagram of $PdP_4^{II}$ as illustrated in \textbf{\textit{Figure}$\,$\textit{\ref{fig;PdP4II}}}.
Let us start with the case of {\bf (a)}.
In this case the instanton factor given by 
\begin{align}
Q_{B1}=\frac{uQ_F}{Q_{m3}},\quad Q_{B2}=\frac{uQ_F}{Q_{m1}Q_{m2}}.
\end{align}
Applying the refined topological vergec formalism to this choice of the preferred direction
yields the following expression
\begin{align}
&Z_{\,PdP_4^{II(a)}}
=\sum_{R_{1,2}}
(-Q_{B1})^{\vert R_1\vert}(-Q_{B2})^{\vert R_2\vert}
f^{-1}_{R_1}(t,q)
\nonumber\\&
\rule{0pt}{3ex}\times
K^{\,[{\boldsymbol 0},{\boldsymbol{-1}},{\boldsymbol{-1}},{\boldsymbol 0}]}_{\,(R_1,R_2)}
(Q_F,Q_{m1},Q_{m2},Q_{m3};t,q)\,
K_{\,(R_2^T,R_1^T)}^{\,[{\boldsymbol 1}]}(Q_F;q,t)
=Z_{\,PdP_4^{II(a)}}^{\,{\textrm{pert.}}}\,
Z_{\,PdP_4^{II(a)}}^{\,{\textrm{inst.}}},
\end{align}
and we can easily show
\begin{align}
&Z_{\,PdP_4^{II(a)}}^{\,{\textrm{pert.}}}=
M\left(\frac{Q_{F}}{Q_{m2}Q_{m3}},\frac{Q_{F}}{Q_{m1}Q_{m3}};t,q\right)\,
Z_{\,\textrm{pert}}^{\,\textrm{vect}}(Q_F)\,
\prod_{f=1}^3Z_{\,\textrm{pert}}^{\,\textrm{matter}}(Q_{mf}),
\\
&Z_{\,PdP_4^{II(a)}}^{\,{\textrm{inst.}}}
=
%M\left(\frac{uQ_{F}}{Q_{m1}Q_{m2}};t,q\right)
Z_{\,PdP_4^{I}}^{\,{\textrm{inst.}}}
.
\end{align}
Since the stacks of parallel external legs lead to the extra contribution
\begin{align}
Z^{\,PdP_4^{II(a)}}_{\,{\textrm{extra}}}=
M\left(\frac{Q_{F}}{Q_{m2}Q_{m3}},\frac{Q_{F}}{Q_{m1}Q_{m3}};t,q\right)\,
M\left(\frac{uQ_{F}}{Q_{m1}Q_{m2}};t,q\right),
\end{align}
we obtain the following relation
\begin{align}
\frac{Z_{\,{PdP}_4^{I}}}{Z^{\,{PdP}_4^{I}}_{\,\textrm{extra}}}
=\frac{Z_{\,{PdP}_4^{II(a)}}}{Z^{\,{PdP}_4^{II(a)}}_{\,\textrm{extra}}}.
\end{align}
Assuming the relation (\ref{conjPdP4I}),
we can thus prove our conjecture for ${PdP}_4^{II(a)}$ in all order in the instanton expansion.

The second case {\bf (b)} in \textbf{\textit{Figure}$\,$\textit{\ref{fig;PdP4II}}}
is subtle because it involves the $T^2$ geometry.
The instanton factor $u$ given by 
\begin{align}
Q_{B1}={uQ_F},\quad Q_{B2}=\frac{uQ_F}{Q_{m1}Q_{m2}},
\end{align}
and the topological string  partition function is
\begin{align}
&Z_{\,PdP_4^{II(b)}}
=\sum_{R_{1,2}}
(-Q_{B1})^{\vert R_1\vert}(-Q_{B2})^{\vert R_2\vert}
f^{-1}_{R_1}(t,q)\,f^{-1}_{R_2}(t,q)
\nonumber\\&
\rule{0pt}{3ex}\times
K^{\,[{\boldsymbol 0},{\boldsymbol{-1}}]}_{\,(R_1,R_2)}
(Q_F,Q_{m1};t,q)\,
K_{\,(R_2^T,R_1^T)}^{\,T^2}(Q_F,Q_{m2},Q_FQ^{-1}_{m2},Q_{m3}^{-1};q,t)
=Z_{\,PdP_4^{II(b)}}^{\,{\textrm{pert.}}}\,
Z_{\,PdP_4^{II(b)}}^{\,{\textrm{inst.}}},
\end{align}
and we can show
\begin{align}
&\frac{Z_{\,PdP_4^{II(b)}}^{\,{\textrm{pert.}}}}{
M\left(\frac{Q_{F}}{Q_{m2}Q_{m3}};t,q\right)\,
M\left(\frac{Q_{m2}}{Q_{m3}};q,t\right)}
=
\frac{Z_{\,PdP_4^{I}}^{\,{\textrm{pert.}}}}{
M\left(\frac{Q_{F}}{Q_{m2}Q_{m3}};q,t\right)},
\\
&Z_{\,PdP_4^{II(b)}}^{\,{\textrm{inst.}}}
\rule{0pt}{3ex}=
\sum_{R_{1,2}}
\left(u\frac{q}{t} \right)^{\vert \vec{R}\vert}
(Q_{m2})^{\vert R_2\vert}\,f^{-1}_{R_2}(t,q)
\,Z_{\,\vec{R}}^{\,\textrm{vect}}(Q_F)\,
{Z}^{\textrm{matt.}}_{\vec{R}}
(Q_{m1})\nonumber\\
&\qquad\qquad\qquad\qquad\quad\qquad\qquad
\times
P_{\,(R_2^T,R_1^T)}(Q_{m2},Q_FQ^{-1}_{m2},Q_{m3}^{-1};q,t)
.
\end{align}
Our conjecture is therefore the following relation between two instanton partition functions
\begin{align}
Z_{\,dP_4}^{\,{\textrm{inst.}}}=
\frac{Z_{\,PdP_4^{II(b)}}^{\,{\textrm{inst.}}}}{M\left(\frac{uQ_F}{Q_{m1}Q_{m2}};q,t\right)\,
M\left(\frac{uQ_F}{Q_{m1}Q_{m3}};q,t\right)}.
\end{align}
In the next subsection
we verify an extended version of this equation in one-instanton order.

\subsection{The toric phases for $\boldsymbol{E_5}$ theory}

Since the local del Pezzo surface $dP_5$
is also non-toric,
we can not apply the refined topological vertex formalism directly to compute its partition function.
However,
we found three toric descriptions as pseudo del Pezzo surfaces.
In this subsection, we verify our conjecture that 
all these pseudo del Pezzo surfaces $PdP_4$
lead to the unique $dP_4$ partition function
after removing extra contribution.

\subsubsection*{Pseudo del Pezzo Phase I: $\boldsymbol{PdP_5^{I}}$}
\begin{figure}[tbp]
 \begin{center}
  \includegraphics[width=52mm, bb=0 0 261 243]{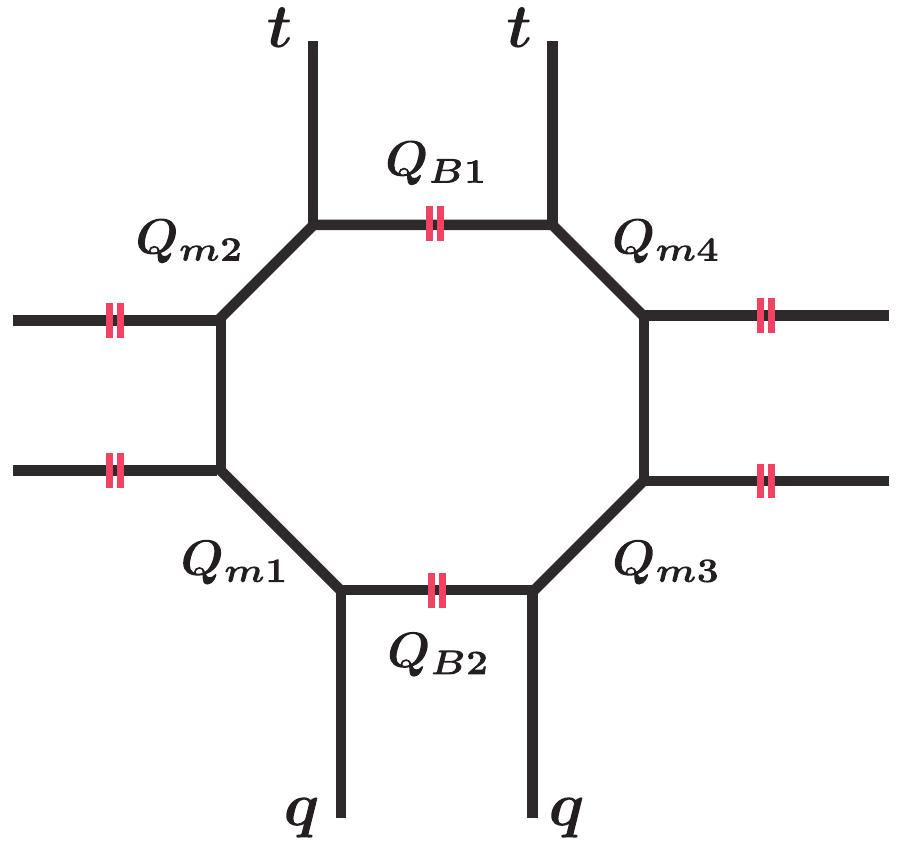}
 \end{center}
 \caption{The $PdP_5^{I}$ toric diagram.}
 \label{fig;PdP5I}
\end{figure}
%%%%%%%%%%%%%%
We start with the simplest phase $PdP_5^I$.
This  case was already studied in \cite{BMPTY,HKN},
and we review computation for readers convenience.

Using the refined topological vertex
gives the $PdP_5^I$ partition function 
\begin{align}
Z_{\,PdP_5^{I}}
=&\sum_{R_{1,2}}
(-Q_{B1})^{\vert R_1\vert}(-Q_{B2})^{\vert R_2\vert}
f_{R_1}(t,q)\,f^{-1}_{R_2}(t,q)
K^{\,[{\boldsymbol 0},{\boldsymbol{-1}},{\boldsymbol 0}]}_{\,(R_1,R_2)}
(Q_F,Q_{m1},Q_{m2};t,q)\,
\nonumber\\&
\rule{0pt}{3ex}\times
K_{\,(R_2^T,R_1^T)}^{\,[{\boldsymbol 0},{\boldsymbol{-1}},{\boldsymbol 0}]}(Q_F,Q_{m4},Q_{m3};q,t)
=Z_{\,PdP_5^{I}}^{\,{\textrm{pert.}}}\,
Z_{\,PdP_5^{I}}^{\,{\textrm{inst.}}},
\end{align}
where the instanton factor is given by
\begin{align}
\label{QBPdPI}
Q_{B1}=\frac{uQ_F}{Q_{m2}Q_{m4}},\quad Q_{B2}=\frac{uQ_F}{Q_{m1}Q_{m3}}.
\end{align}
The perturbative and instanton partition functions are then given by
\begin{align}
&Z_{\,PdP_5^{I}}^{\,{\textrm{pert.}}}=
M\left(
\frac{Q_F}{Q_{m1}Q_{m2}};t,q
\right)\,
M\left(
\frac{Q_F}{Q_{m3}Q_{m4}};q,t
\right)\,
Z_{\,\textrm{pert}}^{\,\textrm{vect}}(Q_F)\,
\prod_{f=1}^4
Z_{\,\textrm{pert}}^{\,\textrm{matter}}(Q_{mf}),\\
&Z_{\,PdP_5^{I}}^{\,{\textrm{inst.}}}\nonumber\\
&=\sum_{R_{1,2}}
\left(u\frac{q}{t} \right)^{\vert \vec{R}\vert}
\,Z_{\,\vec{R}}^{\,\textrm{vect}}(Q_F)\,
{Z}^{\textrm{matt.}}_{\vec{R}}
(Q_{m1})\,
Z^{\,\textrm{matt.}}_{\,\vec{R}}
(Q_{m3})\,
{Z'}^{\textrm{matt.}}_{\vec{R}}
(Q_{m2})\,
{Z'}^{\textrm{matt.}}_{\vec{R}}
(Q_{m4}).
\end{align}
We can see that this instanton partition function is an asymmetric function of the mass parameters $Q_{m1,\cdots,4}$.
Therefore, this partition function can not be that of $dP_5$ because $dP_5$ should have the symmetry with respect to the permutations
of the mass parameters associated with the $E_5$ symmetry.
This asymmetry is actually caused by the extra contribution coming from four stacks in $PdP_5^{I}$ geometry
\begin{align}
&Z^{\,{PdP}_5^I}_{\,\textrm{extra.}}=\nonumber\\&
M\left(
\frac{Q_F}{Q_{m1}Q_{m2}};t,q
\right)
M\left(
\frac{Q_F}{Q_{m3}Q_{m4}};q,t
\right)
M\left(
\frac{uQ_F}{Q_{m2}Q_{m4}};t,q
\right)
M\left(
\frac{uQ_F}{Q_{m1}Q_{m3}};q,t
\right),
\end{align}
and then we can expect that after the following renormalization of the instanton partition function
it becomes a symmetric function of the mass parameters if our conjecture is valid
\begin{align}
\label{renomPdPI}
Z_{\,{dP}_5}^{\,\textrm{inst.}}=
\frac{Z_{\,PdP_5^{I}}^{\,{\textrm{inst.}}}}
{M\left(
\frac{uQ_F}{Q_{m2}Q_{m4}};t,q
\right)\,
M\left(
\frac{uQ_F}{Q_{m1}Q_{m3}};q,t
\right)}.
\end{align}
In fact, we find that the one-instanton part of this renormalized partition function is
\begin{align}
&Z_{\,{dP}_5}^{\,1\textrm{-inst.}}=\nonumber\\
&
\label{dP5oneinst}
\rule{0pt}{3ex}
\frac{q}{t}
\frac{
\left(1+\frac{q}{t}\right)\left(
1+\sum_{f_1\neq f_2}\frac{Q_F}{Q_{mf_1}Q_{mf_2}}
+\frac{Q_F^2}{Q_{m1}Q_{m2}Q_{m3}Q_{m4}}
\right)
-\sqrt{\frac{q}{t}}
\sum_{f=1}^4\left(\frac{1}{Q_{mf}}
+\frac{Q_{mf}Q_F}{
\prod_{g=1}^4Q_{mg}
}\right)\left(1+Q_F\right)
}
{(1-q)(1-t^{-1})(1-Q_Ft^{-1}q)(1-Q_F^{-1}t^{-1}q)},
\end{align}
and it is manifestly symmetric.
This enhancement of symmetry is a non-trivial evidence of our conjecture (\ref{renomPdPI}).

\subsubsection*{Pseudo del Pezzo Phase II: $\boldsymbol{PdP_5^{II}}$}

\begin{figure}[tbp]
 \begin{center}
  \includegraphics[width=120mm, bb=0 0 611 296]{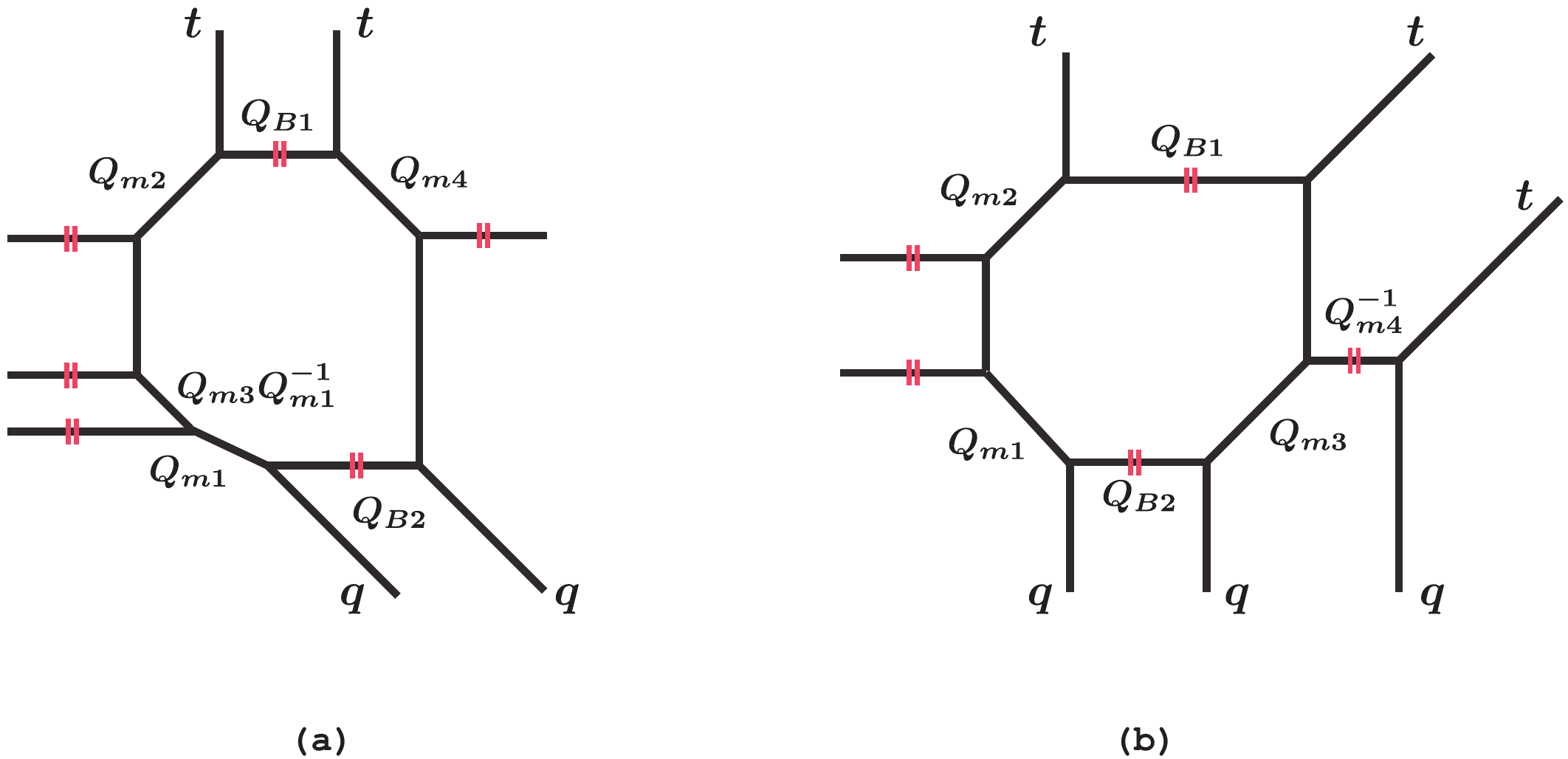}
 \end{center}
 \caption{The two choices of the preferred direction of the $PdP_5^{II}$ toric diagram.}
 \label{fig;PdP5II}
\end{figure}
There are two choices of the preferred direction of the 
web-diagram of $PdP_5^{II}$ as illustrated in \textbf{\textit{Figure}$\,$\textit{\ref{fig;PdP5II}}}.
It is easy to compute the partition function of the case {\bf (a)}
\begin{align}
Z_{\,PdP_5^{II(a)}}
=\sum_{R_{1,2}}&
(-Q_{B1})^{\vert R_1\vert}(-Q_{B2})^{\vert R_2\vert}
f_{R_1}(t,q)\,f^{-1}_{R_2}(t,q)\,
K^{\,[{\boldsymbol 0},{\boldsymbol{-1}},{\boldsymbol{-1}},{\boldsymbol 0}]}_{\,(R_1,R_2)}
(Q_F,Q_{m1},Q_{m3},Q_{m2};t,q)\nonumber\\&
\rule{0pt}{3ex}\times
K_{\,(R_2^T,R_1^T)}^{\,[{\boldsymbol 0},{\boldsymbol 0}]}(Q_F,Q_FQ^{-1}_{m4};q,t)
=Z_{\,PdP_5^{I}}^{\,{\textrm{pert.}}}\,
Z_{\,PdP_5^{I}}^{\,{\textrm{inst.}}}.
\end{align}
The instanton factor $u$ is given by the equations (\ref{QBPdPI}).
The partition function then takes the following form
\begin{align}
&Z_{\,PdP_5^{II(a)}}^{\,{\textrm{pert.}}}=
M\left(
\frac{Q_F}{Q_{m2}Q_{m3}},
\frac{Q_{m3}}{Q_{m1}}
;t,q
\right)\,
Z_{\,\textrm{pert}}^{\,\textrm{vect}}(Q_F)\,
\prod_{f=1}^4
Z_{\,\textrm{pert}}^{\,\textrm{matter}}(Q_{mf}),\\
&Z_{\,PdP_5^{II(a)}}^{\,{\textrm{inst.}}}
=Z_{\,PdP_5^{I}}^{\,{\textrm{inst.}}},
\end{align}
and our factorization conjecture (\ref{conjecturePdP5II}) actually holds for the extra contribution
\begin{align}
Z^{\,{PdP}_5^{II(a)}}_{\,\textrm{extra.}}=
M\left(
\frac{Q_F}{Q_{m2}Q_{m3}},
\frac{Q_{m3}}{Q_{m1}}
;t,q
\right)\,
M\left(
\frac{uQ_F}{Q_{m2}Q_{m4}};t,q
\right)\,
M\left(
\frac{uQ_F}{Q_{m1}Q_{m3}};q,t
\right).
\end{align}
This extra factor is actually associated with the three stacks of external legs in the $PdP_5^{II}$ web-diagram.

The partition function in the second case {\bf (b)} contains the $T^2$ sub-diagram
and its computation is more complicated.
Using the refined topological vertex gives
\begin{align}
Z_{\,PdP_5^{II(b)}}=\sum_{R_{1,2}}&
(-Q_{B1})^{\vert R_1\vert}(-Q_{B2})^{\vert R_2\vert}
\,f^{-1}_{R_2}(t,q)
K^{\,[{\boldsymbol 0},{\boldsymbol{-1}},{\boldsymbol 0}]}_{\,R_1,R_2}
(Q_F,Q_{m1},Q_{m2};t,q)\nonumber\\
&\times
K_{\,R_2^T,R_1^T}^{\,T^2}(Q_F,Q_{m3},Q_FQ_{m3}^{-1},Q_{m4}^{-1},q,t),
\end{align}
where the K\"ahler parameters of the base 2-cycles are given by
the instanton factor as
\begin{align}
Q_{B1}=\frac{uQ_F}{Q_{m2}},\quad Q_{B2}=\frac{uQ_F}{Q_{m1}Q_{m3}}.
\end{align}
We then find the following expression
\begin{align}
&Z_{\,PdP_5^{II(b)}}^{\,{\textrm{pert.}}}=
%M\left(\frac{Q_F}{Q_{m1}Q_{m2}};t,q\right)\,
%M\left(\frac{Q_F}{Q_{m3}Q_{m4}};t,q\right)\,
M\left(
\frac{Q_{m3}}{Q_{m4}};q,t
\right)
Z_{\,PdP_5^{I}}^{\,{\textrm{pert.}}},\\
&Z_{\,PdP_5^{II(b)}}^{\,{\textrm{inst.}}}
=\sum_{R_{1,2}}
\left(u\frac{q}{t} \right)^{\vert \vec{R}\vert}
\,
(Q_{m3})^{\vert R_2\vert}
\,f^{-1}_{R_2}(t,q)\,
Z_{\,\vec{R}}^{\,\textrm{vect}}(Q_F)\,{Z}^{\textrm{matt.}}_{\vec{R}}
(Q_{m1})\,
{Z'}^{\textrm{matt.}}_{\vec{R}}
(Q_{m2})
\nonumber\\
&\qquad\qquad\qquad\qquad\times
P_{\,R_2^T,R_1^T}(Q_{m3},Q_FQ_{m3}^{-1},Q_{m4}^{-1},q,t).
\end{align}
Using (\ref{Poly}), we obtain the one instanton part of this partition function
\begin{align}
Z_{\,{PdP}_5^{II(b)}}^{\,1\textrm{-inst.}}=
\frac{q}{t}
\frac{
N(Q_F,Q_{m1,2,3,4};t,q)
}
{(1-q)(1-t^{-1})(1-Q_Ft^{-1}q)(1-Q_F^{-1}t^{-1}q)}
,
\end{align}
where the numerator is
\begin{align}
&N(Q_F,Q_{m1,2,3,4};t,q)
=
\left(1+\frac{q}{t}\right)
\left(
1+\frac{Q_F}{Q_{m1}Q_{m2}}+\frac{Q_F}{Q_{m3}Q_{m4}}
+\frac{Q_F^2}{Q_{m1}Q_{m2}Q_{m3}Q_{m4}}
\right)\nonumber\\&
+
\left(\frac{1}{Q_{m3}}+\frac{1}{Q_{m4}}\right)\left(
\frac{q}{t}
\frac{Q_F}{Q_{m1}}
\left(-\frac{q}{t}+Q_F+1+\frac{1}{Q_F}\right)
+
\left(1+\frac{q}{t}\right)\frac{Q_F}{Q_{m2}}
\right)
\nonumber\\&
-
\sqrt{\frac{q}{t}}
\sum_{f=1}^4\left(\frac{1}{Q_{mf}}
+\frac{Q_{mf}Q_F}{
\prod_{g=1}^4Q_{mg}
}\right)\left(1+Q_F\right).
\end{align}
This partition function is not a symmetric function of $Q_{m1,2,3,4}$,
and it does not coincide with the $dP_5$ partition function (\ref{dP5oneinst}).
The discrepancy in the one-instanton level, however, takes the following simple form
\begin{align}
Z_{\,{dP}_5}^{\,1\textrm{-inst.}}-Z_{\,{PdP}_5^{II(b)}}^{\,1\textrm{-inst.}}
=-\frac{q
\left(\frac{Q_F}{Q_{m1}Q_{m3}}+\frac{Q_F}{Q_{m1}Q_{m4}}\right)
}{(1-q)(1-t)}.
\end{align}
This precisely corresponds to the one-instanton part of the extra contribution
 \begin{align}
Z^{\,{PdP}_5^{II(b)}}_{\,\textrm{extra.}}=
M\left(
\frac{Q_F}{Q_{m1}Q_{m2}};t,q
\right)\,
&M\left(
\frac{Q_F}{Q_{m3}Q_{m4}};t,q
\right)\,
M\left(
\frac{Q_{m3}}{Q_{m4}};q,t
\right)
\nonumber\\
&\times M\left(
\frac{uQ_F}{Q_{m1}Q_{m3}}
;q,t
\right)\,
M\left(
\frac{uQ_F}{Q_{m1}Q_{m4}};q,t
\right).
\end{align}
This computation is therefore the one-instanton check
of our conjecture (\ref{conjecturePdP5II}).

\subsubsection*{Pseudo del Pezzo Phase III: $\boldsymbol{PdP_5^{III}}$}

\begin{figure}[tbp]
 \begin{center}
  \includegraphics[width=120mm, bb=0 0 611 296]{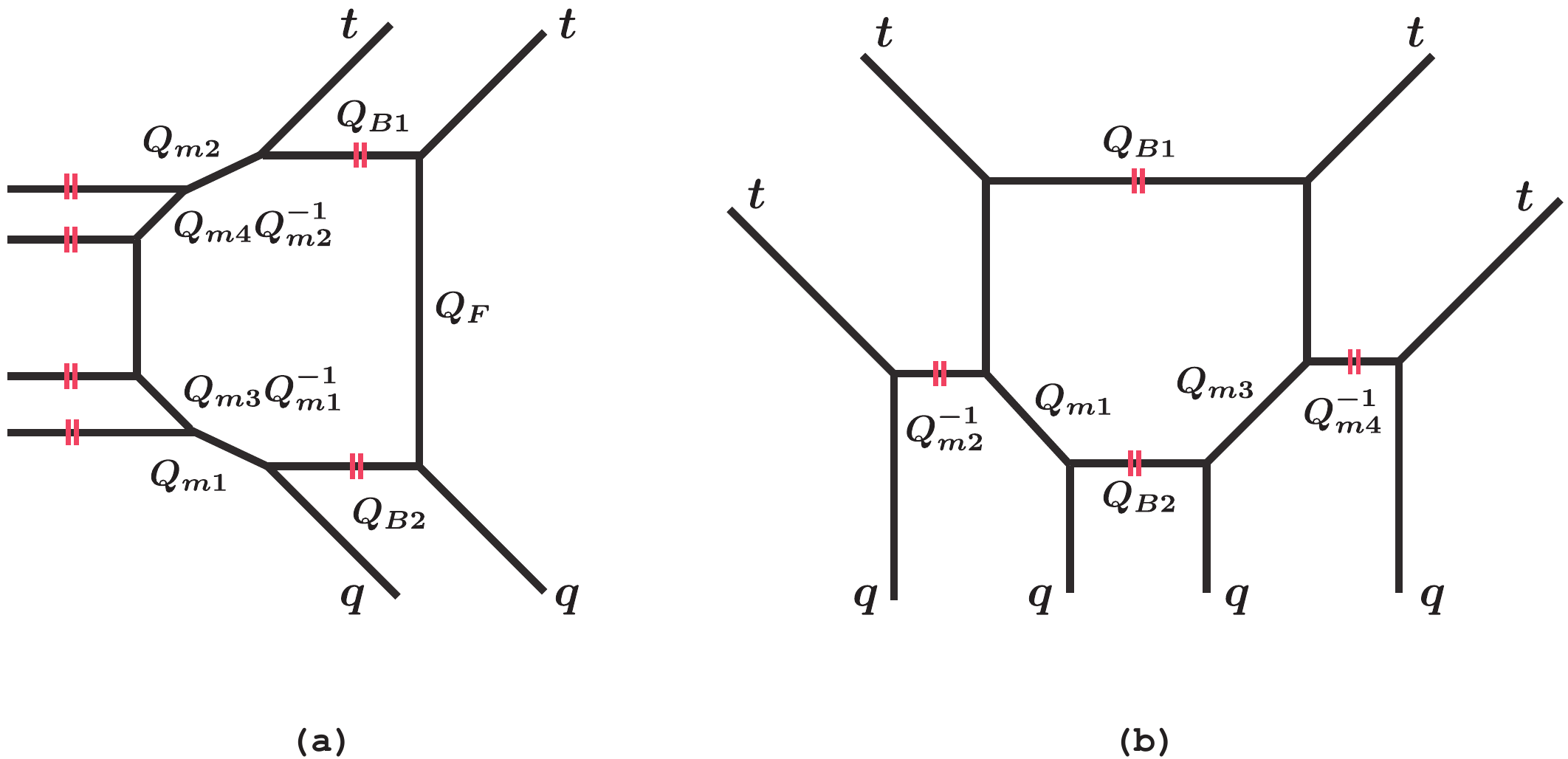}
 \end{center}
 \caption{The two choices of the preferred direction of the $PdP_5^{III}$ toric diagram.}
 \label{fig;PdP5III}
\end{figure}
The 
web-diagram of $PdP_5^{II}$ 
also has
two choices of the preferred direction as illustrated in \textbf{\textit{Figure}$\,$\textit{\ref{fig;PdP5III}}}.
It is easy to compute the partition function of the first case {\bf (a)}
\begin{align}
Z_{\,PdP_5^{III(a)}}
=&\sum_{R_{1,2}}
(-Q_{B1})^{\vert R_1\vert}(-Q_{B2})^{\vert R_2\vert}
f_{R_1}(t,q)\,f^{-1}_{R_2}(t,q)\,
\nonumber\\&
\rule{0pt}{3ex}\times
K^{\,[{\boldsymbol 0},{\boldsymbol{-1}},{\boldsymbol{-1}},{\boldsymbol{-1}},{\boldsymbol 0}]}_{\,(R_1,R_2)}
(Q_F,Q_{m1},Q_{m3},Q_{m4},Q_{m2};t,q)\,
K_{\,(R_2^T,R_1^T)}^{\,[{\boldsymbol 1}]}(Q_F;q,t)
.
\end{align}
Using (\ref{renomPdPI}) and some formulas in Appendix.B,
we can easily show 
\begin{align}
Z_{\,{dP}_5}^{\,\textrm{inst.}}=
\frac{Z_{\,PdP_5^{III(a)}}^{\,{\textrm{inst.}}}}
{M\left(
\frac{uQ_F}{Q_{m2}Q_{m4}};t,q
\right)\,
M\left(
\frac{uQ_F}{Q_{m1}Q_{m3}};q,t
\right)},
\end{align}
and this  partition function actually satisfies the relation (\ref{conjecturePdP5III}).
The extra contribution in this phase is given by
\begin{align}
&Z^{\,{PdP}_5^{III(a)}}_{\,\textrm{extra.}}\nonumber\\&=
M\left(
\frac{Q_{m4}}{Q_{m2}},
\frac{Q_F}{Q_{m3}Q_{m4}},
\frac{Q_{m3}}{Q_{m1}}
;t,q
\right)\,
M\left(
\frac{uQ_F}{Q_{m2}Q_{m4}};t,q
\right)\,
M\left(
\frac{uQ_F}{Q_{m1}Q_{m3}};q,t
\right).
\end{align}
This extra factor is actually associated with the three stacks of external legs in the $PdP_5^{III}$ web-diagram.

The second case {\bf (b)} of the third phase is the most non-trivial case in the pseudo fifth del Pezzo surfaces.
This toric geometry is decomposed into two $T^2$ geometries as  \textbf{\textit{Figure}$\,$\textit{\ref{fig;PdP5III}}}.
The topological string partition function is
\begin{align}
Z_{\,PdP_5^{III}}&=\sum_{R_{1,2}}
(-Q_{B1})^{\vert R_1\vert}(-Q_{B2})^{\vert R_2\vert}
f^{-1}_{R_1}(t,q)\,f^{-1}_{R_2}(t,q)\nonumber\\
&\times
K^{\,T^2}_{\,(R_1,R_2)}
(Q_F,Q_1,Q_2,Q_3,t,q)\,
K_{\,(R_2^T,R_1^T)}^{\,T^2}(Q_F,Q'_1,Q'_2,Q'_3,q,t),
\end{align}
where the gauge theory parameters are introduced by
\begin{align}
&Q_{B1}=uQ_F,\quad Q_{B2}=\frac{uQ_F}{Q_2Q'_1},\quad
Q_F=Q_1Q_2=Q'_1Q'_2,\\
&Q_{2}=Q_{m1},\quad Q_{3}=\frac{1}{Q_{m2}},\quad Q'_{1}=Q_{m3},\quad Q'_{3}=\frac{1}{Q_{m4}}.
\end{align}
Using (\ref{KT2}) gives the following expression
\begin{align}
Z_{\,PdP_5^{III(b)}}&=
M(Q_2Q_3;q,t)\,M(Q_1Q_3;t,q)\,M(Q'_2Q'_3;q,t)\,M(Q'_1Q'_3;t,q)\,
\nonumber\\
&\times
Z_{\,\textrm{pert}}^{\,\textrm{vect}}(Q_1Q_2)\,
\prod_{f=1}^4Z_{\,\textrm{pert}}^{\,\textrm{matter}}(Q_{mf})\,
\sum_{R_{1,2}}
\left(u\frac{q}{t} \right)^{\vert \vec{R}\vert}
\,Z_{\,\vec{R}}^{\,\textrm{vect}}(Q_1Q_2)\,
(Q_2Q'_1)^{-\vert R_2\vert}
\nonumber\\
&\times
f^{-2}_{R_2}(t,q)\,
P_{\,{R_1,R_2}}(Q_{1},Q_{2},Q_{3};t,q)
P_{\,{R_2^T,R_1^T}}(Q'_{1},Q'_{2},Q'_{3};q,t).
\end{align}
The first line $M(Q_2Q_3;q,t)\,M(Q_1Q_3;t,q)\,M(Q'_2Q'_3;q,t)\,M(Q'_1Q'_3;t,q)\equiv Z^{
\,PdP_5^{III}}_{\,{\textrm{extra}},\,{\textrm{pert.}}}$ 
of this equation is the extra factor that does not depend on the instanton factor $u$.
The remaining part of the extra contribution $Z^{\,PdP_5^{III(b)}}_{\,{\textrm{extra}}}=Z^{
\,PdP_5^{III(b)}}_{\,{\textrm{extra}},\,{\textrm{pert.}}}Z^{
\,PdP_5^{III(b)}}_{\,{\textrm{extra}},\,{\textrm{inst.}}}$ is the following function
\begin{align}
&Z^{
\,PdP_5^{III(b)}}_{\,{\textrm{extra}},\,{\textrm{inst.}}}\nonumber\\
&\equiv\label{extraPdP5IIa}
M\left(\frac{uQ_F}{Q_{m1}Q_{m3}};q,t\right)\,
M\left(\frac{uQ_F}{Q_{m2}Q_{m3}};q,t\right)\,
M\left(\frac{uQ_F}{Q_{m1}Q_{m4}};q,t\right)\,
M\left(\frac{uQ_F}{Q_{m2}Q_{m4}};q,t\right).
\end{align}
To check our conjecture (\ref{conjecturePdP5III}),
let us compute the one-instanton part of the $Z_{\,PdP_5^{III(b)}}$ partition function.
Using (\ref{Poly}), we obtain 
\begin{align}
Z_{\,{PdP}_5^{III(b)}}^{\,1\textrm{-inst.}}=
\frac{q}{t}
\frac{
N(Q_F,Q_{m1,2,3,4};t,q)
}
{(1-q)(1-t^{-1})(1-Q_Ft^{-1}q)(1-Q_F^{-1}t^{-1}q)}
,
\end{align}
where the numerator is
\begin{align}
&N(Q_F,Q_{m1,2,3,4};t,q)
=
\left(1+\frac{q}{t}\right)
\left(
1+\frac{Q_F}{Q_{m1}Q_{m2}}+\frac{Q_F}{Q_{m3}Q_{m4}}
+\frac{Q_F^2}{Q_{m1}Q_{m2}Q_{m3}Q_{m4}}
\right)\nonumber\\&
+
\frac{q}{t}
\left(-\frac{q}{t}+Q_F+1+\frac{1}{Q_F}\right)
\left(
\frac{Q_F}{Q_{m1}Q_{m3}}+\frac{Q_F}{Q_{m2}Q_{m3}}+\frac{Q_F}{Q_{m1}Q_{m4}}+\frac{Q_F}{Q_{m2}Q_{m4}}
\right)\nonumber\\&
-
\sqrt{\frac{q}{t}}
\sum_{f=1}^4\left(\frac{1}{Q_{mf}}
+\frac{Q_{mf}Q_F}{
\prod_{g=1}^4Q_{mg}
}\right)\left(1+Q_F\right).
\end{align}
This partition function is not a symmetric function of $Q_{m1,2,3,4}$.
We can, however, show that the difference between this partition function and the symmetric one
(\ref{dP5oneinst}) of $dP_5$ takes the following simple form
\begin{align}
Z_{\,{dP}_5}^{\,1\textrm{-inst.}}-Z_{\,{PdP}_5^{III(b)}}^{\,1\textrm{-inst.}}=
-\frac{q}{t}
\frac{
\frac{Q_F}{Q_{m1}Q_{m3}}+\frac{Q_F}{Q_{m2}Q_{m3}}+\frac{Q_F}{Q_{m1}Q_{m4}}+\frac{Q_F}{Q_{m2}Q_{m4}}
}
{(1-q)(1-t)}
.
\end{align}
This is precisely equal to the one-instanton part of the extra factor (\ref{extraPdP5IIa}).
The relation (\ref{conjecturePdP5III}) at the one-instanton level
is thus satisfied through nontrivial cancellation between two rational functions $Z_{\,{dP}_5,{PdP}_5^{III(b)}}^{\,1\textrm{-inst.}}$.
We expect such cancellation mechanism holds for all $k$-instanton partition functions.

%%%%%%%%%%%%%%%%%%%%%%%%%%%%%%%%%%%%%%%%%%%%%%%%%%%%%%%%
\section{Conclusion}

In this paper,
by extending the findings in \cite{BMPTY,HKN,Taki:2013vka},
we have found duality between 
$(p,q)$-web configurations
that lead to 5d field theories.
This duality enables us to compute the topological string partition function
of a (non-toric) local del Pezzo surface 
by employing a corresponding pseudo del Pezzo surface.
In general,  some pseudo del Pezzo surfaces are associated with a single del Pezzo surface.

There are sixteen inequivalent convex lattice polygons with single internal point,
and this means that there are sixteen inequivalent 5-brane web configurations
with single 5-brane loop.
There are seemingly sixteen  theories in 5d that arise from these configurations,
however, we showed that some of these 5d theories are dual and there are only eight 5d theories.
These eight theories are  associated with
the eight 5d SCFTs that have one dimensional Coulomb branch
and flavor symmetry whose rank is less than 7.
They were precisely the well-studied theories discovered by Seiberg in \cite{Seiberg:1996bd}.
This means that 
no new theory appears because of the non-tivial duality between the corresponding Calabi-Yau singularities.
This result provides the classification of the 5d SCFTs with one dimensional Coulomb branch
that are associated with 5-brane web configurations.
We can extend our discussion to web configurations with multiple 5-brane loops
and classify the 5d SCFTs with higher  dimensional Coulomb branch.

Our conjecture implies
new mathematical relations between 
5d $SU(2)$ Nekrasov partition functions.
At first glance,
two partition functions of different phases
take very different combinatorial forms.
Our conjecture, however, claims that 
the discrepancy between them can be collected into 
a simple prefactor.
In this paper, we check this statement based on  instanton expansion.
It should be possible to verify our conjectural relations rigorously
by employing and developing the mathematical theory of the Macdonald functions.

The 4d Seiberg duality observed in 
\cite{Feng:2000mi,Feng:2001xr,Hanany:2001py,Feng:2001bn,Feng:2002zw,Feng:2002kk,Franco:2002ae,Feng:2002fv,Franco:2002mu,Hanany:2012hi}
is deeply related to our duality.
In these papers,
the authors considered quiver gauge theories that were the world-volume theories
on D3-branes probing local del Pezzo singularities.
Picking up some examples,
they discussed  that 
two theories are Seilerg-dual to each other if the corresponding toric singularities
are related by 7-brane move for the corresponding 7-brane configurations.
They considered the duality acting on the world-volume of probe D3-branes,
but we can also consider the duality between the background singularities.
This relation between toric singularities leads to our duality between 5d field theories.
It would be interesting to study further relation between the 4d Seiberg-duality
and our 5d analogous duality.

It would be also interesting if we can fine clear relation to
the attempt at a non-toric extension of the topological vertex \cite{Diaconescu:2005tr,Diaconescu:2005mv}
and the study on the E-string partition functions \cite{Sakai:2012zq,Sakai:2012ik}.

Another unexplored line of research
is the relation to the AGT conjecture \cite{Alday:2009aq,Wyllard:2009hg,Gaiotto:2009ma,Taki:2009zd,Keller:2011ek}.  
The 5d version of the AGT conjecture \cite{Awata:2009ur,Mironov:2011dk,Nieri:2013yra,BMPTY} 
recasts the 5d Nekrasov partition functions
into the conformal blocks of the 2d $q$-deformed Toda field theories.
Our relation between the topological string partition functions of the dual toric phases
suggests that  some 2d descriptions are associated to a single 5d theory. 
It would be interesting if we can find the role of the extra factors in the 2d side.

%%%%%%%%%%%%%%%%%%%%%%%%%%%%%%%%%%%%%%%%%%%%%%%%%%%%%%%%%%%%%%%
\section*{Acknowledgments}
We are very grateful to Vladimir Mitev, Elli Pomoni and
Futoshi Yagi for fruitful discussions and suggestions during collaborations.

%%%%%%%%%%%%%%%%%%%%%%%%%%%%%%%%%%%%%%%%%%%%%%%%%%%%%%%%

%%%%%%%%%%%%%%%%%%%%%%%%%%%%%%%%%%%%%%%%%%%%%%%%%%%%%%%%
\newpage

\section*{Appendix A : 7-branes and $\boldsymbol{E_n}$ symmetry}

Recall that the $SL(2,\mathbb{Z})$ transformation of a D5-brane
leads to the $(p,q)$ 5-brane with generic NS-NS and  Ramond-Ramond charges.
Similarly, we can define the $[p,q]$ 7-brane
as the $SL(2,\mathbb{Z})$ transformation of a D7-brane.
The $(p,q)$ 5-brane then can terminate on the $[p,q]$ 7-brane.
The symbol ${\bf X}_{[p,q]}$ denotes the $[p,q]$ 7-brane.
Notice that ${\bf X}_{[-p,-q]}$ is equivalent to ${\bf X}_{[p,q]}$. 

We introduce the following three types of 7-branes for convenience sake
\begin{align}
{\bf A}\textrm{-brane}\,:\,{\bf X}_{[1,0]}={\bf A},\quad
{\bf B}\textrm{-brane}\,:\,{\bf X}_{[1,-1]}={\bf B},\quad
{\bf C}\textrm{-brane}\,:\,{\bf X}_{[1,1]}={\bf C}.
\end{align}
We  also define the symplectic inner product between $[p,q]$ charges
\begin{align}
z_i\equiv [p_i,q_i],\quad
z_i\wedge z_j
\equiv\det
\left(\begin{array}{cc}p_i\,\,\, &p_j\\ q_i \,\,\,& q_j\end{array}\right).
\end{align}
A 7-brane ${\bf X}_{[p,q]}$ creates a branch cut in the transversal plane,
and the monodromy matrix around it is given by
\begin{align}
K_{[p,q]}=
\left(\begin{array}{cc}1+pq & -p^2 \\q^2 & 1-pq\end{array}\right)=
1+zz^TS.
\end{align}

%%%%%%%%%%%%%%%%%%%%%%%%%%%%%%%%%%%%%%%%%
\subsection*{A.1$\quad$$\boldsymbol{SL(2,\mathbb{Z})}$ transformation}
\begin{figure}[tb]
\begin{center}
\includegraphics[width=7.5cm, bb=0 0 292 127]{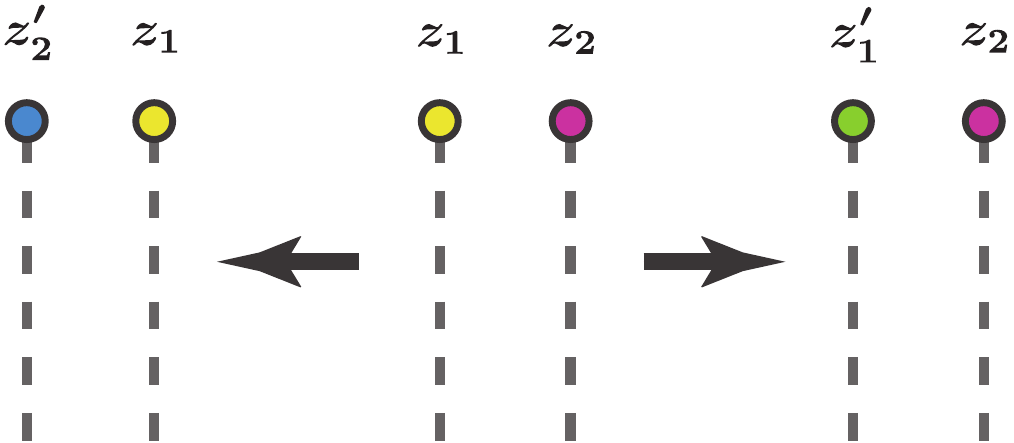}
\end{center}
\caption{
The branch cut move of two adjoining 7-branes ${\bf X}_{z_{1,2}}$.
By moving ${\bf X}_{z_{2}}$ across the branch cut of ${\bf X}_{z_{1}}$,
the 7-brane charge $z_2$ changes into $z'_2={z_2+(z_1\wedge z_2)z_1}$.
Moving ${\bf X}_{z_{1}}$ yields the 7-brane with charge $z'_1={z_1+(z_1\wedge z_2)z_2}$
on the right side.
 }
\label{fig;bcm}
\end{figure}

$SL(2,\mathbb{Z})$ is generated by the two generators $S$ and $U$, which are defined by
\begin{align}
S= \left(\begin{array}{cc}0\,\,\, &-1\\ 1 \,\,\,& 0\end{array}\right),\quad
T= \left(\begin{array}{cc}1\,\,\, &0\\ 1 \,\,\,& 1\end{array}\right),\quad
U=ST=
\end{align}

Type IIB superstring enjoys the $SL(2,\mathbb{Z})$ S-duality.
The action of the duality change the dilaton and the axion as
\begin{align}
SL(2,\mathbb{Z})\ni \left(\begin{array}{cc}a\,\,\, &b\\ c \,\,\,& d\end{array}\right) \,\,: \,\,
\tau\,\,\longmapsto\,\,\frac{a\tau+b}{c\tau+d}.
\end{align}
This  also acts on the $(p,q)$-charges and the 7-brane monodormies as
\begin{align}
SL(2,\mathbb{Z})\ni g \,\,: \,\,
 \left(\begin{array}{c}p\\ q\end{array}\right)
 \,\,\longmapsto\,\,
 g \left(\begin{array}{c}p\\ q\end{array}\right),\quad
 K_{[p,q]}
  \,\,\longmapsto\,\,
 g K_{[p,q]}g^{-1}.
\end{align}

%%%%%%%%%%%%%%%%%%%%%%%%%%%%%%%%%%%%%%%%%
\subsection*{A.2$\quad$7-brane move}
Let us consider a 7-brane configuration ${\bf X}_{z_1}{\bf X}_{z_2}$.
In our convention,
the branch cuts go downward.
We can consider two basic reordering procedure  \textbf{\textit{Figure}$\,$\textit{\ref{fig;bcm}}}.
When a 7-brane passes a branch cut,
its charge chances with obeying the following rule
\begin{align}
{\bf X}_{z_1}{\bf X}_{z_2}=
{\bf X}_{z_2+(z_1\wedge z_2)z_1}{\bf X}_{z_1}=
{\bf X}_{z_2}{\bf X}_{z_1+(z_1\wedge z_2)z_2}.
\end{align}

%%%%%%%%%%%%%%%%%%%%%%%%%%%%%%%%%%%%%%%%%
\subsection*{A.3$\quad$The collapsible 7-branes}
The total monodromy $K=K_{z_n}\cdots K_{z_2}K_{z_1}$ around a collapsible 7-brane configuration
satisfies the following condition
\begin{align}
\textrm{Tr}K=-2,-1,0,1,2.
\end{align}

%%%%%%%%%%%%%%%%%%%%%%%%%%%%%%%%%%%%%%%%%
%\subsection*{A.4$\quad$Kodaira classification and their affinization}

%%%%%%%%%%%%%%%%%%%%%%%%%%%%%%%%%%%%%%%%%
\section*{Appendix B : Nekrasov partition functions}

\subsection*{B.1$\quad$The refined topological vertex}

In this section, we collect definitions and conventions of the refined topological vertex formalism.
The vertex function is
\begin{align}
&\parbox{\bw}{\usebox{\boxv}}\hspace{-55mm}=C_{R_1R_2R_3}(t,q)\nonumber\\
&=(-1)^{\vert R_2\vert}
\,f_{R_2^T}(t,q)
\,q^{\frac{\parallel R_3\parallel^2}{2}}
\tilde{Z}_{R_2}(t,q)
\sum_Y
S_{R_1^T/Y}(t^{-\rho}q^{R_3^T})
S_{R_2/Y}(q^{-\rho}t^{R_3}).
\end{align}
The leg marked with the red double line
is the preferred direction.
The framing factor is defined by 
\begin{align}
f_{R}(t,q)=
(-1)^{\vert R\vert}
\,t^{\frac{\parallel R^T\parallel^2}{2}}
\,q^{-\frac{\parallel R\parallel^2}{2}},\quad
f_{R}(t,q)=\left( \frac{t}{q}\right)^{\frac{\vert R\vert}{2}}
f_{R}(t,q).
\end{align}
$\widetilde{Z}_R$ is the following specialized Macdonald function \cite{MacDonnaldSymmetric}
\begin{align}
\widetilde{Z}_R(t,q)=\prod_{i=1}^{d(R)}\prod_{j=1}^{R_i} (1-q^{R_i-j}t^{R^t_j-i+1})^{-1}.
\end{align}
See \cite{BMPTY} for the basic rules of the refined topological vertex formalism.

To compute topological string partition functions,
we need to calculate summations of symmetric functions over the Young diagrams.
The Cauchy formulas play a key role in this calculation 
\begin{align}
&\sum_{R}
S_{R/Y_1}(x)\,S_{R/Y_2}(y)
=\prod_{i,j}(1-x_iy_j)^{-1}\sum_{R}
S_{Y_1/R}(y)\,S_{Y_2/R}(x),\\
&\sum_{R}S_{R^T/Y_1}(x)\,S_{R/Y_2}(y)
=\prod_{i,j}(1+x_iy_j)\sum_{R}
S_{Y_1^T/R}(y)\,S_{Y_2^T/R^T}(x).
\end{align}

\subsection*{B.2$\quad$$SU(2)$ Nekrasov partition functions}

Let us introduce the following combinatorial factor 
\begin{align}
N_{R_\alpha R_\beta}(Q;t,q)
&=
\prod_{s\in R_\alpha}
\left( 1-Q\,t^{\ell_{R_\beta}(s)}q^{a_{R_\alpha}(s)+1} \right)
\prod_{t\in Y_\beta}
\left( 1-Q\,t^{-(\ell_{R_\alpha}(t)+1)}q^{-a_{R_\beta}(t)} \right)\nonumber\\
&=
\prod_{i,j=1}^\infty
\frac{1-Q\,t^{-R^t_{\alpha,j}+i-1}q^{-R_{\beta,i}+j}}{1-Q\,t^{i-1}q^{j}}.
\end{align}
The $SU(2)$ vector multiplet contribution to the instanton part of the Nekrasov partition functions is
\begin{align}
Z^{\,\textrm{vect.}}_{\,\vec{R}}
(Q_{21};t,q)
=\left(\frac{q}{t}\right)^{{\vert\vec{R}\vert}}
\frac{1}{\prod_{\alpha,\beta=1,2}N_{R_\alpha R_\beta}(Q_{\beta\alpha};t,q)},
\end{align}
where $Q_{\alpha\beta}=Q_\alpha Q_\beta^{-1}$, $Q_\alpha=e^{-Ra_\alpha}$ and $a_1=-a_2$.
The contribution of the Chern-Simons term with the effective level $m$ takes the following form \cite{Tachikawa:2004ur} 
\begin{align}
Z^{\,\textrm{CS},m}_{\,\vec{R}}(Q_{21};t,q)=
\prod_\alpha Q_\alpha^{-m\vert R_\alpha\vert}
t^{-m\frac{\parallel R_\alpha^T\parallel^2}{2}}q^{m\frac{\parallel R_\alpha\parallel^2}{2}}.
\end{align}

The (anti)fundamental matter contribution is
\begin{align}
&Z_{\,\vec{R}}^{\,\textrm{matt.}}(Q_{21},Q_m;t,q)
=\prod_{(i,j)\in R_1}
\left(1-\frac{Q_{21}}{Q_m}\,t^{-i+\frac{1}{2}}q^{j-\frac{1}{2}} \right)
\prod_{(i,j)\in R_2}\left(1-\frac{1}{Q_m} \,t^{-i+\frac{1}{2}}q^{j-\frac{1}{2}}\right),\\
&{Z'}_{\vec{R}}^{\textrm{matt.}}(Q_{21},Q_m;t,q)
=\prod_{(i,j)\in R_1}\left(1-\frac{1}{Q_m}\,t^{i-\frac{1}{2}}q^{-j+\frac{1}{2}} \right)
\prod_{(i,j)\in R_2}\left(1-\frac{Q_{21}}{Q_m} \,t^{i-\frac{1}{2}}q^{-j+\frac{1}{2}}\right).
\end{align}
Notice that they satisfy
\begin{align}
Z_{\,{R}_2^TR_1^T}^{\,\textrm{matt.}}(Q_{21},Q_m;q,t)
={Z'}_{{R}_1R_2}^{\textrm{matt.}}(Q_{21},{Q_m};t,q).
\end{align}

We can also write down the perturbative contributions to the Nekrasov partition function.
The vector multiplet contribution is
\begin{align}
Z_{\,\textrm{pert.}}^{\,\textrm{vect.}}(Q_{21};t,q)=\prod_{i,j=1}^\infty
\frac{1}{(1-Q_{21}t^{i-1}q^j)(1-Q_{21}t^{i}q^{j-1})}
,
\end{align}
and the (anti)fundamental matter contribution is
\begin{align}
Z_{\,\textrm{pert.}}^{\,\textrm{vect.}}(Q_{21},Q_m;t,q)=\prod_{i,j=1}^\infty
\left(1-Q_{m}t^{i-\frac{1}{2}}q^{j-\frac{1}{2}}\right)
\left(1-\frac{Q_{21}}{Q_m}t^{i-\frac{1}{2}}q^{j-\frac{1}{2}}\right)
.
\end{align}

The full Nekrasov partition function is then
\begin{align}
&Z(Q_{21},u,Q_m;t,q)\nonumber\\
&=\rule{0pt}{4ex}
Z_{\,\textrm{pert}}^{\,\textrm{vect.}}(Q_{21};t,q)
\prod_{\textrm{matters}}{Z}_{\,\textrm{pert}}^{\,\textrm{matt.}}(Q_{21},Q_m;t,q)\nonumber\\
&\times
\sum_{\vec{R}}\,
\left(u\frac{q}{t}\right)^{\vert\vec{R}\vert  }
Z^{\,\textrm{CS},m}_{\,\vec{R}}(Q_{21};t,q)
Z^{\,\textrm{vect.}}_{\,\vec{R}}
(Q_{21};t,q)
\prod_{\textrm{matters}}Z_{\,\vec{R}}^{\,\textrm{matt.}}(Q_{21},Q_m;t,q),
\end{align}
where $m$ is the Chern-Simons level of this theory.
The third line of this equation is the instanton partition function $Z^{\,\textrm{inst.}}$ of this theory,
and we introduce the following instanton expansion 
\begin{align}
Z(Q_{21},u,Q_m;t,q)=1+\sum_{k=1}^\infty
u^k\,Z^{\,k\textrm{-inst.}}(Q_{21},Q_m;t,q),
\end{align}
and $Z^{\,k\textrm{-inst.}}$ is the $k$-instanton partition function.
In this paper, we use the following symbol to parametrize the Coulomb branch parameter
\begin{align}
Q_F=Q_{21}.
\end{align}

\subsection*{B.3$\quad$Building blocks of Nekrasov partition functions}
\begin{figure}[tbp]
 \begin{center}
  \includegraphics[width=40mm, bb=0 0 146 164]{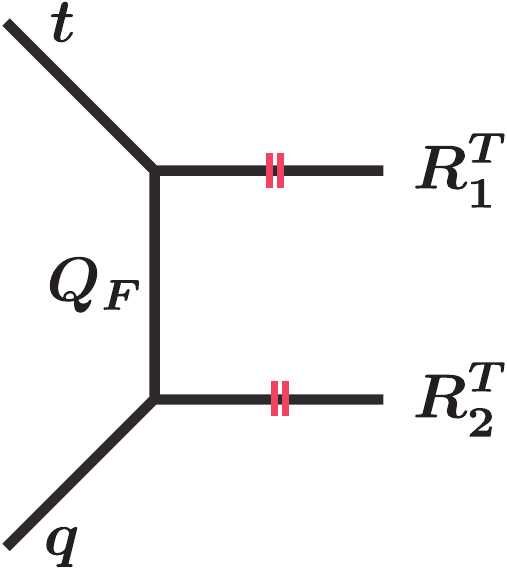}
 \end{center}
 \caption{A sub-diagram for $SU(2)$ geometry.
 We label this diagram by the framing number of the two-cycle $Q_F$ as $\,[{\boldsymbol{1}}]$.}
 \label{fig;sub1}
\end{figure}
In this subsection, we compute building blocks of the topological string partition functions 
for the toric phases of the local del Pezzo surfaces $dP_{1,2,\cdots,6}$.
\textbf{\textit{Figure}$\,$\textit{\ref{fig;sub1}}} is the half geometry which gives the $\widetilde{dP}_1$
web diagram.
The refined topological vertex on the geometry \textbf{\textit{Figure}$\,$\textit{\ref{fig;sub1}}}
gives the following partition function
\begin{align}
K_{R_1R_2}^{\,[{\boldsymbol{1}}]}
(Q_F;t,q)
&=\sum_{Y}(-Q_F)^{\vert Y\vert}\,\tilde{f}_Y(q,t)\,C_{\emptyset YR_1^T}(q,t)\,C_{Y^T\emptyset R_2^T}(q,t)\nonumber\\
&=\prod_{\alpha=1,2}\left(t^{\frac{\parallel R_a^T\parallel}{2}}\tilde{Z}_{R_a^T}(q,t)\right)
\,\frac{Z_{\,\textrm{pert.}}^{\,\textrm{vect.(L)}}(Q_F;t,q)}{N_{R_1R_2}(Q_F,t,q)},
\end{align}
where the perturbative part is defined by
\begin{align}
Z_{\,\textrm{pert}}^{\,\textrm{vect.(L)}}(Q_F;t,q)=\prod_{i,j=1}^\infty
(1-Q_Ft^{i-1}q^j)^{-1}.
\end{align}
The perturbative partition function of the vector multiplet is given by
\begin{align}
Z_{\,\textrm{pert.}}^{\,\textrm{vect.}}(Q_F;t,q)
=Z_{\,\textrm{pert.}}^{\,\textrm{vect.(L)}}(Q_F;t,q)\,
Z_{\,\textrm{pert.}}^{\,\textrm{vect.(L)}}(Q_F;q,t).
\end{align}
We can show the following identity
\begin{align}
&K_{R_1R_2}^{\,[{\boldsymbol{1}}]}(Q_F;t,q)
K_{R_2^TR_1^T}^{\,[{\boldsymbol{1}}]}(Q_F;q,t)\nonumber\\
&=Z_{\,\textrm{pert}}^{\,\textrm{vect.}}(Q_F;t,q)\left(-\frac{1}{Q_F}\frac{q}{t}\right)^{\vert\vec{R}\vert}\,
f_{R_1}(t,q)\,f_{R_2}^{-1}(t,q)\,
\prod_{\alpha,\beta=1,2}
\frac{1}
{N_{R_\alpha R_\beta}(Q_{\beta\alpha};t,q)},
\end{align}
which gives the $SU(2)$ vector multiplet contribution to the Nekrasov partition function.
In our convention, the Coulomb branch parameter is given by $Q_{21}=Q_F$.

\begin{figure}[tbp]
 \begin{center}
  \includegraphics[width=45mm, bb=0 0 180 181]{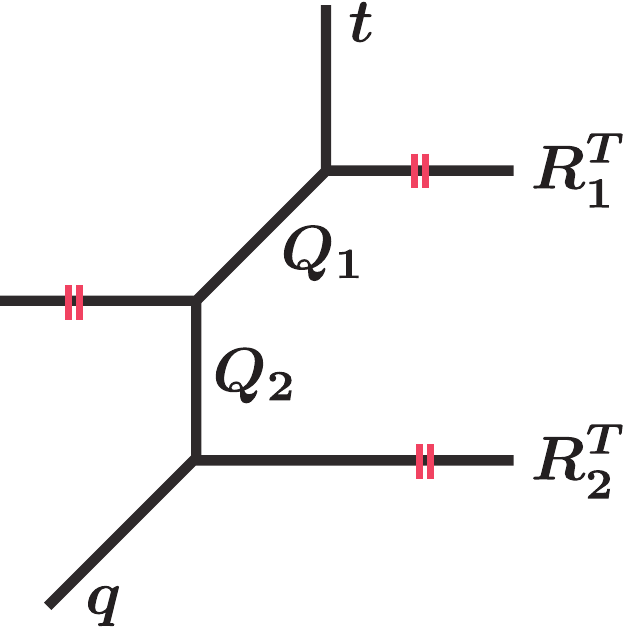}
 \end{center}
 \caption{A sub-diagram for $SU(2)$ geometry.
 We label this diagram by the framing numbers of the two-cycles as $\,[{\boldsymbol{0}},{\boldsymbol{0}}]$.}
 \label{fig;sub2}
\end{figure}
The sub-diagram \textbf{\textit{Figure}$\,$\textit{\ref{fig;sub2}}} is used to construct the $dP_2$ geometry for instance.
Using the refined topological vertex formalism gives
\begin{align}
&K_{R_1R_2}^{\,[{\boldsymbol{0}},{\boldsymbol{0}}]}
(Q_1Q_2,Q_1;t,q)
=\sum_{Y_{1,2}}\left(-{Q_1}\right)^{\vert Y_1\vert}\left(-{Q_2}\right)^{\vert Y_2\vert}\,
C_{ Y_1^T\emptyset R_1^T}(q,t)\,
C_{Y_1 Y_2^T\emptyset }(t,q)\,
C_{\emptyset Y_2 R_2^T}(q,t)\nonumber\\
&=K_{R_1R_2}^{\,[{\boldsymbol{1}}]}
(Q_1Q_2;t,q)
Z_{\,\textrm{pert.}}^{\,\textrm{matt.}}(Q_1Q_2,Q_1;t,q)
\,Q_1^{\vert R_1\vert}\,f_{R_1}^{-1}(t,q)
{Z'}_{\vec{R}}^{\textrm{matt.}}(Q_1Q_2,Q_1;t,q)
\nonumber\\
&=\rule{0pt}{4ex}K_{R_1R_2}^{\,[{\boldsymbol{1}}]}
(Q_1Q_2;t,q)
Z_{\,\textrm{pert.}}^{\,\textrm{matt.}}(Q_1Q_2,Q_2;t,q)
\,Q_2^{\vert R_1\vert}\,f_{R_2}(t,q)
Z_{\,\vec{R}}^{\,\textrm{matt.}}(Q_1Q_2,Q_2;t,q)
.
\end{align}
This local structure thus creates single matter multiplet whose mass is given by $Q_2$.

\begin{figure}[htbp]
 \begin{center}
  \includegraphics[width=40mm, bb=0 0 180 231]{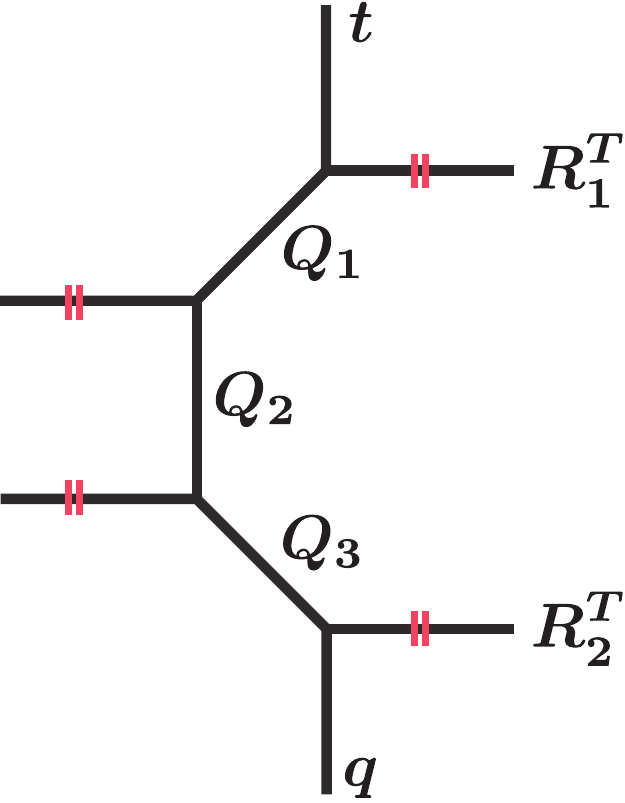}
 \end{center}
 \caption{The sub-diagram for $SU(2)$ geometry 
 whose  framing numbers of the two-cycles are $\,[{\boldsymbol{0}},{\boldsymbol{-1}},{\boldsymbol{0}}]$.}
 \label{fig;sub3}
\end{figure}
We come upon the geometry  \textbf{\textit{Figure}$\,$\textit{\ref{fig;sub3}}} 
in the computation of $PdP_3$ partition function. 
The refined topological vertex formalism leads to the following expression 
\begin{align}
&K_{R_1R_2}^{\,[{\boldsymbol{0}},{\boldsymbol{-1}},{\boldsymbol{0}}]}
(Q_F,Q_3,Q_1;t,q)\nonumber\\
&=\rule{0pt}{4ex}\sum_{Y_{1,2,3}}\left(-{Q_1}\right)^{\vert Y_1\vert}\left(-{Q_2}\right)^{\vert Y_2\vert}\left(-{Q_3}\right)^{\vert Y_3\vert}\,
\tilde{f}_{Y_2}(t,q)\nonumber\\
&\times
C_{ Y_1^T\emptyset R_1^T}(q,t)\,
C_{ Y_1Y_2^T \emptyset}(t,q)\,
C_{ Y_2Y_3^T \emptyset}(t,q)\,
C_{\emptyset Y_3 R_2^T}(q,t)\nonumber\\
&=\rule{0pt}{4ex}
M(Q_2;t,q)
K_{R_1R_2}^{\,[{\boldsymbol{1}}]}
(Q_F;t,q)\,
Z_{\,\textrm{pert.}}^{\,\textrm{matt.}}(Q_F,Q_1;t,q)\,
Z_{\,\textrm{pert.}}^{\,\textrm{matt.}}(Q_F,Q_3;t,q)\nonumber\\
&\rule{0pt}{4ex}\times Q_1^{\vert R_1\vert} Q_3^{\vert R_2\vert}
f_{R_1}^{-1}(t,q)f_{R_2}(t,q)
{Z'}_{\vec{R}}^{\textrm{matt.}}(Q_F,Q_1;t,q)
Z_{\,\vec{R}}^{\,\textrm{matt.}}(Q_F,Q_3;t,q)
,
\end{align}
where we introduce $Q_F=Q_1Q_2Q_3$.
We can see that this local structure gives two matter multiplets.

\begin{figure}[tbp]
 \begin{center}
  \includegraphics[width=48mm, bb=0 0 218 222]{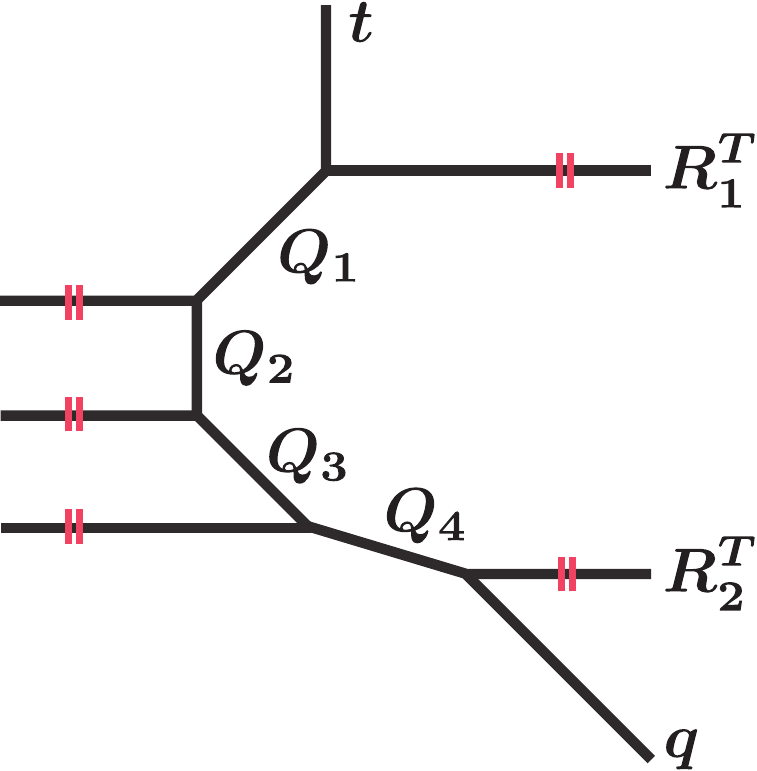}
 \end{center}
 \caption{A sub-diagram for $SU(2)$ geometry whose  framing numbers of the two-cycles are 
 $\,[{\boldsymbol{0}},{\boldsymbol{-1}},{\boldsymbol{-1}},{\boldsymbol{0}}]$.}
 \label{fig;sub5}
\end{figure}

The geometry \textbf{\textit{Figure}$\,$\textit{\ref{fig;sub5}}} is
used to compute $PdP_4$ partition function.
By using the topological vertex, we obtain
\begin{align}
&K_{R_1R_2}^{\,[{\boldsymbol{0}},{\boldsymbol{-1}},{\boldsymbol{-1}},{\boldsymbol{0}}]}
(Q_F,Q_5,Q_4Q_5,Q_1,Q_1Q_2;t,q)\nonumber\\
&=\rule{0pt}{4ex}\sum_{Y_{1,2,3,4}}\left(-{Q_1}\right)^{\vert Y_1\vert}\left(-{Q_2}\right)^{\vert Y_2\vert}\left(-{Q_3}\right)^{\vert Y_3\vert}
\left(-{Q_4}\right)^{\vert Y_4\vert}\,
\tilde{f}_{Y_2}(t,q)\tilde{f}_{Y_3}(t,q)\nonumber\\
&\times
C_{ Y_1^T\emptyset R_1^T}(q,t)\,
C_{ Y_1Y_2^T \emptyset}(t,q)\,
C_{ Y_2Y_3^T \emptyset}(t,q)\,
C_{ Y_3Y_4^T \emptyset}(t,q)\,
C_{\emptyset Y_4 R_2^T}(q,t).
\end{align}
With some algebra, we find the following expression
%%%%%%%% 
\begin{align}
&K_{R_1R_2}^{\,[{\boldsymbol{0}},{\boldsymbol{-1}},{\boldsymbol{-1}},{\boldsymbol{0}}]}
(Q_F,Q_5,Q_4Q_5,Q_1,Q_1Q_2;t,q)=
M(Q_2,Q_3;t,q)
\,
\prod_{Q_m=Q_{1,4},Q_3Q_4}
Z_{\,\textrm{pert.}}^{\,\textrm{matt.}}(Q_F,Q_m;t,q)
\nonumber\\
&%\rule{0pt}{4ex}
\times 
Q_1^{\vert R_1\vert} (Q_3Q_4^2)^{\vert R_2\vert}
f_{R_1}^{-1}(t,q)f^2_{R_2}(t,q)\,
K_{R_1R_2}^{\,[{\boldsymbol{1}}]}
(Q_F;t,q)\nonumber\\
&\rule{0pt}{4ex}\times
{Z'}_{\vec{R}}^{\textrm{matt.}}(Q_F,Q_1;t,q)
Z_{\,\vec{R}}^{\,\textrm{matt.}}(Q_F,Q_3Q_4;t,q)
Z_{\,\vec{R}}^{\,\textrm{matt.}}(Q_F,Q_4;t,q)
,
\end{align}
where we introduce $Q_F=Q_1Q_2Q_3Q_4$.
Three matter multiplets are associated with this local structure.

\begin{figure}[tbp]
 \begin{center}
  \includegraphics[width=48mm, bb=0 0 240 424]{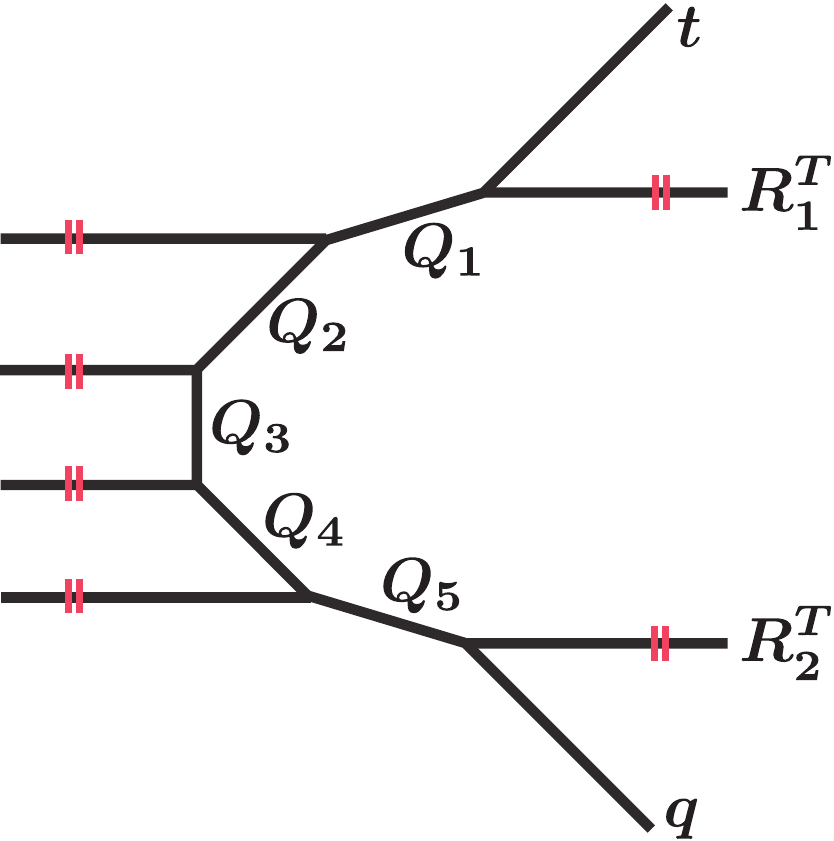}
 \end{center}
 \caption{A sub-diagram for $SU(2)$ geometry whose  framing numbers of the two-cycles are 
 $\,[{\boldsymbol{0}},{\boldsymbol{-1}},{\boldsymbol{-1}},{\boldsymbol{-1}},{\boldsymbol{0}}]$.}
 \label{fig;sub6}
\end{figure}

The $PdP_5^{III}$ diagram involves the sub-diagram \textbf{\textit{Figure}$\,$\textit{\ref{fig;sub6}}}.
The refined topological vertex gives
\begin{align}
&K_{R_1R_2}^{\,[{\boldsymbol{0}},{\boldsymbol{-1}},{\boldsymbol{-1}},{\boldsymbol{-1}},{\boldsymbol{0}}]}
(Q_F,Q_4,Q_4Q_3,Q_1,Q_1Q_2;t,q)\nonumber\\
&=\rule{0pt}{4ex}\sum_{Y_{1,2,3,4,5}}\left(-{Q_1}\right)^{\vert Y_1\vert}\left(-{Q_2}\right)^{\vert Y_2\vert}\left(-{Q_3}\right)^{\vert Y_3\vert}
\left(-{Q_4}\right)^{\vert Y_4\vert}\left(-{Q_5}\right)^{\vert Y_5\vert}\,
\tilde{f}_{Y_2}(t,q)\tilde{f}_{Y_3}(t,q)\tilde{f}_{Y_4}(t,q)\nonumber\\
&\times
C_{ Y_1^T\emptyset R_1^T}(q,t)\,
C_{ Y_1Y_2^T \emptyset}(t,q)\,
C_{ Y_2Y_3^T \emptyset}(t,q)\,
C_{ Y_3Y_4^T \emptyset}(t,q)\,
C_{ Y_4Y_5^T \emptyset}(t,q)\,
C_{\emptyset Y_5 R_2^T}(q,t).
\end{align}
Using the Cauchy formulas, we obtain the following expression
\begin{align}
&K_{R_1R_2}^{\,[{\boldsymbol{0}},{\boldsymbol{-1}},{\boldsymbol{-1}},{\boldsymbol{-1}},{\boldsymbol{0}}]}
%(Q_F,Q_4,Q_4Q_3,Q_1,Q_1Q_2;t,q)
=
M(Q_2,Q_3,Q_4;t,q)\,
\prod_{Q_m=Q_{1,5},Q_1Q_2,Q_4Q_5}
Z_{\,\textrm{pert.}}^{\,\textrm{matt.}}(Q_F,Q_m;t,q)
\nonumber\\
&\rule{0pt}{4ex}\times
(Q_1^2Q_2)^{\vert R_1\vert} (Q_4Q_5^2)^{\vert R_2\vert}
f_{R_1}^{-2}(t,q)f^2_{R_2}(t,q)
K_{R_1R_2}^{\,[{\boldsymbol{1}}]}
(Q_F;t,q)\,
{Z'}_{\vec{R}}^{\textrm{matt.}}(Q_F,Q_1;t,q)
\nonumber\\
&\rule{0pt}{4ex}\times 
{Z'}_{\vec{R}}^{\textrm{matt.}}(Q_F,Q_1Q_2;t,q)
Z_{\,\vec{R}}^{\,\textrm{matt.}}(Q_F,Q_4Q_5;t,q)
Z_{\,\vec{R}}^{\,\textrm{matt.}}(Q_F,Q_5;t,q)
,
\end{align}
where we introduce $Q_F=Q_1Q_2Q_3Q_4Q_5$.
This local structure of a Calabi-Yau manifold leads to two fundamental matter multiplets 
and two anti-fundamental matter multiplets.
We can therefore use this diagram to compute the partition function for $E_5$ SCFT.

\subsubsection*{$T^2$ subdiagram}
In contrast,
we can not give a closed expression of the $T^2$ sub-diagram \textbf{\textit{Figure}$\,$\textit{\ref{fig;T2}}}.
This is because the above cases are strip geometries \cite{Iqbal:2004ne,Taki:2007dh} whose
partition functions are given by recursive applications of the Cauchy formulas.
Using the topological vertex formalism yields the expression
\begin{align}
&K^{\,T^2}_{\,\vec{R}}
(Q_1,Q_2,Q_3,t,q)\nonumber\\&
=\frac{Z_{\,T^2}(Q_1,Q_2,Q_3,t,q)}{M(Q_1Q_2;q,t)}
{K_{\,\vec{R}}^{\,[{\boldsymbol 1}]}(Q_1Q_2,t,q)}
P_{\,\vec{R}}(Q_1,Q_2,Q_3,t,q),
\end{align}
where
\begin{align}
&P_{\,\vec{R}}(Q_1,Q_2,Q_3,t,q)=
\frac{
\prod_{i,j=1}^\infty
\left(1-Q_2Q_3t^{i-1}q^{j}\right)
\left(1-Q_1Q_3t^{i}q^{j-1}\right)}{\prod_{i,j=1}^\infty
\left(1-Q_1Q_2Q_3t^{i-\frac{1}{2}}q^{j-\frac{1}{2}}\right)
\left(1-Q_{3}t^{i-\frac{1}{2}}q^{j-\frac{1}{2}}\right)}\nonumber\\
&\times\rule{0pt}{3ex}
\sum_{Y}(-Q_3)^{\vert Y\vert}\,t^{\frac{\parallel Y^T\parallel^2}{2}}
q^{\frac{\parallel Y\parallel^2}{2}}
\widetilde{Z}_{Y}(t,q)\,
\widetilde{Z}_{Y^T}(q,t)
\nonumber\\
&\rule{0pt}{3ex}\times\prod_{s\in Y}
\left(1-Q_1t^{-\ell_{R_1}-\frac{1}{2}}q^{-a_Y-\frac{1}{2}}\right)
\left(1-Q_2t^{\ell_{R_2}+\frac{1}{2}}q^{a_Y+\frac{1}{2}}\right)\nonumber\\
&
\times
\prod_{s\in R_1}
\left(1-Q_1t^{\ell_Y+\frac{1}{2}}q^{a_{R_1}+\frac{1}{2}}\right)
\prod_{s\in R_2}
\left(1-Q_2t^{-\ell_Y-\frac{1}{2}}q^{-a_{R_2}-\frac{1}{2}}\right).
\end{align}
We can observe that this function is a polynomial in $Q_3$ despite its appearances.
It would be interesting to prove this observation.
Notice that this function satisfies
\begin{align}
P_{R_1,R_2}(Q_1,Q_2,Q_3,t,q)
=P_{R_2,R_1}(Q_2,Q_1,Q_3,t^{-1},q^{-1}).
\end{align}
%%%%%%%%%%%%%%%%%%%%%%%%%%%%%%%%%%%%%%%%%%%%%%%%%%%%%%%%%%%%

%%%%%%%%%%%%%%%%%%%%%%%%%%%%%%%%%%%%%%%%%%%%%%%%%%%%%%%%%%%%

\end{document}